\documentclass[journal,draftclsnofoot,onecolumn,12pt]{IEEEtran}
\usepackage{amsthm,amssymb,graphicx,multirow,amsmath,color,amsfonts,cuted}%,ulem}
\usepackage[caption=false,font=footnotesize]{subfig}
\usepackage[update,prepend]{epstopdf}
\usepackage[noadjust]{cite}
\usepackage[latin1]{inputenc}
\usepackage{tikz}
\usepackage{bbm} % for \mathbbm{1}
\usepackage{pdfpages}
\usepackage{multirow}
\usepackage{tabulary}
\usepackage{comment}
\usepackage{algorithm}
\usepackage{url}
\usepackage{algpseudocode}
\def\BSTATE{\STATE\hskip-\ALG@thistlm}
\makeatother

\def\nbc{{\mathbf{c}}}

\def\nbe{{\mathbf{e}}}

\def\nbi{{\mathbf{i}}}

\def\nbm{{\mathbf{m}}}

\def\nbs{{\mathbf{s}}}

\def\nbv{{\mathbf{v}}}

\def\nbx{{\mathbf{x}}}

\def\nbz{{\mathbf{z}}}
\def\nb0{{\mathbf{0}}}
\def\nb1{{\mathbf{1}}}

\def\nbA{{\mathbf{A}}}
\def\nbB{{\mathbf{B}}}

\def\nbD{{\mathbf{D}}}

\def\nbI{{\mathbf{I}}}

\def\nbP{{\mathbf{P}}}

\def\nbV{{\mathbf{V}}}

\def\nbX{{\mathbf{X}}}

\def\ncalC{{\mathcal{C}}}

\def\ncalL{{\mathcal{L}}}

\def\ncalQ{{\mathcal{Q}}}
\def\ncalR{{\mathcal{R}}}
\def\ncalS{{\mathcal{S}}}

\def\ncalV{{\mathcal{V}}}

\def\nbbE{{\mathbb{E}}}

\def\nbbP{{\mathbb{P}}}

\def\nbbR{{\mathbb{R}}}

\def\nrmK{{\rm K}}
\def\nrmP{{\rm P}}
\def\nrmZ{{\rm Z}}
\def\nrmX{{\rm X}}
\def\nrmM{{\rm M}}
\def\nrmI{{\rm I}}
\newtheorem{lemma}{Lemma}

\newtheorem{nrem}{Remark}
\newtheorem{theorem}{Theorem}
\newtheorem{prop}{Proposition}
\newtheorem{cor}{Corollary}

\def\a{\overset{({\rm a})}{=}}
\def\b{\overset{({\rm b})}{=}}
\def\c{\overset{({\rm c})}{=}}
\def\d{\overset{({\rm d})}{=}}

\allowdisplaybreaks % Allows breaking of eqnarray over multiple pages (avoids unnecessary blanks in the document before eqnarray)

\begin{document}
%\pagenumbering{gobble}
\graphicspath{{./Figures/}}
\title{
%A Stochastic Hybrid Systems Framework for Characterizing 
Joint Distribution of Ages of Information in Networks
}
\author{
Mohamed A. Abd-Elmagid and Harpreet S. Dhillon
\thanks{The authors are with Wireless@VT, Bradley Department of Electrical and Computer Engineering, Virginia Tech, Blacksburg, VA. Email: \{maelaziz,\ hdhillon\}@vt.edu. The support of the U.S. NSF (Grant CNS-1814477) is gratefully acknowledged. This work will be presented in part at the IEEE/IFIP WiOpt 2022 \cite{wiopt22_joint}. 
%This paper was presented in part at the IEEE Globecom, 2019 \cite{AbdElmagid2019Globecom_a}. 
%\hfill Manuscript updated: \today.
}
%\vspace{-5mm}
}

\maketitle

\begin{abstract}
We study a general setting of status updating systems in which a set of source nodes provide status updates about some physical process(es) to a set of monitors. The freshness of information available at each monitor is quantified in terms of the Age of Information (AoI), and the vector of AoI processes at the monitors (or equivalently the \textit{age vector}) models the \textit{continuous} state of the system. While the \textit{marginal} distributional properties of each AoI process have been studied for a variety of settings using the stochastic hybrid system (SHS) approach, we lack a counterpart of this approach to systematically study their \textit{joint} distributional properties. Developing such a framework is the main contribution of this paper. In particular, we model the discrete state of the system as a finite-state continuous-time Markov chain, and describe the coupled evolution of the continuous and discrete states of the system by a piecewise linear SHS with linear reset maps. Using the notion of tensors, we first derive first-order linear differential equations for the temporal evolution of both the joint moments and the joint moment generating function (MGF) for an arbitrary set of age processes. We then characterize the conditions under which the derived differential equations are asymptotically stable. The generality of our framework is demonstrated by recovering several existing results as its special cases. Finally, we apply our framework to derive closed-form expressions of the stationary joint MGF in a multi-source updating system under non-preemptive and source-agnostic/source-aware preemptive in service queueing disciplines.
\end{abstract}

\section{Introduction} \label{sec:intro}
The ongoing massive deployment of the Internet of Things (IoT) will enable many critical real-time status updating systems that fundamentally rely on the timely delivery of status updates \cite{abd2018role}. The authors of \cite{kaul2012real} introduced the concept of AoI which provides a rigorous way of quantifying the freshness of information at a {\it destination node} as a result of receiving status updates over time from a {\it transmitter node}. In particular, for a single-source queueing-theoretic model in which status updates are generated randomly at a transmitter with a single source of information and a single server, the AoI at the destination was defined in \cite{kaul2012real} as the following random process: $x(t) = t - u(t)$, where $u(t)$ is the generation time instant of the latest status update received at the destination by time $t$. A key assumption in the analysis of \cite{kaul2012real} was the ergodicity of the AoI process. This allowed the authors to derive the stationary average value of AoI under first-come-first-served (FCFS) queueing discipline by leveraging the properties of its sample functions and applying appropriate geometric arguments. Although this geometric approach has been considered in a series of subsequent prior works to analyze the marginal distributional properties of AoI for adaptations of the queueing model studied in \cite{kaul2012real}, it often requires tedious calculations of joint moments. Thus, it is very challenging (if not intractable) to use such geometric arguments in the AoI analysis of more sophisticated queueing models/disciplines including the ones that allow preemption between the status updates in service/waiting. Motivated by this, the authors of \cite{yates2018age} and \cite{yates2020age} have developed an SHS-based framework (building on \cite{hespanha2006modelling}) for characterizing the marginal distributional properties of each AoI process in a network with multiple AoI processes. The results of \cite{yates2018age} and \cite{yates2020age} have then been applied to characterize the marginal distributional properties of AoI under a variety of queueing disciplines. On the other hand, a systematic approach to the joint analysis of an arbitrary set of AoI processes in a network is an open problem. In this paper, we develop an SHS-based general framework to facilitate the analysis of the joint distributional properties of an arbitrary set of AoI processes in a network through the characterization of their stationary joint moments and MGFs. Therefore, this paper can be thought of as the joint distributional counterpart of \cite{yates2018age} and \cite{yates2020age}. We demonstrate the generality of our framework by recovering several existing results as its special cases.
% clarify that almost all the literature did not study the joint analysis of AoIs in a network. Clarify that our results are unified/general in the sense that the results in Yates' papers hold as special cases.
\subsection{Related Work}\label{sub:Rework}
The relevant literature to this paper can be categorized into the following three categories: i) prior analyses of AoI applying the geometric approach, ii) prior analyses of AoI applying the SHS approach, and iii) prior analysis of the joint distributional properties of AoI processes. Each of these categories is discussed next.

{\it Geometric approach to the AoI analysis.} Following \cite{kaul2012real}, the geometric approach has been widely adopted to analyze AoI or peak AoI (an AoI-related metric introduced in \cite{costa2016age} to capture the peak values of AoI over time) in a series of subsequent prior works \cite{kaul2012status,soysal,costa2016age,chen2016age,kam2018age,kavitha2018controlling,zou2019waiting,Inoue19,kosta2020non,Champati19,Chiariotti_dist,ayan2020probability,olga20,yates2012real,Moltafet_multisource,pappas2015age,yates2017status,kosta2019age,najm2018status,huang2015optimizing,Xu_21,Akar21,Ozancan_2021}. In particular, the average of AoI or peak AoI in single-source systems was characterized under several queueing disciplines in \cite{kaul2012status,soysal,costa2016age,chen2016age,kam2018age,kavitha2018controlling,zou2019waiting}. Further, a handful of recent works aimed to characterize the distribution (or some distributional properties) of AoI/peak AoI in single-source systems \cite{Inoue19,kosta2020non,Champati19,Chiariotti_dist,ayan2020probability,olga20}. On the other hand, the AoI analysis in multi-source systems is quite challenging, and hence the prior work in this direction is relatively sparse \cite{yates2012real,Moltafet_multisource,pappas2015age,yates2017status,kosta2019age,najm2018status,huang2015optimizing,Xu_21,Akar21,Ozancan_2021}. Note that a multi-source system refers to the setup where a transmitter has multiple sources generating status updates about multiple physical processes. For the multi-source systems, the average AoI was characterized for the M/M/1 FCFS queueing model in \cite{yates2012real}, the M/G/1 FCFS queueing model in \cite{Moltafet_multisource}, and the M/M/1 FCFS with preemption in waiting queueing model (where the transmitter has a buffer that only keeps the latest generated status update from each source) in \cite{pappas2015age}. The authors of \cite{yates2017status} and \cite{kosta2019age} analyzed the average AoI under scheduled and random multiaccess strategies for delivering the status updates generated from different sources at the transmitter. The average peak AoI was derived for the M/G/1 last-come-first-served (LCFS) queueing model with (without) preemption in service in \cite{najm2018status} (in \cite{huang2015optimizing}), and for the priority FCFS and LCFS queueing models (where the sources of information are prioritized at the transmitter) in \cite{Xu_21}. Further, the distributions of AoI and PAoI were numerically characterized for various discrete time queues in \cite{Akar21}, and for a probabilistically preemptive queueing model in \cite{Ozancan_2021} where a new arriving status update preempts the one in service with some probability.
%(depends on the indices of the sources that the two packets belong to).
Note that the analyses of the above works studying multi-source system settings (i.e., there are multiple AoI or age processes in the system)
have been limited to the characterization of the marginal distributional properties of the AoI process of each source.

{\it SHS approach to the AoI analysis.} The SHS approach has been applied to characterize the marginal distributional properties of AoI under a variety of system settings/queueing disciplines \cite{SHS_2,SHS_3,SHS_1,SHS_4,SHS_5,SHS_6,SHS_7,abdelmagid_2021a,abd2021distributional,SHS_8,abdelmagid_2021b,abd2022dist_ICC,abd2022dist_INFOCOM}. In particular, the average AoI was characterized for single-source systems in \cite{SHS_2,SHS_3} and multi-source systems in \cite{SHS_1,SHS_4,SHS_5,SHS_6,SHS_7}, whereas the MGF of AoI was derived for single-source systems in \cite{abdelmagid_2021a,abd2021distributional}, two-source systems in \cite{SHS_8}, and multi-source systems in \cite{abdelmagid_2021b,abd2022dist_ICC,abd2022dist_INFOCOM}. The authors of \cite{SHS_2} derived the average AoI under the LCFS with preemption in service queueing discipline when the transmitter contains multiple parallel servers. Further, the authors of \cite{SHS_3} derived the average AoI under the LCFS with preemption in service queueing discipline when the transmitter contains multiple servers in series or there exists a series of nodes between the transmitter and destination nodes. In \cite{SHS_1}, the average AoI was characterized under the priority LCFS with preemption in service/waiting queueing model. The authors of \cite{SHS_4} derived the average AoI in the presence of packet delivery errors under stationary randomized and round-robin scheduling policies. In \cite{SHS_5}, the average AoI was characterized under the LCFS with preemption in service queueing discipline when the transmitter contains multiple parallel servers. The authors of \cite{SHS_6} analyzed the average AoI for a network in which multiple transmitter-destination pairs contend for the channel using the carrier sense multiple access scheme. In \cite{SHS_7} (In \cite{SHS_8}), the average AoI (the MGF of AoI) was derived under several source-aware packet management scheduling policies at the transmitter. For the case where the transmitter is powered by energy harvesting (EH), the authors of \cite{abdelmagid_2021a} and \cite{abdelmagid_2021b} derived the MGF of AoI under several queueing disciplines including the LCFS with and without preemption in service/waiting strategies.

\begin{table*}
\centering
{\caption{A Summary of the Queueing Theory-based Analyses of AoI in the Existing Literature.} 
\label{table:summary}
%\scalebox{}
{ \begin{tabular}{ |c |c|c|}
\hline
%\multicolumn{2}{|c||}{\textbf{System  variables }} & \multicolumn{2}{c|}{\textbf{Performance metrics}}\\ \hline 
    & The geometric approach  & The SHS approach\\ \hline
Marginal distributional properties of AoI/peak AoI& \cite{kaul2012real,kaul2012status,soysal,costa2016age,chen2016age,kam2018age,kavitha2018controlling,zou2019waiting,Inoue19,kosta2020non,Champati19,Chiariotti_dist,ayan2020probability,olga20,yates2012real,Moltafet_multisource,pappas2015age,yates2017status,kosta2019age,najm2018status,huang2015optimizing,Xu_21,Akar21,Ozancan_2021} & \cite{yates2018age,yates2020age,SHS_2,SHS_3,SHS_1,SHS_4,SHS_5,SHS_6,SHS_7,abdelmagid_2021a,abd2021distributional,SHS_8,abdelmagid_2021b,abd2022dist_ICC,abd2022dist_INFOCOM} \\ \hline
Joint distributional properties of AoI/peak AoI& \cite{jiang2020correlation,jiang2020joint} &  This paper\\ \hline
\end{tabular}}} 
\end{table*} 
{\it Joint analysis of AoI processes.} A very recent prior work \cite{jiang2020correlation,jiang2020joint} has also analyzed the joint distributional properties of all the AoI processes in a particular bufferless multi-source single-server system setting using tools from Palm calculus. In contrast, our framework: i) enables one to analyze the joint distributional properties of an arbitrary set of AoI processes in a network, and ii) is applicable to any generic queuing discipline including the ones with buffers and/or multiple servers. In fact, we will recover a key result of \cite{jiang2020correlation} as a special case of our analysis in Section \ref{sec:MGF_analysis}. Table \ref{table:summary} further highlights the gap in the literature that we aim to fill in this paper.

Before going into more details about our contributions, it is worth noting that besides the above queueing theory-based analyses of AoI, there have been efforts to evaluate and optimize AoI or some other AoI-related metrics in a variety of communication systems that deal with time-sensitive information (see \cite{roy_survey} for a comprehensive survey). For instance, AoI has been studied in the context of age-optimal transmission scheduling policies \cite{sun2017update,bedewy2016optimizing,Qing_he,lu2018age,jiang2019timely,huang2020age,Dong20,Vaze,han2020fairness}, multi-hop networks \cite{talak2017,bedewy2017age,Buyukates_ulu}, broadcast networks \cite{kadota2016,hsu2019scheduling}, ultra-reliable low-latency vehicular networks \cite{abdel2018ultra}, unmanned aerial vehicle (UAV)-assisted communication systems \cite{abd2018average,AbdElmagid2019Globecom_b,ferdowsi2021neural}, Internet of Underwater Things networks \cite{Zhengru22}, reconfigurable intelligent surface (RIS)-assisted communication systems \cite{AoI_RIS_a, AoI_RIS_b}, EH systems \cite{Baran_EH,Jing_EH,arafa2019age,AbdElmagid2019Globecom_a,hatami2020age,abd2019tcom,AbdElmagid_joint,Elvina21,khorsandmanesh2020average,Quanjia22}, large-scale analysis of IoT networks \cite{emara2019spatiotemporal,mankar2020stochastic_GC2,Praful_GC1}, remote estimation \cite{ornee2019sampling,Tsai}, information-theoretic analysis \cite{Sun_IT,bastopcu2020partial,wang2020value,shisher2021age}, timely source coding \cite{zhong2016timeliness,feng2019age}, cache updating systems \cite{tang2020age,ma2020age,bastopcu2020information}, economic systems \cite{zhang2019price}, and timely communication in federated learning \cite{yang2020age,buyukates2020timely}.
%\textit{This paper develops a counterpart SHS-based framework to the one of \cite{yates2020age} allowing the analysis of the joint distributional properties of AoI processes in networks through the characterization of the stationary joint moments and MGFs}. 
%It is worth highlighting that our framework can be applied to investigate the joint distributional properties of AoI processes under any arbitrary queueing discipline in the multi-source multi-server general system setting (where the transmitter has multiple servers serving status updates). 
%
%
%\textit{In fact, our framework is quite general in the sense that it can be applied to analyze the joint distributional properties of AoI processes in a broad range of system settings} (including the one considered in \cite{jiang2020correlation}, as will be demonstrated in Section 
%\ref{sec:MGF_analysis} by recovering a key result of \cite{jiang2020correlation}) \textit{under any arbitrary queueing discipline}.    
\subsection{Contributions}
A general setting of status updating systems is studied in this paper, where a set of source nodes provide status updates about some physical process(es) to a set of monitors. We quantify the freshness of information available at each monitor in terms of AoI. The continuous state of the system is then formed by the AoI/age processes at different monitors, and the discrete state of the system is modeled using a finite-state continuous-time Markov chain. For this setup, our main contributions are listed next.

{\it An SHS-based framework for the joint analysis of AoI processes in networks.} We formulate the coupled evolution of the continuous and discrete states of the system as a piecewise linear SHS with linear reset maps.
%\footnote{The definition of such type of SHSs will be presented in Section \ref{sec:Model}. We will also elaborate shortly on the technical challenges in developing our SHS-based general framework with respect to the SHS-based framework developed in \cite{yates2018age} and \cite{yates2020age} for the analysis of the marginal distributional properties of each AoI process in a network.}. 
We then define two classes of test functions which account for the correlation between an arbitrary set $\nrmK$ of AoI processes of interest and the state of the Markov chain. By applying Dynkin's formula to each test function and using the notion of tensors, we derive a system of first-order ordinary differential equations characterizing the temporal evolution of the joint moments and MGFs for $\nrmK$. Afterwards, we characterize the conditions for asymptotic stability of the differential equations, which in turn enables the characterization of the stationary joint  moments and joint MGFs for the AoI processes forming set $\nrmK$. An interesting insight obtained from our analysis is that the existence of the stationary joint first moments guarantees the existence of the stationary joint higher order moments and MGFs. Further, when $\nrmK$ is a singleton set, we recover the results of \cite{yates2018age} and \cite{yates2020age}. We will also elaborate shortly on the precise technical challenges involved in generalizing the SHS approach to the joint analysis of age processes, specifically with respect to \cite{yates2018age} and \cite{yates2020age} where it was developed for the analysis of the marginal distributional properties of each AoI process.

{\it Analysis of the stationary joint MGF in multi-source updating systems.} We apply our developed SHS-based framework to study the joint distributional properties in multi-source updating systems. The status updates generated from each source are assumed to arrive at the transmitter according to a Poisson process, and the service time of each status update is assumed to be exponentially distributed. We derive closed-form expressions of the stationary MGF for an arbitrary set $\nrmK$ of AoI processes under several queueing disciplines including non-preemptive and source-agnostic/source-aware preemptive in service queueing disciplines. It is worth emphasizing that our paper is the first to derive the stationary joint MGF expressions for these queueing disciplines, which is a key outcome of the proposed framework. 

{\it System design insights.} Using the MGF expression derived for each queueing discipline considered in this paper, we obtain a closed-form expression of the correlation coefficient between any two arbitrary AoI processes. A key insight drawn from the correlation coefficient expressions is that while any two AoI processes are negatively correlated under (source-agnostic/source-aware) preemptive in service queueing disciplines for any choice of values of the system parameters, they may be positively correlated under the non-preemptive queueing discipline. In particular, for a two-source updating system, there exists a threshold value of server utilization above which the two age processes are positively correlated under the non-preemptive queueing discipline. Further, we numerically demonstrate that the source-aware preemption in service slightly reduces the negative correlation of the two age processes compared to the source-agnostic one. 
\subsection{Challenges in Generalizing the SHS-based Framework for the Joint Analysis of Age Processes}
As already conveyed above, the marginal distributional properties of each age/AoI process in a network have been studied for a variety of settings using the SHS-based framework developed in \cite{yates2018age} and \cite{yates2020age}. However, we lack a counterpart of this framework to systematically study the joint distributional properties for an arbitrary set of age processes. Since the current paper solves this open problem, it is obvious to wonder about the technical challenges involved in generalizing the SHS-based framework of \cite{yates2018age} and \cite{yates2020age} to the joint analysis for an arbitrary set of age processes.
%Different from the analyses of \cite{kaul2012real,kaul2012status,soysal,costa2016age,chen2016age,kam2018age,kavitha2018controlling,zou2019waiting,Inoue19,kosta2020non,Champati19,Chiariotti_dist,ayan2020probability,olga20,yates2012real,Moltafet_multisource,pappas2015age,yates2017status,kosta2019age,najm2018status,huang2015optimizing,Xu_21,Akar21,Ozancan_2021,yates2018age,yates2020age,SHS_2,SHS_3,SHS_1,SHS_4,SHS_5,SHS_6,SHS_7,abdelmagid_2021a,abd2021distributional,SHS_8,abdelmagid_2021b,abd2022dist_ICC,abd2022dist_INFOCOM} that were focused on characterizing the marginal distributional properties of AoI, this paper develops an SHS-based general framework to facilitate the study of joint distributional properties of an arbitrary set of AoI processes in a network. When the set of AoI processes under consideration is a singleton set, our results reduce to the ones obtained in \cite{yates2018age} and \cite{yates2020age} which enable one to analyze the marginal distributional properties of each AoI process in the network. 

In order to understand these challenges, it is useful to first recall that an SHS models the coupled evolution over time between the discrete and continuous states of the system. In the context of the AoI analysis in this paper, the discrete state of the system is modeled using a finite-state continuous-time Markov chain and the continuous state of the system is modeled by a vector containing the age processes in the network. To derive a system of differential equations using the SHS framework for the characterization of the temporal evolution of elements in the continuous state vector (i.e., the age/AoI processes), a key step in the analysis is to carefully construct an appropriate set of test functions (i.e., functions whose expected values are quantities of interest) and then apply Dynkin's formula \cite{hespanha2006modelling}. For the purpose of characterizing the marginal moments/MGF of each age process in \cite{yates2018age} and \cite{yates2020age}, it was sufficient that each test function includes only a single age process from the continuous state vector. In other words, it was not required to capture higher-order dependencies between the age processes in this analysis. Consequently, the results of \cite{yates2018age} and \cite{yates2020age} are not applicable to the joint analysis of two or more age processes in a network. 

As will be evident from the technical discussion shortly, the analysis of joint moments/MGFs for an arbitrary set of age processes requires us to explicitly consider higher-order dependencies by constructing test functions that depend on all the age processes, which complicates the SHS analysis significantly. Under these dependencies, it becomes challenging to keep track of the updated values of each test function (resulting from updating each age process inside it) after each transition in the Markov chain modeling the system discrete state. Therefore, we need to devise a new way that facilitates expressing the updated value of each test function in closed-form. We achieve this by introducing the idea of tensors in this framework, which naturally departs from the framework of \cite{yates2018age} and \cite{yates2020age} where such higher-order dependencies did not appear (because the focus was on the marginal analysis). Using the proposed tensor notations, we also express the differential equations appearing in the SHS framework in closed-form, which further allows us to characterize the conditions for their asymptotic stability. Interestingly, the use of tensors also enables us to present a unified SHS-based framework for the AoI analysis in the sense that when the set of age processes is a singleton set, the tensors will become vectors and the results of \cite{yates2018age} and \cite{yates2020age} can be recovered.
\subsection{Organization}
The rest of the paper is organized as follows. Section \ref{sec:Model} presents the system model, the SHS-based formulation, and the problem statement. Afterwards, in Section \ref{sec:joint_analysis}, we develop the SHS-based framework for the joint analysis of an arbitrary set of age processes in a network through the characterization of their stationary joint moments and MGFs. In particular, we start by deriving a system of differential equations for the characterization of the temporal evolution of the joint moments and MGFs, and then characterize the conditions under which these differential equations are asymptotically stable. Section \ref{sec:MGF_analysis} applies the SHS-based framework developed in Section \ref{sec:joint_analysis} to derive the joint MGF in a multi-source updating system under both non-preemptive and preemptive in service queueing disciplines. %Section \ref{sec:insights} is dedicated to a discussion about the key insights obtained from our analytical/numerical results. 
Finally, Section \ref{sec:con} concludes the paper.
\subsection{Notations}
%We use almost similar notations to that of \cite{yates2020age} to make it easy for the readers to relate the connection between the marginal analysis of AoI in \cite{yates2020age} and the joint analysis in this paper. 
A set $\nrmX \in \nbbR^n$ has a cardinality of $|\nrmX| = n$ and its $j$-th element is denoted by $\nrmX(j)$, where $1 \leq j \leq n$. A vector $\nbx \in \nbbR^{1 \times n}$ is a $1 \times n$ row vector with $[\nbx]_j$ or $[\nbx]_{\nrmK = \{j\}}$ denoting its $j$-th element ($1 \leq j \leq n$). A matrix $\nbX \in  \nbbR^{n_1 \times n_2}$ has $(i,j)$-th element $[\nbX]_{i,j}$ or $[\nbX]_{\nrmK = \{i,j\}}$, and $j$-th column $[\nbX]_j$, where $1 \leq i \leq n_1$ and $1 \leq j \leq n_2$. The vectors ${\bf 0}_n$ and ${\bf 1}_n$ are the row vectors containing all zeros and ones in $\nbbR^{1 \times n}$, respectively, the vector ${\rm Tra}(\nbx)$ or $\nbx^{\rm T}$ is the transpose of $\nbx$, the matrix ${\rm Tra}(\nbX)$ or $\nbX^{\rm T}$ is the transpose of $\nbX$, and $\nbI_n$ is the $n \times n$ identity matrix. Whenever subscript $n$ is dropped, the dimensions of ${\bf 0}$, ${\bf 1}$, and $\nbI$ will be clear from the context. The Kronecker delta function $\delta_{i,j}$ equals
1 if $i = j$ and 0 otherwise. The vector $\nbe_i$ denotes the $i$-th Cartesian unit vector satisfying $[\nbe_i]_j = \delta_{i,j}$. A tensor is a multi-dimensional array whose order defines the number of dimensions of the array. For instance, a vector is a one-dimensional array or first-order tensor, and a matrix is a two-dimensional array or second-order tensor. An $I$-th order tensor $\mathcal{X} \in \nbbR^{n_1 \times n_2 \times \cdots \times n_{I}}$ has $\nrmK$-th element $[\mathcal{X}]_{\nrmK}$, where $|\nrmK| = I$ and $1 \leq \nrmK(j) \leq n_j$ for all $1 \leq j \leq I$. The product of the tensor $\mathcal{X}$ and a matrix along its $j$-th dimension is denoted by $\times_{j}$ and known as the $j$-mode product. In particular, the $j$-mode product of $\mathcal{X}$ and a matrix $\nbX \in \nbbR^{m \times n_j}$ is represented as: $\mathcal{Y} = \mathcal{X} \times_{j} \nbX$, where $\mathcal{Y} \in \nbbR^{n_1 \times \cdots \times n_{j-1} \times m \times n_{j+1} \times \cdots \times n_{I}}$. For a process $\nbx(t)$, $\nbX(t)$ or $\mathcal{X}(t)$, $\dot{\nbx}(t)$, $\dot{\nbX}(t)$ or $\dot{\mathcal{X}}(t)$ denote the derivative ${\rm d}\nbx(t)/{\rm d}t$, ${\rm d}\nbX(t)/{\rm d}t$ or ${\rm d}\mathcal{X}(t)/{\rm d}t$. For a scalar function $f(\cdot)$ and a vector $\nbx = [x_1\;x_2\;\cdots\;x_n]$, $f(\nbx) = \left[f(x_1)\; f(x_2)\;\cdots\;f(x_n)\right]$. For integers $m \leq n, m:n$ is the set $\{m,m+1,\cdots,n\}$, and $\nrmX(m:n) = \{\nrmX(m),\nrmX(m+1),\cdots,\nrmX(n)\}$. The set of all permutations of a set $\nrmX$ is denoted by $\mathcal{P}(\nrmX)$, and the set of all subsets of a set $\nrmX$ is denoted by $2^{\nrmX}$. For instance, when $\nrmX = \{x_1,x_2,x_3\}$, we have:
\begin{align*}
&\mathcal{P}(\nrmX) = \{\{x_1,x_2,x_3\},\{x_1,x_3,x_2\},\{x_2,x_1,x_3\},\{x_2,x_3,x_1\},\{x_3,x_1,x_2\},\{x_3,x_2,x_1\}\}, \\
&2^{\nrmX} = \{\varnothing,\{x_1\},\{x_2\},\{x_3\},\{x_1,x_2\},\{x_1,x_3\},\{x_2,x_3\},\{x_1,x_2,x_3\}\},
\end{align*}
 where the symbol $\varnothing$  denotes the empty set. The indicator function $\nb1(\cdot)$ is 1 if the condition inside the brackets is satisfied and 0 otherwise. 
\section{System Model and Problem Statement}\label{sec:Model}
\subsection{Network Model}
We consider a general setting of status updating systems where a set of source nodes provide status updates about some physical process(es) to a set of monitors. The freshness of information available at each monitor is quantified in terms of AoI. The AoI processes (or equivalently the age processes) in the system are modeled using the row vector $\nbx(t) = [x_1(t) \;\cdots\; x_n(t)]$, which is also referred to as the continuous state of the system. In particular, $x_j(t)$ is the age process at monitor $j$, which may refer to a node, a position in a queue, or a server in a multi-server system. 
Further, the discrete state of the system is modeled using a finite-state continuous-time Markov chain $q(t) \in \ncalQ = \{0,\cdots,q_{max}\}$, where $\ncalQ$ is the discrete state space. This Markov chain governs the dynamics of the system discrete state, e.g., $q(t)$ may describe the system occupancy with respect to the status updates generated by each source node. In the graphical representation of the Markov chain $q(t)$, each state $q \in \ncalQ$ is a node and each transition $l$ is a directed edge $(q_l, q'_l)$ with fixed transition rate $\lambda^{(l)}(q(t)) = \lambda^{(l)} \delta_{q_l,q(t)}$, where the Kronecker delta function $\delta_{q_l,q(t)}$ ensures that transition $l$ occurs only in state $q_l$. We denote the set of all transitions $\{l\}$ by $\ncalL$, and the sets of incoming and outgoing transitions for state $\bar{q} \in \ncalQ$ by $\ncalL'_{\bar{q}} = \{l\in\ncalL: q'_l = \bar{q}\}$ and $\ncalL_{\bar{q}} = \{l\in\ncalL: q_l = \bar{q}\}$, respectively. 
%Note that $\ncalL'_{\bar{q}} = \cup_{i}{\ncalL_{i,\bar{q}}}$ such that $\ncalL_{i,\bar{q}} = \{l\in\ncalL: q_l = i, q'_l = \bar{q}\}$.
%As a consequence of the occurrence of transition $l$, the discrete state of the system changes from state $q_l$  to state $q'_l$, and the continuous state $\nbx$ is reset to $\nbx'$ according to a binary reset map matrix $\nbA_l \in \nbbB^{n\times n}$ as $\nbx'=\nbx \nbA_l$. Further, $\overset{\cdot}{\nbx}(t) \triangleq \dfrac{\partial\nbx(t)}{\partial t} = {\bf 1}$ holds as long as the state $q(t)$ is unchanged, where ${\bf 1}$ is the row vector $[1, \cdots,1] \in \nbbR^{1\times n}$. Different from ordinary continuous-time Markov chains, an inherent feature of SHSs is the possibility of having self-transitions in the Markov chain modeling the system discrete state. In particular, although a self-transition keeps $q(t)$ unchanged, it causes a change in the continuous process $x(t)$.
\subsection{An SHS-based Formulation and Problem Statement}
The coupled evolution of  the continuous state $\nbx(t)$ and the discrete state $q(t)$ is modeled using a piecewise linear SHS with linear reset maps \cite{yates2020age}. In particular, when a transition $l$ occurs in the Markov chain $q(t)$, the continuous state $\nbx$ is reset to $\nbx'$ according to a reset map matrix $\nbA_l$ as $\nbx'=\nbx \nbA_l$. Further, as long as the state $q(t)$ is unchanged, each element in the age vector $\nbx(t)$ grows at a unit rate with time (which yields piecewise linear age processes over time), i.e., $\overset{\cdot}{\nbx}(t) \triangleq \dfrac{{\rm d}\nbx(t)}{{\rm d}t} = {\bf 1}$. To capture the temporal evolution of the age processes, it is sufficient to assume that $\nbA_l$ is a binary matrix with no more than a single 1 in a column. Since column $[\nbA_l]_j$ determines the value that will be assigned to $x'_j$, we have two different cases given the assumed structure of $\nbA_l$. In the first case, $[\nbA_l]_j = {\bf 0}^{\rm T}$ and so $x'_j = 0$, whereas the second case corresponds to $[\nbA_l]_j = \nbe_i^{\rm T}$ where $x'_j$ is reset to $x_i$. Different from ordinary continuous-time Markov chains, an inherent feature of SHS is the possibility of having self-transitions in the Markov chain $q(t)$ modeling the system discrete state. In particular, although a self-transition keeps $q(t)$ unchanged, it causes a change in the continuous state $\nbx(t)$. Further, there may be multiple transitions between any two states in $\ncalQ$ such that their associated reset map matrices are different.

For the above SHS formulation, our prime objective in this paper is to develop a framework that allows understanding/analyzing the joint distributional properties of an arbitrary set of elements in the age vector $\nbx(t)$. Formally, we aim at characterizing the stationary joint moments and joint MGFs for an arbitrary set $\nrmK \subseteq 1:n$ of age processes, which are respectively of the following forms: $\underset{t \rightarrow \infty}{\rm lim} \nbbE\Big[\prod_{j=1}^{|\nrmK|}{x_{\nrmK(j)}^{[\nbm]_j}(t)}\Big]$ and $\underset{t \rightarrow \infty}{\rm lim} \nbbE\left[{\rm exp}\Big[\sum_{j=1}^{|\nrmK|}{[\nbs]_j x_{\nrmK(j)}(t)}\Big]\right]$, where the length of vector $\nbm$ or $\nbs$ is $|\nrmK|$, $[\nbm]_j \in \{0,1,2,\cdots\}$, and $[\nbs]_j \in \nbbR, \forall j \in 1 : |\nrmK|$. As already discussed in Section \ref{sec:intro}, when $|\nrmK| = 1$, the problem at hand reduces to the one studied in \cite{yates2018age} and \cite{yates2020age}, where the goal was to characterize the marginal distributional properties of each age element in $\nbx(t)$. Clearly, the characterization of such joint moments and joint MGFs allows one to derive the correlation coefficient between all possible pairwise combinations of the age vector elements. Towards this objective, we first derive a system of first-order ordinary differential equations for the temporal evolution of both the joint moments and joint MGFs. We then derive the conditions under which these differential equations are stable, which in turn enables the evaluation of the stationary joint moments and joint MGFs. Given the generality of the system setting considered in this paper, the importance of our framework lies in the fact that it is applicable to the joint analysis of AoIs in a broad range of status updating system setups under arbitrary queueing disciplines.
\section{Joint Analysis of Age Processes in Networks}\label{sec:joint_analysis}
\subsection{Differential Equations for the Temporal Evolution of the Joint Moments and Joint MGFs}
In order to characterize the temporal evolution of the joint moments and joint MGFs for a set $\nrmK$ of age processes, $\nbbE\left[\prod_{j=1}^{|\nrmK|}{x_{\nrmK(j)}^{[\nbm]_j}(t)}\right]$ and $\nbbE\left[{\rm exp}\Big[\sum_{j=1}^{|\nrmK|}{[\nbs]_j x_{\nrmK(j)}(t)}\Big]\right]$, it is useful to define the following quantities that express different forms of correlation between $q(t)$ and the age processes in $\nbx(t)$:
%\begin{align}\label{exp_marginal_moms_q}
%v_{\bar{q},j}^{(m)}(t) = \nbbE[x_j^m(t)\delta_{\bar{q},q(t)}],
%\end{align}
%\begin{align}\label{exp_marginal_mgfs_q}
%v_{\bar{q},j}^{(s)}(t) = \nbbE\left[{\rm exp}\left[s x_j(t)\right]\delta_{\bar{q},q(t)}\right],
%\end{align}
\begin{align}\label{exp_joint_moms_q}
v_{\bar{q},\nrmK}^{(\nbm)}(t) = \nbbE\left[\prod_{j=1}^{|\nrmK|}{x_{\nrmK(j)}^{[\nbm]_j}}(t) \delta_{\bar{q},q(t)}\right],
\end{align}
\begin{align}\label{exp_joint_mgfs_q}
v_{\bar{q},\nrmK}^{(\nbs)}(t) = \nbbE\left[{\rm exp}\Big[\sum_{j=1}^{|\nrmK|}{[\nbs]_j x_{\nrmK(j)}(t)}\Big] \delta_{\bar{q},q(t)}\right],
\end{align}
for all states $\bar{q} \in \ncalQ$ and $\nrmK \subseteq \{1,2,\cdots,N\}$.
%, and $m_1, m_2 \geq 1$. 
To see this, note that we have %$\nbbE\Big[\prod_{j=1}^{|\nrmK|}{x_{\nrmK(j)}^{[\nbm]_j}(t)}\Big]$ and $\nbbE\left[{\rm exp}\Big[\sum_{j=1}^{|\nrmK|}{[\nbs]_j x_{\nrmK(j)}(t)}\Big]\right]$ can be respectively expressed as
\begin{align}\label{exp_joint_moms}
\nbbE\left[\prod_{j=1}^{|\nrmK|}{x_{\nrmK(j)}^{[\nbm]_j}(t)}\right] = \sum_{\bar{q} \in \ncalQ} {\nbbE\left[\prod_{j=1}^{|\nrmK|}{x_{\nrmK(j)}^{[\nbm]_j}}(t) \delta_{\bar{q},q(t)}\right]} = \sum_{\bar{q} \in \ncalQ}{v^{(\nbm)}_{\bar{q},\nrmK}(t)},
\end{align}
\begin{align}\label{exp_joint_mgfs}
\nbbE\left[{\rm exp}\Big[\sum_{j=1}^{|\nrmK|}{[\nbs]_j x_{\nrmK(j)}(t)}\Big]\right] = \sum_{\bar{q} \in \ncalQ}{\nbbE\left[{\rm exp}\Big[\sum_{j=1}^{|\nrmK|}{[\nbs]_j x_{\nrmK(j)}(t)}\Big] \delta_{\bar{q},q(t)}\right]} = \sum_{\bar{q} \in \ncalQ}{v^{(\nbs)}_{\bar{q},\nrmK}(t)}.
\end{align}

Thus, according to (\ref{exp_joint_moms}) and (\ref{exp_joint_mgfs}), characterizing the temporal evolution of $v^{(\nbm)}_{\bar{q},\nrmK}(t)$ and $v^{(\nbs)}_{\bar{q},\nrmK}(t)$ directly characterizes the temporal evolution of $\nbbE\left[\prod_{j=1}^{|\nrmK|}{x_{\nrmK(j)}^{[\nbm]_j}(t)}\right]$ and $\nbbE\left[{\rm exp}\Big[\sum_{j=1}^{|\nrmK|}{[\nbs]_j x_{\nrmK(j)}(t)}\Big]\right]$, respectively. 
%Similarly, computing the expectations in (\ref{exp_marginal_moms_q}) and (\ref{exp_marginal_mgfs_q}) is sufficient for the characterization of the marginal $m$-th moments and marginal MGFs, $\nbbE[x_j^m(t)]$ and $\nbbE[e^{sx_j(t)}]$, as has been demonstrated in \cite{yates2020age}.
Some key notes about the notations in (\ref{exp_joint_moms_q}) and (\ref{exp_joint_mgfs_q}) are provided next. First, $v_{\bar{q},\nrmK}^{({\bf 1})}(t)$ may generally refer to $v_{\bar{q},\nrmK}^{(\nbm)}(t)|_{\nbm={\bf 1}}$ or $v_{\bar{q},\nrmK}^{(\nbs)}(t)|_{\nbs={\bf 1}}$. To eliminate this conflict, the convention that $v_{\bar{q},\nrmK}^{(\nbi)}$, for any set of integers $\{[\nbi]_j \geq 1\}_{j \in 1:|\nrmK|}$, refers to $v_{\bar{q},\nrmK}^{(\nbm)}$ at $\nbm = \nbi$ is maintained here. Further, we have $v_{\bar{q},\nrmK}^{(\nbm)}(t)|_{\nbm={\bf 0}} = v_{\bar{q},\nrmK}^{(\nbs)}(t)|_{\nbs={\bf 0}} = \nbbE[\delta_{\bar{q},q(t)}] = \nbbP[q(t) = \bar{q}]$, i.e., $v_{\bar{q},\nrmK}^{({\bf 0})}(t) = \nbbP[q(t) = \bar{q}]$ (the probability that $q(t)$ is equal to $\bar{q}$) regardless of $\nrmK$. For $\bar{q} \in \mathcal{Q}$, we define $\nbv_{\bar{q}}^{(0)}(t) \in \nbbR^{1 \times n}$ as:
\begin{align}
[\nbv_{\bar{q}}^{(0)}(t)]_k = v_{\bar{q}}^{(0)}(t) = \nbbP[q(t) = \bar{q}], \forall k \in 1:n.
\end{align}

It will also be useful in our subsequent analysis and exposition to define the following tensors in $\nbbR^{n_1 \times n_2 \times \cdots \times n_{|\nrmK|}} (n_j = n, \forall j \in 1 : |\nrmK|)$ containing the scalars in (\ref{exp_joint_moms_q}) and (\ref{exp_joint_mgfs_q}): $\left[\mathcal{V}_{\bar{q},|\nrmK|}^{(\nbm)}(t)\right]_{\nrmK} = v_{\bar{q},\nrmK}^{(\nbm)}(t)$ and $\left[\mathcal{V}_{\bar{q},|\nrmK|}^{(\nbs)}(t)\right]_{\nrmK} = v_{\bar{q},\nrmK}^{(\nbs)}(t),\forall \bar{q} \in \ncalQ$. In other words, the tensors $\mathcal{V}_{\bar{q},|\nrmK|}^{(\nbm)}(t)$ and $\mathcal{V}_{\bar{q},|\nrmK|}^{(\nbs)}(t)$ contain the scalars $\{v_{\bar{q},\nrmM}^{(\nbm)}(t)\}_{\nrmM \subseteq 1:n, |\nrmM| = |\nrmK|}$ and  $\{v_{\bar{q},\nrmM}^{(\nbs)}(t)\}_{\nrmM \subseteq 1:n, |\nrmM| = |\nrmK|}$, respectively. For instance, $\mathcal{V}_{\bar{q},1}^{(\nbm)}(t)$ and  $\mathcal{V}_{\bar{q},2}^{(\nbm)}(t)$ can be respectively expressed as: 
\begin{align}
\mathcal{V}_{\bar{q},1}^{(\nbm)}(t) = [v_{\bar{q},\{1\}}^{(\nbm)}(t)\; v_{\bar{q},\{2\}}^{(\nbm)}(t)\; \cdots\; v_{\bar{q},\{n\}}^{(\nbm)}(t)], \;\;\forall \bar{q} \in \mathcal{Q},
\end{align}
\begin{align}
\mathcal{V}_{\bar{q},2}^{(\nbm)}(t) = \begin{bmatrix}
v_{\bar{q},\{1,1\}}^{(\nbm)}(t) & v_{\bar{q},\{1,2\}}^{(\nbm)}(t) & \dots &  v_{\bar{q},\{1,n\}}^{(\nbm)}(t)\\ v_{\bar{q},\{2,1\}}^{(\nbm)}(t) & v_{\bar{q},\{2,2\}}^{(\nbm)}(t) & \dots & v_{\bar{q},\{2,n\}}^{(\nbm)}(t)\\ \vdots & \vdots & \ddots & \vdots \\ v_{\bar{q},\{n,1\}}^{(\nbm)}(t) & v_{\bar{q},\{n,2\}}^{(\nbm)}(t) & \dots & v_{\bar{q},\{n,n\}}^{(\nbm)}(t)
\end{bmatrix}
, \;\; \forall \bar{q} \in \mathcal{Q}.
\end{align}

The following Lemma shows that $\{v^{(\nbm)}_{\bar{q},\nrmK}(t)\}_{\bar{q}\in\mathcal{Q}}$ and $\{v^{(\nbs)}_{\bar{q},\nrmK}(t)\}_{\bar{q}\in \mathcal{Q}}$ obey a system of first-order ordinary differential equations.
%According to the ergodicity assumption of the continuous-time Markov chain modeling $q(t)$ in the AoI analysis \cite{yates2018age,yates2020age}, the state probability vector $\pi(t) = [\pi_0(t),\cdots,\pi_m(t)]$ converges uniquely to the stationary vector $\bar{v}^{(0)} = [\bar{v}^{(0)}_{0},\cdots,\bar{v}^{(0)}_m]$ satisfying
%\begin{align}\label{gen_steady}
%\bar{v}^{(0)}_q\sum_{l\in\ncalL_q}{\lambda^{(l)}} = \sum_{l\in\ncalL'_q}{\lambda^{(l)}\bar{v}^{(0)}_{q_l}},\; q \in \ncalQ,\;\; \sum_{q\in\ncalQ}{\bar{v}^{(0)}_q} = 1,
%\end{align}
%where $\ncalL'_q = \{l\in\ncalL: q'_l = q\}$ and $\ncalL_q = \{l\in\ncalL: q_l = q\}$ denote the sets of incoming and outgoing transitions for state $q, \forall q \in \ncalQ$. 
\begin{lemma}\label{lemma:1}
For state $\bar{q} \in \ncalQ$ in the piecewise linear SHS with linear reset maps under consideration,
\begin{align}\label{Lemma1_eq1}
\dot{v}^{(\nbm)}_{\bar{q},\nrmK}(t) = \sum_{j=1}^{|\nrmK|}{[\nbm]_j v^{(\nbm - \nbe_j)}_{\bar{q},\nrmK}(t)} + \sum_{l \in \ncalL'_{\bar{q}}}{\lambda^{(l)}\Big[\ncalV_{\bar{q}_l,|\nrmK|}^{(\nbm)}(t) \times_1 \nbA_l \times_2 \nbA_l \cdots \times_{|\nrmK|} \nbA_l\Big]_{\nrmK}} - v^{(\nbm)}_{\bar{q},\nrmK}(t)\sum_{l \in \ncalL_{\bar{q}}}{\lambda^{(l)}},
\end{align}
\begin{align}\label{Lemma1_eq2}
\dot{v}^{(\nbs)}_{\bar{q},\nrmK}(t) = \left[\sum_{j=1}^{|\nrmK|}{[\nbs]_j} - \sum_{l \in \ncalL_{\bar{q}}}{\lambda^{(l)}}\right] v^{(\nbs)}_{\bar{q},\nrmK}(t) + \sum_{l \in \ncalL'_{\bar{q}}}{\lambda^{(l)}\Big[\ncalV_{\bar{q}_l,|\nrmK|}^{(\nbs)}(t) \times_1 \nbA_l \times_2 \nbA_l \cdots \times_{|\nrmK|} \nbA_l\Big]_{\nrmK}} + c_{\bar{q},\nrmK}(t),
\end{align}
%where $[\nbC_{\bar{q}}(t)]_{j,k} = c_{\bar{q},jk}(t) $ is defined as  
%\begin{align}\label{c_app}
%c_{\bar{q},jk}(t) \nonumber&= \sum_{l \in \ncalL'_{\bar{q}}}\lambda^{(l)}\Big[\nb1\big([\nbx \nbA_l]_j \neq 0 \;\&\; [\nbx \nbA_l]_k = 0\big)[\nbv^{(s_1)}_{q_l}(t) \nbA_l]_j \\ \nonumber&+ \nb1\big([\nbx \nbA_l]_j = 0 \;\&\; [\nbx \nbA_l]_k \neq 0\big)[\nbv^{(s_2)}_{q_l}(t) \nbA_l]_k \\&+ \nb1\big([\nbx \nbA_l]_j = 0 \;\&\; [\nbx \nbA_l]_k = 0\big)[\nbv^{(0)}_{q_l}]_j\Big],
%\end{align}
%where $\&$ is the logical AND operator and $\nb1(\cdot)$ is the indicator function.
such that $c_{\bar{q},\nrmK}(t)$ is defined as 
\begin{align}\label{c_app}
c_{\bar{q},\nrmK}(t) = \sum_{l \in \ncalL'_{\bar{q}}}{\lambda^{(l)} \sum_{\nrmZ \in 2^{\nrmK}\setminus\varnothing}{\nb1\big(\nrmZ_l = \nrmZ\big)\Big[\ncalV_{\bar{q}_l,|\nrmK\setminus\nrmZ_l|}^{(\nbs')}(t) \times_1 \nbA_l \times_2 \nbA_l \cdots \times_{|\nrmK \setminus \nrmZ_l|} \nbA_l\Big]_{\nrmK \setminus \nrmZ_l}}},
\end{align}
where the set $\nrmZ_l = \{j \in \nrmK: [\nbA_l]_j = {\bf 0}^{\rm T}\}$, the vector $\nbs' = \left[[\nbs]_{\nrmI_l(1)}\;[\nbs]_{\nrmI_l(2)}\;\cdots\;[\nbs]_{\nrmI_l(|\nrmK \setminus \nrmZ_l|)}\right]$, and the set $\nrmI_l$ contains the indices of the elements of $\nrmK \setminus \nrmZ_l$ inside $\nrmK$. When $\nrmZ_l = \nrmK$, we also define:
\begin{align}
   \Big[\ncalV_{\bar{q}_l,|\nrmK\setminus\nrmZ_l|}^{(\nbs')}(t) \times_1 \nbA_l \times_2 \nbA_l \cdots \times_{|\nrmK \setminus \nrmZ_l|} \nbA_l\Big]_{\nrmK \setminus \nrmZ_l} = v_{\bar{q}_l}^{({0})}(t).
\end{align}
\end{lemma}
\begin{IEEEproof}
See Appendix \ref{app:lemma:1}.
\end{IEEEproof}
It is worth noting that (\ref{Lemma1_eq1}) and (\ref{Lemma1_eq2}) in Lemma \ref{lemma:1} can be expressed in a tensor form as
\begin{align}\label{Lemma1_eq1_tensor}
\dot{\ncalV}^{(\nbm)}_{\bar{q},|\nrmK|}(t) = \sum_{j=1}^{|\nrmK|}{[\nbm]_j \ncalV^{(\nbm - \nbe_j)}_{\bar{q},|\nrmK|}(t)} + \sum_{l \in \ncalL'_{\bar{q}}}{\lambda^{(l)}\Big[\ncalV_{\bar{q}_l,|\nrmK|}^{(\nbm)}(t) \times_1 \nbA_l \times_2 \nbA_l \cdots \times_{|\nrmK|} \nbA_l\Big]} - \ncalV^{(\nbm)}_{\bar{q},|\nrmK|}(t)\sum_{l \in \ncalL_{\bar{q}}}{\lambda^{(l)}},
\end{align}
\begin{align}\label{Lemma1_eq2_tensor}
\dot{\ncalV}^{(\nbs)}_{\bar{q},|\nrmK|}(t) = \left[\sum_{j=1}^{|\nrmK|}{[\nbs]_j} - \sum_{l \in \ncalL_{\bar{q}}}{\lambda^{(l)}}\right] \ncalV^{(\nbs)}_{\bar{q},|\nrmK|}(t) + \sum_{l \in \ncalL'_{\bar{q}}}{\lambda^{(l)}\Big[\ncalV_{\bar{q}_l,|\nrmK|}^{(\nbs)}(t) \times_1 \nbA_l \times_2 \nbA_l \cdots \times_{|\nrmK|} \nbA_l\Big]} + \ncalC_{\bar{q},|\nrmK|}(t),
\end{align}
where $\ncalC_{\bar{q},|\nrmK|}(t) \in \nbbR^{n_1 \times n_2 \times \cdots \times n_{|\nrmK|}} (n_j = n, \forall j \in 1 : |\nrmK|)$ such that $[\ncalC_{\bar{q},|\nrmK|}(t)]_{\nrmK} = c_{\bar{q},\nrmK}(t)$.
%is given by 
%\begin{align}\label{c_app_tensor}
%c_{\bar{q},\nrmK}(t) = \sum_{l \in \ncalL'_{\bar{q}}}{\lambda^{(l)} \sum_{\nrmZ \in 2^{\nrmK}/\varnothing}{\nb1\big(\nrmZ_l = \nrmZ\big)\Big[\ncalV_{\bar{q}_l,\nrmK/\nrmZ_l}^{(\nbs')}(t) \times_1 \nbA_l \times_2 \nbA_l \cdots \times_{|\nrmK/\nrmZ_l|} \nbA_l\Big]_{\nrmK / \nrmZ_l}}}.
%\end{align}
In order to clearly see that Lemma \ref{lemma:1} characterizes the trajectories of $v^{(\nbm)}_{\bar{q},\nrmK}(t)$ and $v^{(\nbs)}_{\bar{q},\nrmK}(t)$ over time, it is useful to first state the following Corollaries.
\begin{cor}\label{cor:systemdiff_marginal}
When $\nrmK = \{k\}$, $\nbm = [m_1]$ and $\nbs = [s_1]$, the system of first-order ordinary differential equations in Lemma \ref{lemma:1} reduces to:
\begin{align}\label{cor:systemdiff_marginal_eq1}
\dot{v}^{([m_{1}])}_{\bar{q},\{k\}}(t) = m_1 v^{([m_1-1])}_{\bar{q},\{k\}}(t) + \sum_{l \in \ncalL'_{\bar{q}}}{\lambda^{(l)}\big[\nbv_{\bar{q}_l}^{(m_1)}(t) \nbA_l\big]_{k}} - v^{([m_1])}_{\bar{q},\{k\}}(t)\sum_{l \in \ncalL_{\bar{q}}}{\lambda^{(l)}},
\end{align}
\begin{align}\label{cor:systemdiff_marginal_eq2}
\dot{v}^{([s_1])}_{\bar{q},\{k\}}(t) = \Big[s_1 - \sum_{l \in \ncalL_{\bar{q}}}{\lambda^{(l)}}\Big] v^{([s_1])}_{\bar{q},\{k\}}(t) + \sum_{l \in \ncalL'_{\bar{q}}}{\lambda^{(l)}\left[\nbv_{\bar{q}_l}^{(s_1)}(t) \nbA_l + \nb1\big(\nrmZ_l = \{k\}\big) \nbv_{\bar{q}_l}^{(0)}(t) \right]_{k}}.
\end{align}
\end{cor}
\begin{IEEEproof}
The expressions in (\ref{cor:systemdiff_marginal_eq1}) and (\ref{cor:systemdiff_marginal_eq2}) directly follow from Lemma \ref{lemma:1} along with noting that $\ncalV_{\bar{q}_l,|\nrmK|}^{(\nbm)}(t)$ and $\ncalV_{\bar{q}_l,|\nrmK|}^{(\nbs)}(t)$ are vectors in $\nbbR^{1 \times n}$ when $|\nrmK| = 1$, and hence we define $\nbv^{(m_1)}_{\bar{q}_l}(t) = \ncalV_{\bar{q}_l,1}^{([m_1])}(t)$ and $\nbv^{(s_1)}_{\bar{q}_l}(t) = \ncalV_{\bar{q}_l,1}^{([s_1])}(t)$.
\end{IEEEproof}
\begin{nrem}
Note that the system of differential equations in Corollary \ref{cor:systemdiff_marginal} is identical to the one derived in \cite[Lemma 1]{yates2020age} for the temporal evolution of the marginal moments and MGF of each age element in $\nbx(t)$.
\end{nrem}
\begin{cor}\label{cor:systemdiff_jointtwo}
When $\nrmK = \{k_1,k_2\}$, $\nbm = [m_1\;m_2]$ and $\nbs = [s_1\;s_2]$, the system of first-order ordinary differential equations in Lemma \ref{lemma:1} reduces to:
\begin{align}\label{cor:systemdiff_jointtwo_eq1}
\dot{v}_{\bar{q},\{k_1,k_2\}}^{([m_{1}\;m_{2}])}(t) = m_{1} v_{\bar{q},\{k_1,k_2\}}^{([m_{1}-1\;m_{2}])}(t) + m_{2} v_{\bar{q},\{k_1,k_2\}}^{([m_{1}\;m_{2}-1])}(t) \nonumber&+ \sum_{l \in \ncalL'_{\bar{q}}}{\lambda^{(l)} \big[\nbA_l^{{\rm T}} \nbV_{q_l}^{(m_{1},m_{2})}(t) \nbA_l\big]_{k_1,k_2}} \\&- v_{\bar{q},\{k_1,k_2\}}^{([m_{1}\;m_{2}])}(t) \sum_{l \in \ncalL_{\bar{q}}}{\lambda^{(l)}},
\end{align}
\begin{align}\label{cor:systemdiff_jointtwo_eq2}
\dot{v}^{([s_{1}\;s_{2}])}_{\bar{q},\{k_1,k_2\}}(t) = \Big[s_{1} + s_{2} - \sum_{l \in \ncalL_{\bar{q}}}{\lambda^{(l)}}\Big] v^{([s_{1}\;s_{2}])}_{\bar{q},\{k_1,k_2\}}(t) +\sum_{l \in \ncalL'_{\bar{q}}}{\lambda^{(l)}[\nbA_l^{\rm T} \nbV^{(s_{1},s_{2})}_{\bar{q}_l}(t)\nbA_l]_{k_1,k_2}} + c_{\bar{q},\{k_1,k_2\}}(t),
\end{align}
where $c_{\bar{q},\{k_1,k_2\}}(t)$ is given by
\begin{align}\label{c_two}
c_{\bar{q},\{k_1,k_2\}}(t) = \sum_{l \in \ncalL'_{\bar{q}}}\lambda^{(l)}\Bigg[\nb1\Big(\nrmZ_l = \{k_2\}\Big)\big[\nbv^{(s_{1})}_{q_l}(t) \nbA_l\big]_{k_1} \nonumber&+ \nb1\Big(\nrmZ_l = \{k_1\}\Big)\big[\nbv^{(s_{2})}_{q_l}(t) \nbA_l\big]_{k_2} \\&+ \nb1\Big(\nrmZ_l = \{k_1,k_2\}\Big) v^{(0)}_{q_l}(t)\Bigg],    
\end{align}
\end{cor}
\begin{IEEEproof}
The expressions in (\ref{cor:systemdiff_jointtwo_eq1}) and (\ref{cor:systemdiff_jointtwo_eq2}) directly follow from Lemma \ref{lemma:1} along with noting that $\ncalV_{\bar{q}_l,|\nrmK|}^{(\nbm)}(t)$ and $\ncalV_{\bar{q}_l,|\nrmK|}^{(\nbs)}(t)$ are matrices in $\nbbR^{n \times n}$ when $|\nrmK| = 2$, and hence we define $\nbV^{(m_1,m_2)}_{\bar{q}_l}(t) = \ncalV_{\bar{q}_l,2}^{([m_1\;m_2])}(t)$ and $\nbV^{(s_1,s_2)}_{\bar{q}_l}(t) = \ncalV_{\bar{q}_l,2}^{([s_1\;s_2])}(t)$.
\end{IEEEproof}
%differential equations of \cite[Lemma~1]{yates2020age} characterizing the temporal evolution of the marginal $m$-th moments and marginal MGFs:
%\begin{align}\label{diff_marginal_0}
%\dot{\nbv}^{(0)}_{\bar{q}}(t) = \sum_{l \in \ncalL'_{\bar{q}}}{\lambda^{(l)} \nbv^{(0)}_{q_l}(t)} - \nbv^{(0)}_{\bar{q}}(t) \sum_{l \in \ncalL_{\bar{q}}}{\lambda^{(l)}}, 
%\end{align}
%\begin{align}\label{diff_marginal_mom}
%\dot{\nbv}^{(m)}_{\bar{q}}(t) = m \nbv^{(m-1)}_{\bar{q}}(t) \nonumber&+ \sum_{l \in \ncalL'_{\bar{q}}}{\lambda^{(l)}\nbv^{(m)}_{q_l}(t)\nbA_l} \\&- \nbv^{(m)}_{\bar{q}}(t) \sum_{l \in \ncalL_{\bar{q}}}{\lambda^{(l)}}, \forall m \geq 1,
%\end{align}
%\begin{align}\label{diff_marginal_mgf}
%\dot{\nbv}^{(s)}_{\bar{q}}(t) = s \nbv^{(s)}_{\bar{q}}(t) \nonumber&+ \sum_{l \in \ncalL'_{\bar{q}}}{\lambda^{(l)}\big[\nbv^{(s)}_{q_l}(t)\nbA_l + \nbv^{(0)}_{q_l}(t)\hat{\nbA}_l\big] } \\&- \nbv^{(s)}_{\bar{q}}(t) \sum_{l \in \ncalL_{\bar{q}}}{\lambda^{(l)}},
%\end{align}
%where $\hat{\nbA}_l$ can be expressed as
%\begin{align}
%[\hat{\nbA}]_{i,j} = \begin{cases}
%1 & i = j,[\nbA_l]_j = {\bf 0}^{\rm T},\\
%0 & {\rm otherwise}.
%\end{cases}
%\end{align}

We are now ready to elaborate on the use of Lemma \ref{lemma:1} to obtain the trajectories of $\ncalV^{(\nbm)}_{\bar{q},|\nrmK|}(t)$ and $\ncalV^{(\nbs)}_{\bar{q},|\nrmK|}(t)$ for an arbitrary set $\nrmK$ starting from a given initial condition at $t = 0$. We start this discussion with the case of $|\nrmK| = 2$ for which the trajectories can be characterized using Corollaries \ref{cor:systemdiff_marginal} and \ref{cor:systemdiff_jointtwo}. When $|\nrmK| = 2$ and for all $\bar{q} \in \ncalQ$, we observe from Corollary \ref{cor:systemdiff_jointtwo} that $\ncalV^{([m_{1}\;m_{2}])}_{\bar{q},2}(t) = \nbV^{(m_{1},m_{2})}_{\bar{q}}(t)$ and $\ncalV^{([s_{1}\;s_{2}])}_{\bar{q},2}(t) = \nbV^{(s_{1},s_{2})}_{\bar{q}}(t)$ can be evaluated using (\ref{cor:systemdiff_jointtwo_eq1}) and (\ref{cor:systemdiff_jointtwo_eq2}), respectively. In particular, we note from (\ref{cor:systemdiff_jointtwo_eq1}) that in order to compute $\nbV^{(1,1)}_{\bar{q}}(t)$, we need to first compute $\nbV^{(0,1)}_{\bar{q}}(t) = {\rm Tra}\big(\nbV^{(1,0)}_{\bar{q}}(t)\big)$ and $\nbV^{(1,0)}_{\bar{q}}(t) = \big[{\rm Tra}\big(\nbv^{(1)}_{\bar{q}}(t)\big)\;{\rm Tra}\big(\nbv^{(1)}_{\bar{q}}(t)\big)\;\cdots\;{\rm Tra}\big(\nbv^{(1)}_{\bar{q}}(t)\big)\big]$ by using (\ref{cor:systemdiff_marginal_eq1}) in Corollary \ref{cor:systemdiff_marginal} to evaluate $\nbv^{(1)}_{\bar{q}}$. From (\ref{cor:systemdiff_marginal_eq1}), we note that $\nbv^{(1)}_{\bar{q}}$ is obtained from $\nbv^{(0)}_{\bar{q}}(t)$, which can be computed from \cite[Lemma~1]{yates2020age} as:
\begin{align}\label{diff_marginal_0}
\dot{\nbv}^{(0)}_{\bar{q}}(t) = \sum_{l \in \ncalL'_{\bar{q}}}{\lambda^{(l)} \nbv^{(0)}_{q_l}(t)} - \nbv^{(0)}_{\bar{q}}(t) \sum_{l \in \ncalL_{\bar{q}}}{\lambda^{(l)}}, \forall \bar{q} \in \mathcal{Q}.
\end{align}

Afterwards, $\nbV^{(2,1)}_{\bar{q}}(t)$ is computed from $\nbV^{(2,0)}_{\bar{q}}(t) = \big[{\rm Tra}\big(\nbv^{(2)}_{\bar{q}}(t)\big)\;{\rm Tra}\big(\nbv^{(2)}_{\bar{q}}(t)\big)\;\cdots\;{\rm Tra}\big(\nbv^{(2)}_{\bar{q}}(t)\big)\big]$ and $\nbV^{(1,1)}_{\bar{q}}(t)$ such that $\nbv^{(2)}_{\bar{q}}(t)$ can be evaluated from $\nbv^{(1)}_{\bar{q}}(t)$ using (\ref{cor:systemdiff_marginal_eq1}). The process can be repeated to compute $\nbV^{(m_1,m_2)}_{\bar{q}}(t)$ for the desired $m_1, m_2 \geq 2$ using $\nbV^{(m_1-1,m_2)}_{\bar{q}}(t)$ and $\nbV^{(m_1,m_2-1)}_{\bar{q}}(t)$ evaluated in the previous steps. Further, by inspecting the structure of $c_{\bar{q},\{k_1,k_2\}}(t)$ in (\ref{c_two}), we note that $\nbV^{(s_1,s_2)}_{\bar{q}}(t)$ can be computed from $\nbv^{(s_k)}_{\bar{q}}(t)$ and $\nbv^{(0)}_{\bar{q}}(t)$, where $\nbv^{(s_k)}_{\bar{q}}(t)$ can be evaluated from $\nbv^{(0)}_{\bar{q}}(t)$ using (\ref{cor:systemdiff_marginal_eq2}). Now, one can clearly see from Lemma \ref{lemma:1} that $\ncalV^{([m_{1}\;m_{2}\;m_{3}])}_{\bar{q},3}(t)$ can be computed from $\ncalV^{([m_{1}\;m_{2}])}_{\bar{q},2}(t) = \nbV^{(m_{1},m_{2})}_{\bar{q}}(t)$, and $\ncalV^{([s_{1}\;s_{2}\;s_{3}])}_{\bar{q},3}(t)$ can be computed from $\ncalV^{([s_{1}\;s_{2}])}_{\bar{q},2}(t) = \nbV^{(s_{1},s_{2})}_{\bar{q}}(t)$ and $\ncalV^{([s_{1}])}_{\bar{q},1}(t) = \nbv^{(s_{1})}_{\bar{q}}(t)$. Thus, through the repeated application of Lemma \ref{lemma:1}, we can evaluate $\ncalV^{(\nbm)}_{\bar{q},|\nrmK|}$ and $\ncalV^{(\nbs)}_{\bar{q},|\nrmK|}$ for an arbitrary set $\nrmK$ with $|\nrmK| \geq 3$.
\subsection{Stationary Joint Moments and Joint MGFs}
While Lemma \ref{lemma:1} holds for any collection of reset map matrices $\{\nbA_l\}_{l \in \ncalL}$, the set of differential equations in Lemma \ref{lemma:1} can be unstable for some choices of $\{\nbA_l\}_{l \in \ncalL}$. Thus, it is essential to investigate the conditions under which the differential equations in Lemma \ref{lemma:1} are stable. While there are several notions of stability including Lyapunov, Lagrange, and exponential stability, we are interested here in the asymptotic stability under which $v^{(\nbm)}_{\bar{q},\nrmK}(t)$ and $v^{(\nbs)}_{\bar{q},\nrmK}(t)$ respectively converge to the limits $\bar{v}^{(\nbm)}_{\bar{q},\nrmK}$ and $\bar{v}^{(\nbs)}_{\bar{q},\nrmK}$ as $t \rightarrow \infty$. The limiting values can then be evaluated as the solution of the equations resulting from setting the derivatives in Lemma \ref{lemma:1} to zero. To clearly see why we are concerned about the asymptotic stability in this paper, recall that our prime objective is to characterize the stationary joint moments and joint MGFs: $\underset{t \rightarrow \infty}{\rm lim} \nbbE\Big[\prod_{j=1}^{|\nrmK|}{x_{\nrmK(j)}^{[\nbm]_j}(t)}\Big]$ and $\underset{t \rightarrow \infty}{\rm lim} \nbbE\left[{\rm exp}\Big[\sum_{j=1}^{|\nrmK|}{[\nbs]_j x_{\nrmK(j)}(t)}\Big]\right]$. Under the asymptotic stability, these quantities can simply be evaluated from (\ref{exp_joint_moms}) and (\ref{exp_joint_mgfs}) as
\begin{align}
 \underset{t \rightarrow \infty}{\rm lim} \nbbE\Big[\prod_{j=1}^{|\nrmK|}{x_{\nrmK(j)}^{[\nbm]_j}(t)}\Big] = \sum_{\bar{q} \in \ncalQ}{\underset{t \rightarrow \infty}{\rm lim} v^{(\nbm)}_{\bar{q},\nrmK}(t)} = \sum_{\bar{q} \in \ncalQ}{\bar{v}^{(\nbm)}_{\bar{q},\nrmK}}, 
\end{align}
\begin{align}
 \underset{t \rightarrow \infty}{\rm lim} \nbbE\left[{\rm exp}\Big[\sum_{j=1}^{|\nrmK|}{[\nbs]_j x_{\nrmK(j)}(t)}\Big]\right] = \sum_{\bar{q} \in \ncalQ}{\underset{t \rightarrow \infty}{\rm lim} v^{(\nbs)}_{\bar{q},\nrmK}(t)} = \sum_{\bar{q} \in \ncalQ}{\bar{v}^{(\nbs)}_{\bar{q},\nrmK}}.
\end{align}

We now proceed to characterizing the conditions under which the differential equations in Lemma \ref{lemma:1} are asymptotically stable. Let us first recall the asymptotic stability theorem for linear systems. The linear system
\begin{align}\label{LSs_stab_theor}
\dot{\nbv}(t) = \nbv(t) \nbP, \; \nbv(0) = \nbv_0    
\end{align}
is asymptotically stable if and only if the eigenvalues of $\nbP$ have strictly negative real parts. Thus, according to (\ref{LSs_stab_theor}), it is always useful to write the differential equations at hand in a vector form to test the asymptotic stability. 
%This was also done in \cite{yates2020age} to characterize the conditions under which the differential equations describing the temporal evolution of the marginal $m$-th moments and marginal MGFs (given by (\ref{diff_marginal_mom}) and (\ref{diff_marginal_mgf})) are asymptotically stable. 
For all $\bar{q} \in \ncalQ$, let $\bar{\nbv}_{\bar{q}}^{(0)}$, $\bar{\ncalV}^{(\nbm)}_{\bar{q},|\nrmK|}$ and $\bar{\ncalV}^{(\nbs)}_{\bar{q},|\nrmK|}$ respectively denote the limiting values of $\nbv_{\bar{q}}^{(0)}(t)$, $\ncalV^{(\nbm)}_{\bar{q},|\nrmK|}(t)$ and $\ncalV^{(\nbs)}_{\bar{q},|\nrmK|}(t)$, when $t \rightarrow \infty$. Clearly, $\bar{\nbv}_{\bar{q}}^{(0)}$, $\bar{\ncalV}^{(\nbm)}_{\bar{q},|\nrmK|}$ and $\bar{\ncalV}^{(\nbs)}_{\bar{q},|\nrmK|}$ are the fixed points of (\ref{diff_marginal_0}), (\ref{Lemma1_eq1}) and (\ref{Lemma1_eq2}), respectively, which can be obtained after setting the derivatives to zero. The next theorem characterizes the conditions for asymptotic stability for the differential equations in Lemma \ref{lemma:1}, which in turn enables the characterization of stationary joint moments and joint MGFs for an arbitrary set $\nrmK$.
\begin{theorem}\label{theorem 1}
If the Markov chain $q(t)$ is ergodic with stationary distribution $\{\bar{\nbv}_{\bar{q}}^{(0)} > {\bf 0}: \bar{q} \in \mathcal{Q}\}$, and there exist positive fixed points $\{\bar{\ncalV}^{({\bf 1}_{|\nrmZ|})}_{\bar{q},|\nrmZ|} : \bar{q} \in \ncalQ \}_{|\nrmZ| \in \{1,2,\cdots, |\nrmK|\}}$ of (\ref{Lemma1_eq1}), then:
\begin{itemize}
    \item (i) For all $\bar{q} \in \ncalQ$, $v^{(\nbm)}_{\bar{q},\nrmK}(t)$ converges to $\bar{v}^{(\nbm)}_{\bar{q},\nrmK}$ satisfying
    \begin{align}\label{theorem 1_eq1}
  \bar{v}^{(\nbm)}_{\bar{q},\nrmK}\sum_{l \in \ncalL_{\bar{q}}}{\lambda^{(l)}} = \sum_{j=1}^{|\nrmK|}{[\nbm]_j \bar{v}^{(\nbm - \nbe_j)}_{\bar{q},\nrmK}} + \sum_{l \in \ncalL'_{\bar{q}}}{\lambda^{(l)}\Big[\bar{\ncalV}_{\bar{q}_l,|\nrmK|}^{(\nbm)} \times_1 \nbA_l \times_2 \nbA_l \cdots \times_{|\nrmK|} \nbA_l\Big]_{\nrmK}}.
\end{align}
    \item (ii) There exists $s_0 > 0$ such that for all $\nbs \in \ncalS = \{\nbs: \sum_{j=1}^{|\nrmK|}{[\nbs]_{j}} < s_0 \}$ and $\bar{q} \in \ncalQ$, $v^{(\nbs)}_{\bar{q},\nrmK}(t)$ and $c_{\bar{q},\nrmK}(t)$ respectively converge to $\bar{v}^{(\nbs)}_{\bar{q},\nrmK}$ and $\bar{c}_{\bar{q},\nrmK}$ satisfying
    \begin{align}\label{theorem 1_eq2}
 \bar{v}^{(\nbs)}_{\bar{q},\nrmK} \sum_{l \in \ncalL_{\bar{q}}}{\lambda^{(l)}} =  \bar{v}^{(\nbs)}_{\bar{q},\nrmK} \sum_{j=1}^{|\nrmK|}{[\nbs]_j} + \sum_{l \in \ncalL'_{\bar{q}}}{\lambda^{(l)}\Big[\bar{\ncalV}_{\bar{q}_l,|\nrmK|}^{(\nbs)} \times_1 \nbA_l \times_2 \nbA_l \cdots \times_{|\nrmK|} \nbA_l\Big]_{\nrmK}} + \bar{c}_{\bar{q},\nrmK},
\end{align}
where $\bar{c}_{\bar{q},\nrmK}$ is given by
\begin{align}
\bar{c}_{\bar{q},\nrmK} = \sum_{l \in \ncalL'_{\bar{q}}}{\lambda^{(l)} \sum_{\nrmZ \in 2^{\nrmK}\setminus\varnothing}{\nb1\big(\nrmZ_l = \nrmZ\big)\Big[\bar{\ncalV}_{\bar{q}_l,|\nrmK\setminus\nrmZ_l|}^{(\nbs')} \times_1 \nbA_l \times_2 \nbA_l \cdots \times_{|\nrmK \setminus \nrmZ_l|} \nbA_l\Big]_{\nrmK \setminus \nrmZ_l}}}.    
\end{align}
\end{itemize}
\end{theorem}
\begin{IEEEproof}
See Appendix \ref{app:theorem 1}.
\end{IEEEproof}

Theorem $\ref{theorem 1}$ is a generalization of \cite[Theorem 1]{yates2020age} which was focused on the characterization of the marginal stationary moments and MGFs, i.e., the fixed points of the differential equations in Corollary \ref{cor:systemdiff_marginal}. In particular, when $|\nrmK| = 1$, Theorem \ref{theorem 1} directly reduces to \cite[Theorem 1]{yates2020age}. A useful insight provided by Theorem \ref{theorem 1} is that the existence of the stationary joint first moments guarantees the existence of the stationary joint higher order moments and MGFs. It is worth emphasizing that the generality of Theorem \ref{theorem 1} lies in the fact that it allows the investigation of the stationary joint moments and MGFs for an arbitrary set of age processes under any arbitrary queueing discipline. This opens the door for the use of Theorem \ref{theorem 1} to study the joint analysis of age processes in networks for different queueing disciplines/status updating system settings in the literature, which have only been analyzed in terms of the marginal moments and MGFs until now.
\section{Stationary Joint MGF in Multi-source Updating Systems}\label{sec:MGF_analysis}
\begin{figure}[t!]
\centering
\includegraphics[width=0.8\textwidth]{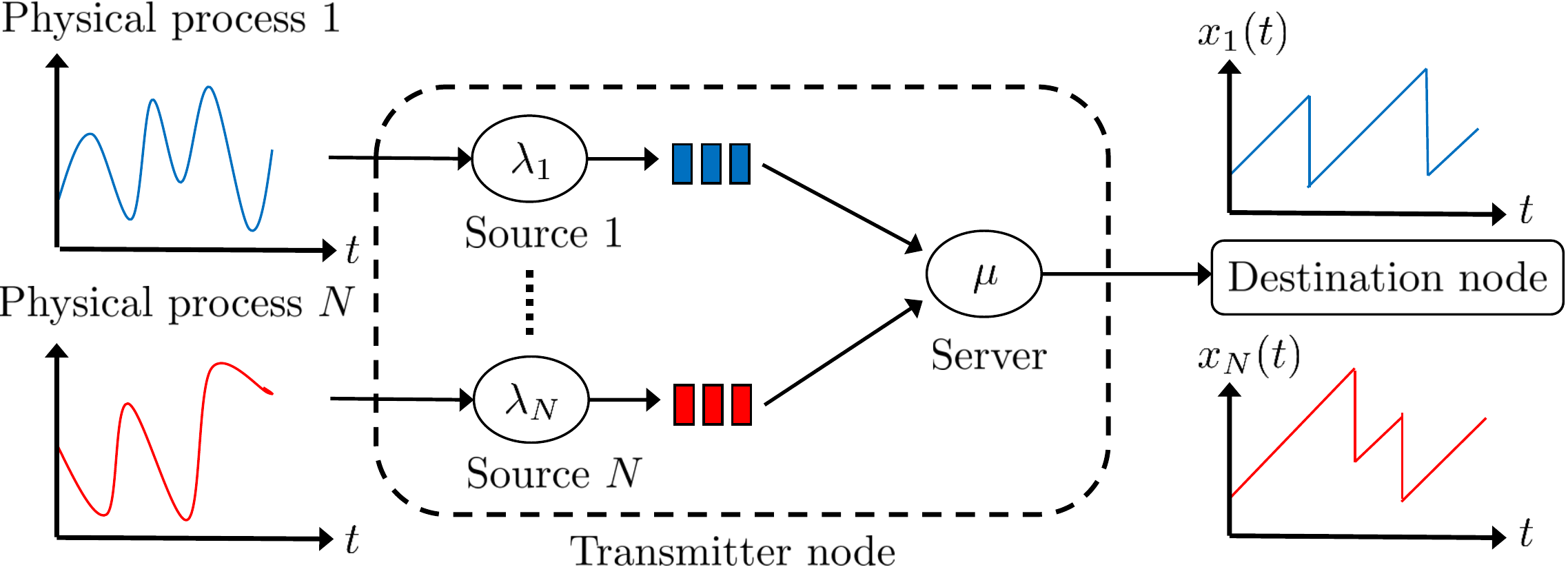}
\caption{An illustration of a multi-source status updating system.}
\label{f:multi_source_SU}
\end{figure}

In this section, we use Theorem \ref{theorem 1} to analyze the stationary joint MGF of the age processes in a multi-source status updating system, where a transmitter monitors $N$ physical processes, and sends its measurements to a destination in the form of status updates. As shown in Fig. \ref{f:multi_source_SU}, the transmitter consists of $N$ sources and a single server; each source generates status updates about one physical process, and the server delivers the status updates generated from the sources to the destination. Status updates generated by the $i$-th source are assumed to follow a Poisson process with rate $\lambda_i$. Further, the time needed by the server to send a status update is assumed to be a rate $\mu$ exponential random variable. Let $\rho = \frac{\lambda}{\mu}$ denote the server utilization factor, where $\lambda = \sum_{j=1}^{N}{\lambda_j}$. Further, we define $\lambda_{\nrmZ} = \sum_{j=1}^{|\nrmZ|}{\lambda_{\nrmZ(j)}}, \lambda_{-\nrmZ} = \sum_{j=1,\;j\notin \nrmZ}^{N}{\lambda_{\nrmZ(j)}}, \rho_{\nrmZ} = \frac{\lambda_{\nrmZ}}{\mu}$ and $\rho_{-\nrmZ} = \frac{\lambda_{-\nrmZ}}{\mu}$. Thus, we have $\rho_i = \frac{\lambda_i}{\mu}$, $\lambda_{-i} = \sum_{j=1,\;j\neq i}^{N}{\lambda_j}$, and $\rho_{-i} = \frac{\lambda_{-i}}{\mu}$. To maintain generality, we derive the joint MGF of an arbitrary set $\nrmK \subseteq \{1,2,\cdots,N\}$ of age processes (associated with the observed $N$ physical processes) at the destination under three different queueing disciplines for managing status update arrivals at the transmitter, which are described next.
\begin{itemize}
    \item {\it LCFS with no preemption (LCFS-NP)}: Under this queueing discipline, a new arriving status update at the transmitter (from any of the sources) enters service upon its arrival if the server is idle (i.e., there is no status update in service); otherwise, the new arriving status update is discarded.
    \item {\it LCFS with source-agnostic preemption in service (LCFS-PS)}: When the server is idle, the management of a new arriving status update under this queueing discipline is similar to the LCFS-NP one. However, when the server is busy, a new arriving status update replaces the current update being served (regardless of the index of its generating source) and the old update in service is discarded. 
    \item {\it LCFS with source-aware preemption in service (LCFS-SA)}: This queueing discipline is similar to the LCFS-PS one with the only difference being that a new arriving status update preempts the update in service only if the two updates (the new arriving update and the one in service) are generated from the same source.
\end{itemize}

\begin{figure}[t!]
\centering
\includegraphics[width=0.75\textwidth]{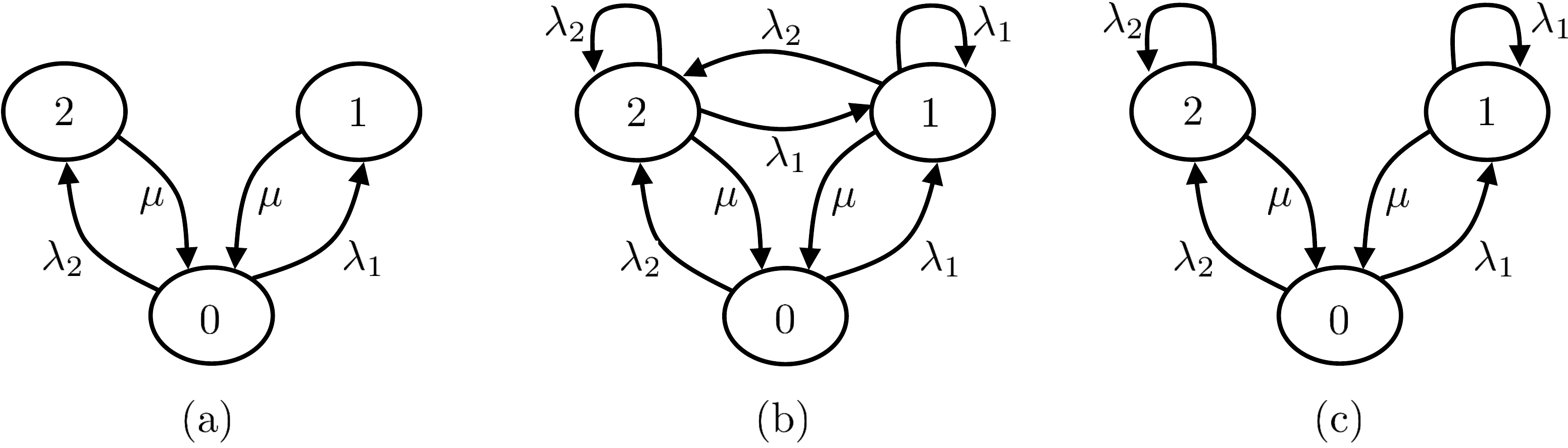}
\caption{Markov chains modeling the system discrete state $q(t)$ for $N = 2$ under different queueing disciplines: (a) LCFS-NP, (b) LCFS-PS, and (c) LCFS-SA.}
\label{f:MCs}
\end{figure}

 Using the notations of the SHS framework, the continuous state $\nbx(t)$ in each queueing discipline is given by $\nbx(t)=[x_0(t) \;x_1(t)\;\cdots x_N(t)]$, where $x_i(t), i \in 1:N$, represents the value of the source $i$'s AoI at the destination node, and $x_0(t)$ is the age of the status update in service. Further, the discrete state space in each queueing discipline is given by $\ncalQ = \{0,1,\cdots,N\}$, where $q(t)=0$ indicates that the system is empty and hence the server is idle, and $q(t)=i, i \in 1:N$, indicates that the server is serving a status update generated from the $i$-th source. Further, the continuous-time Markov chain modeling the system discrete state $q(t) \in \ncalQ$ under each of the queueing disciplines is depicted in Fig. \ref{f:MCs}.
\subsection{LCFS-NP Queueing Discipline}\label{sub:NP}
Table \ref{table:NP} presents the set of transitions $\ncalL$ and their impact on the values of both $q(t)$ and $\nbx(t)$. Before proceeding into evaluating $\bar{v}^{(\nbs)}_{\bar{q},\nrmK}$, $\forall \bar{q} \in \ncalQ$, satisfying (\ref{theorem 1_eq2}), we first describe the set of transitions as follows:

$l = 2i-1$: This transition occurs if there is a new arriving status update of source $i$ at the transmitter node when the server is idle. Note that the age of this new arriving status update is 0 and it does not have any impact on the AoI processes of the $N$ sources at the destination. Thus, as a result of this transition, the age process $x_0$ in the updated age vector is reset to 0 (i.e., $[\nbx \nbA_{2i-1}]_1 = 0$) whereas the other age processes remain the same.

$l = 2i$: This transition occurs when the source $i$'s status update in service is delivered to the destination. Thus, as a result of this transition, the source $i$'s AoI is reset to the age of the status update received at the destination whereas the AoI values of the other sources do not change. In addition, since the system becomes empty after the occurrence of this transition, the first element of the age vector $\nbx(t)$ becomes irrelevant. Following the convention of \cite{yates2018age}, we set the value corresponding to such irrelevant elements in the updated age value to 0, and thus we observe that $[\nbx \nbA_{2i}]_{1} = 0$.
\begin{table}
\centering
{\caption{Transitions of the LCFS-NP queueing discipline in Fig. 2a $(2 \leq i \leq N)$.} 
\label{table:NP}
\scalebox{.8}
{ \begin{tabular}{ |c |c|c|c|c|}
\hline
 $l$   & $q_l\rightarrow q'_l$  & $\lambda^{(l)}$ & $\nbx \nbA_l$ & $\nbA_l$\\ \hline
1& 0 $\rightarrow$ 1& $\lambda_1$&$[0\;x_1\;x_2\;\cdots\;x_N]$&
$\begin{bmatrix}
0 & 0 & 0 & \dots &  0\\ 0 & 1 & 0 & \dots & 0\\0 & 0 & 1 & \dots & 0\\ \vdots & \vdots & \vdots & \ddots & \vdots \\ 0 & 0 & 0 & \dots & 1
\end{bmatrix}$\\ \hline
2& 1 $\rightarrow$ 0&$\mu$&$[0\;x_0\;x_2\;\cdots\;x_N]$& $\begin{bmatrix}
0 & 1 & 0 & \dots &  0\\ 0 & 0 & 0 & \dots & 0\\0 & 0 & 1 & \dots & 0\\ \vdots & \vdots & \vdots & \ddots & \vdots \\ 0 & 0 & 0 & \dots & 1
\end{bmatrix}$\\ \hline
$2i-1$& 0 $\rightarrow$ $i$&$\lambda_i$&$[0\;x_1\;x_2\;\cdots\;x_N]$& $[{\bf 0}_{N+1}^{\rm T}\;\nbe_2^{\rm T}\;\nbe_3^{\rm T}\;\cdots\;\nbe_{N+1}^{\rm T}]$\\ \hline
$2i$& $i$ $\rightarrow$ 0&$\mu$&$[0\;x_1\;x_2\;\cdots\;x_{i-1}\;x_0\;\cdots\;x_N]$ & $[{\bf 0}_{N+1}^{\rm T}\;\nbe_2^{\rm T}\;\nbe_3^{\rm T}\;\cdots\;\nbe_i^{\rm T}\;\nbe_1^{\rm T}\;\cdots\;\nbe_{N+1}^{\rm T}]$
 \\ \hline
\end{tabular}}} 
\end{table}

Using Table \ref{table:NP}, we are now ready to derive $\{\bar{v}^{(\nbs)}_{\bar{q},\nrmK}\}_{\bar{q} \in \ncalQ}$ satisfying (\ref{theorem 1_eq2}), from which the stationary joint MGF of set $\nrmK$ is characterized in the following theorem.
\begin{theorem}\label{theorem:MGF_NP}
Under the LCFS-NP queueing discipline and for $\nbs = [s_{\nrmK(1)}\;s_{\nrmK(2)}\;\cdots\;s_{\nrmK(|\nrmK|)}]$, the stationary joint MGF of a set $\nrmK\subseteq \{1,2,\cdots,N\}$ of age processes is given by
\begin{align}\label{theorem:MGF_NP_eq1}
\overset{{\rm NP} }{M} (\nbs) = \sum_{\bar{q} \in \ncalQ}{\bar{v}^{(\nbs)}_{\bar{q},\nrmK}} = \mu^{|\nrmK|} \left(\prod_{i=1}^{|\nrmK|}{\lambda_{\nrmK(i)}}\right)\left(\dfrac{\mu}{\lambda + \mu}\right)\left(1 + \dfrac{\lambda}{\mu - \sum_{j=1}^{|\nrmK|}{s_{\nrmK(j)}}}\right) \sum_{\nrmP \in \mathcal{P}(\nrmK)}{\dfrac{1}{C(\nrmP)}},
\end{align}
where $C(\nrmP) = \prod_{i=1}^{|\nrmP|}{c_{\nrmP\left(i:|\nrmP|\right)}}$ such that $c_{\nrmZ}$ is defined for a set $\nrmZ \subseteq \{1,2,\cdots,N\}$ as 
 \begin{align}\label{theorem:MGF_NP_eq2}
   c_{\nrmZ} = \left(\lambda - \sum_{j=1}^{|\nrmZ|}{s_{\nrmZ(j)}}\right) \left(\mu - \sum_{j=1}^{|\nrmZ|}{s_{\nrmZ(j)}}\right) - \mu \sum_{j=1,j\notin\nrmZ}^{N}{\lambda_j}.
 \end{align}
%Under the LCFS-NP queueing discipline, the stationary joint MGF  of the $N$ age processes $\{x_1(t),x_2(t),\cdots,x_N(t)\}$ is given by
%\begin{align}\label{theorem:MGF_NP_eq1}
%\overset{{\rm NP} }{M} (\nbs)= \mu^N \left(\prod_{i=1}^{N}{\lambda_i}\right)\left(\dfrac{\mu}{\lambda + \mu}\right)\left(1 + \dfrac{\lambda}{\mu - \sum_{j=1}^{N}{s_j}}\right) \sum_{\nrmP \in \mathcal{P}(1:N)}{\dfrac{1}{C(\nrmP)}},
%\end{align}
%where $C(\nrmP) = \prod_{i=1}^{N}{c_{\nrmP(i:N)}}$ such that $c_{\nrmP(i:N)}$ is given by
%\begin{align}\label{theorem:MGF_NP_eq2}
% c_{\nrmP(i:N)} = \begin{cases}
 %\left(\lambda - \sum_{j=1}^{N}{s_{j}}\right) \left(\mu - \sum_{j=1}^{N}{s_{j}}\right),\; & i = 1, \\\left(\lambda - \sum_{j=i}^{N}{s_{\nrmP(j)}}\right) \left(\mu - \sum_{j=i}^{N}{s_{\nrmP(j)}}\right) - \mu  \sum_{j=1}^{i-1}{\lambda_{\nrmP(j)}},\; &i \in 2:N.
 %\end{cases}
%\end{align}
\end{theorem}
\begin{IEEEproof}
See Appendix \ref{app:theorem:MGF_NP}.
\end{IEEEproof}
\begin{cor}\label{cor:NP_marginal}
Under the LCFS-NP queueing discipline, the marginal stationary MGF of source $k$'s AoI is given by
\begin{align}\label{cor:NP_marginal_eq1}
\overset{{\rm NP} }{M} (\bar{s}_{k}) = \dfrac{\rho_k \left(1 + \rho - \bar{s}_k\right)}{\left(1 + \rho\right)\left(1 - \bar{s}_k\right)\left[\left(1 - \bar{s}_k\right)\left(\rho - \bar{s}_k\right) - \rho_{-k}\right]}, 
\end{align}
where $k \in 1:N$ and $\bar{s}_k = \frac{s_k}{\mu}$.
\end{cor}
\begin{IEEEproof}
This result follows from Theorem \ref{theorem:MGF_NP} by setting $\nrmK = \{k\}$.
\end{IEEEproof}
\begin{cor}\label{cor:NP_twosources}
For $k_1, k_2 \in 1:N$, the stationary joint MGF of the two AoI processes $x_{k_1}(t)$ and $x_{k_2}(t)$ under the LCFS-NP queueing discipline is given by
\begin{align}\label{cor:NP_twosources_eq1}
\overset{{\rm NP} }{M} (\bar{s}_{k_1},\bar{s}_{k_2})= \nonumber& \dfrac{\rho_{k_1} \rho_{k_2}\big[1 + \rho - (\bar{s}_{k_1} + \bar{s}_{k_2})\big]}{(1 + \rho)\Big[\big[\rho - (\bar{s}_{k_1} + \bar{s}_{k_2})\big] \big[1 - (\bar{s}_{k_1} + \bar{s}_{k_2})\big] - \rho_{-\{k_1,k_2\}}\Big]\Big[1 - (\bar{s}_{k_1} + \bar{s}_{k_2})\Big]}\\ &\times \sum_{i\in\{k_1,k_2\}}{\dfrac{1}{(1 - \bar{s}_i)(\rho - \bar{s}_i) - \rho_{-i}}}.
\end{align}
\end{cor}
\begin{IEEEproof}
This result follows from Theorem \ref{theorem:MGF_NP} by setting $\nrmK = \{k_1,k_2\}$.
\end{IEEEproof}
\begin{prop}\label{prop:MGF_NP_twosources}
For $k_1, k_2 \in 1:N$, the correlation coefficient of the two AoI processes $x_{k_1}(t)$ and $x_{k_2}(t)$ under the LCFS-NP queueing discipline is given by
\begin{align}\label{prop:MGF_NP_twosources_eq1}
\overset{\rm NP}{\rm Cor} = \dfrac{\rho\left(\rho_{k_1} + \rho_{k_2}\right) \left[\rho_{k_1}\rho_{k_2}\left(\rho + 2\right) - \rho_{-\{k_1,k_2\}} \left(1 + \rho\right)^2\right] - 2 \rho_{k_1}\rho_{k_2}\left(1+\rho\right)^2}{\left(\rho_{k_1} + \rho_{k_2}\right) \prod_{i\in\{k_1,k_2\}}{\sqrt{(1 + \rho)^2 [\rho^2 + 2 \rho_{-i} + 1] + \rho_{i}^2\rho(\rho + 2)}}}.
\end{align}
\end{prop}
\begin{IEEEproof}
See Appendix \ref{app:prop:MGF_NP_twosources}.
\end{IEEEproof}
\begin{cor}\label{cor_NP}
When $N=2$, the correlation coefficient of the two AoI processes $x_1(t)$ and $x_2(t)$ under the LCFS-NP queueing discipline is given by
\begin{align}\label{cor_NP_eq1}
\overset{\rm NP}{\rm Cor} = \dfrac{\rho_1 \rho_2 \big[\rho^3 - 2 (2\rho + 1)\big]}{\rho \prod_{i=1}^{2}{\sqrt{(1 + \rho)^2 [\rho^2 + 2 \rho_{-i} + 1] + \rho_{i}^2\rho(\rho + 2)}}}.
\end{align}
\end{cor}
\begin{IEEEproof}
This result follows from Proposition \ref{prop:MGF_NP_twosources} by setting $\rho = \rho_1 + \rho_2$.
\end{IEEEproof}
\subsection{LCFS-PS Queueing Discipline}\label{sub:PS}
The set of transitions under the LCFS-PS queueing discipline is listed in Table \ref{table:PS}. Different from the LCFS-NP queueing discipline, we note from transition $l = (2 + N)i - N + j$ that under the LCFS-PS queueing discipline, a new arriving status update at the transmitter preempts the packet in service regardless of its generating source (i.e., source-agnostic preemption). The stationary joint MGF of set $\nrmK$ for this case is provided in the next theorem.
\begin{table}
\centering
{\caption{Transitions of the LCFS-PS queueing discipline in Fig. 2b $(2 \leq i \leq N,\; 1 \leq j \leq N)$.} 
\label{table:PS}
\scalebox{.8}
{ \begin{tabular}{ |c |c|c|c|c|}
\hline
 $l$   & $q_l\rightarrow q'_l$  & $\lambda^{(l)}$ & $\nbx \nbA_l$ & $\nbA_l$\\ \hline
1& 0 $\rightarrow$ 1& $\lambda_1$&$[0\;x_1\;x_2\;\cdots\;x_N]$&
$\begin{bmatrix}
0 & 0 & 0 & \dots &  0\\ 0 & 1 & 0 & \dots & 0\\0 & 0 & 1 & \dots & 0\\ \vdots & \vdots & \vdots & \ddots & \vdots \\ 0 & 0 & 0 & \dots & 1
\end{bmatrix}$\\ \hline
2& 1 $\rightarrow$ 0&$\mu$&$[0\;x_0\;x_2\;\cdots\;x_N]$& $\begin{bmatrix}
0 & 1 & 0 & \dots &  0\\ 0 & 0 & 0 & \dots & 0\\0 & 0 & 1 & \dots & 0\\ \vdots & \vdots & \vdots & \ddots & \vdots \\ 0 & 0 & 0 & \dots & 1
\end{bmatrix}$\\ \hline
$2+j$& 1 $\rightarrow$ $j$& $\lambda_j$&$[0\;x_1\;x_2\;\cdots\;x_N]$&
$\begin{bmatrix}
0 & 0 & 0 & \dots &  0\\ 0 & 1 & 0 & \dots & 0\\0 & 0 & 1 & \dots & 0\\ \vdots & \vdots & \vdots & \ddots & \vdots \\ 0 & 0 & 0 & \dots & 1
\end{bmatrix}$\\ \hline
$(2+N)i - (N+1)$& 0 $\rightarrow$ $i$&$\lambda_i$&$[0\;x_1\;x_2\;\cdots\;x_N]$& $[{\bf 0}_{N+1}^{\rm T}\;\nbe_2^{\rm T}\;\nbe_3^{\rm T}\;\cdots\;\nbe_{N+1}^{\rm T}]$\\ \hline
$(2+N)i - N$& $i$ $\rightarrow$ 0&$\mu$&$[0\;x_1\;x_2\;\cdots\;x_{i-1}\;x_0\;\cdots\;x_N]$& $[{\bf 0}_{N+1}^{\rm T}\;\nbe_2^{\rm T}\;\nbe_3^{\rm T}\;\cdots\;\nbe_i^{\rm T}\;\nbe_1^{\rm T}\;\cdots\;\nbe_{N+1}^{\rm T}]$\\ \hline
$(2+N)i - N + j$& $i$ $\rightarrow$ $j$& $\lambda_j$&$[0\;x_1\;x_2\;\cdots\;x_N]$&
$[{\bf 0}_{N+1}^{\rm T}\;\nbe_2^{\rm T}\;\nbe_3^{\rm T}\;\cdots\;\nbe_{N+1}^{\rm T}]$\\ \hline
\end{tabular}}} 
\end{table}
\begin{theorem}\label{theorem:MGF_PS}
Under the LCFS-PS queueing discipline and for $\nbs = [s_{\nrmK(1)}\;s_{\nrmK(2)}\;\cdots\;s_{\nrmK(|\nrmK|)}]$, the stationary joint MGF of a set $\nrmK\subseteq \{1,2,\cdots,N\}$ of age processes is given by
\begin{align}\label{theorem:MGF_PS_eq1}
\overset{{\rm PS} }{M} (\nbs) = \sum_{\bar{q} \in \ncalQ}{\bar{v}^{(\nbs)}_{\bar{q},\nrmK}} = \mu^{|\nrmK|} \left(\prod_{i=1}^{|\nrmK|}{\lambda_{\nrmK(i)}}\right) \sum_{\nrmP \in \mathcal{P}(\nrmK)}{\dfrac{1}{C(\nrmP)}}.
\end{align}
\end{theorem}
\begin{IEEEproof}
See Appendix \ref{app:theorem:MGF_PS}.
\end{IEEEproof}
%\begin{IEEEproof}
%The proof follows along similar lines to the proof of Theorem \ref{theorem:MGF_NP}. In particular, Tables \ref{table:NP} and \ref{table:PS} will be used to derive $\bar{v}^{(s_1,s_2)}_{\bar{q},12}$ satisfying (\ref{theorem 1_eq2}), form which the stationary joint MGF  in (\ref{theorem:MGF_PS_eq1}) can be obtained. 
%\end{IEEEproof}
\begin{cor}\label{cor:PS_marginal}
Under the LCFS-PS queueing discipline, the marginal stationary MGF of source $k$'s AoI is given by
\begin{align}\label{cor:PS_marginal_eq1}
\overset{{\rm PS} }{M} (\bar{s}_{k}) = \dfrac{\rho_k}{\left(1 - \bar{s}_k\right)\left(\rho - \bar{s}_k\right) - \rho_{-k}}, 
\end{align}
where $k \in 1:N$.
\end{cor}
\begin{cor}\label{cor:PS_twosources}
For $k_1, k_2 \in 1:N$, the stationary joint MGF of the two AoI processes $x_{k_1}(t)$ and $x_{k_2}(t)$ under the LCFS-PS queueing discipline is given by
\begin{align}\label{cor:PS_twosources_eq1}
\overset{{\rm PS} }{M} (\bar{s}_{k_1},\bar{s}_{k_2})=  \dfrac{\rho_{k_1} \rho_{k_2}}{\big[\rho - (\bar{s}_{k_1} + \bar{s}_{k_2})\big] \big[1 - (\bar{s}_{k_1} + \bar{s}_{k_2})\big] - \rho_{-\{k_1,k_2\}}} \sum_{i\in\{k_1,k_2\}}{\dfrac{1}{(1 - \bar{s}_i)(\rho - \bar{s}_i) - \rho_{-i}}}.
\end{align}
\end{cor}
\begin{prop}\label{prop:MGF_PS_twosources}
For $k_1, k_2 \in 1:N$, the correlation coefficient of the two AoI processes $x_{k_1}(t)$ and $x_{k_2}(t)$ under the LCFS-PS queueing discipline is given by
\begin{align}\label{prop:MGF_PS_twosources_eq1}
\overset{\rm PS}{\rm Cor} = \dfrac{-2 \rho_{k_1} \rho_{k_2} }{\left(\rho_{k_1} + \rho_{k_2}\right)\sqrt{\left(\rho^2 + 2 \rho_{-k_1} + 1\right) \left(\rho^2 + 2 \rho_{-k_2} + 1\right)}}.
\end{align}
\end{prop}
\begin{IEEEproof}
See Appendix \ref{app:prop:MGF_PS_twosources}.
\end{IEEEproof}
\begin{cor}\label{cor_PS}
When $N=2$, the correlation coefficient of the two AoI processes $x_1(t)$ and $x_2(t)$ under the LCFS-PS queueing discipline is given by
\begin{align}\label{cor_PS_eq1}
\overset{\rm PS}{\rm Cor} = \dfrac{-2 \rho_1 \rho_2 }{\rho\sqrt{(\rho^2 + 2 \rho_1 + 1) (\rho^2 + 2 \rho_2 + 1)}}.
\end{align}
\end{cor}
\begin{nrem}
Note that the expression in (\ref{cor_PS_eq1}) is identical to the correlation coefficient expression derived in \cite[Theorem 2]{jiang2020correlation} using tools from Palm calculus (for a two-source system setting under the LCFS-PS queueing discipline).
\end{nrem}
\subsection{LCFS-SA Queueing Discipline}\label{sub:SA}
The set of transitions under the LCFS-SA queueing discipline is listed in Table \ref{table:SA}. From transition $l = 3i$, we note that the LCFS-SA queueing discipline only allows preemption in service between the status updates generated from the same source (i.e., source-aware preemption). In the next theorem, we provide the stationary joint MGF of set $\nrmK$ under this queueing discipline.
\begin{table}
\centering
{\caption{Transitions of the LCFS-SA queueing discipline in Fig. 2C $(2 \leq i \leq N)$.} 
\label{table:SA}
\scalebox{.8}
{ \begin{tabular}{ |c |c|c|c|c|}
\hline
 $l$   & $q_l\rightarrow q'_l$  & $\lambda^{(l)}$ & $\nbx \nbA_l$ & $\nbA_l$\\ \hline
1& 0 $\rightarrow$ 1& $\lambda_1$&$[0\;x_1\;x_2\;\cdots\;x_N]$&
$\begin{bmatrix}
0 & 0 & 0 & \dots &  0\\ 0 & 1 & 0 & \dots & 0\\0 & 0 & 1 & \dots & 0\\ \vdots & \vdots & \vdots & \ddots & \vdots \\ 0 & 0 & 0 & \dots & 1
\end{bmatrix}$\\ \hline
2& 1 $\rightarrow$ 0&$\mu$&$[0\;x_0\;x_2\;\cdots\;x_N]$& $\begin{bmatrix}
0 & 1 & 0 & \dots &  0\\ 0 & 0 & 0 & \dots & 0\\0 & 0 & 1 & \dots & 0\\ \vdots & \vdots & \vdots & \ddots & \vdots \\ 0 & 0 & 0 & \dots & 1
\end{bmatrix}$\\ \hline
3& 1 $\rightarrow$ 1& $\lambda_1$&$[0\;x_1\;x_2\;\cdots\;x_N]$&
$\begin{bmatrix}
0 & 0 & 0 & \dots &  0\\ 0 & 1 & 0 & \dots & 0\\0 & 0 & 1 & \dots & 0\\ \vdots & \vdots & \vdots & \ddots & \vdots \\ 0 & 0 & 0 & \dots & 1
\end{bmatrix}$\\ \hline
$3i - 2$& 0 $\rightarrow$ $i$&$\lambda_i$&$[0\;x_1\;x_2\;\cdots\;x_N]$& $[{\bf 0}_{N+1}^{\rm T}\;\nbe_2^{\rm T}\;\nbe_3^{\rm T}\;\cdots\;\nbe_{N+1}^{\rm T}]$\\ \hline
$3i - 1$& $i$ $\rightarrow$ 0&$\mu$&$[0\;x_1\;x_2\;\cdots\;x_{i-1}\;x_0\;\cdots\;x_N]$& $[{\bf 0}_{N+1}^{\rm T}\;\nbe_2^{\rm T}\;\nbe_3^{\rm T}\;\cdots\;\nbe_i^{\rm T}\;\nbe_1^{\rm T}\;\cdots\;\nbe_{N+1}^{\rm T}]$\\ \hline
$3i$& $i$ $\rightarrow$ $i$& $\lambda_i$&$[0\;x_1\;x_2\;\cdots\;x_N]$&
$[{\bf 0}_{N+1}^{\rm T}\;\nbe_2^{\rm T}\;\nbe_3^{\rm T}\;\cdots\;\nbe_{N+1}^{\rm T}]$\\ \hline
\end{tabular}}} 
\end{table}
\begin{theorem}\label{theorem:MGF_SA}
Under the LCFS-SA queueing discipline and for $\nbs = [s_{\nrmK(1)}\;s_{\nrmK(2)}\;\cdots\;s_{\nrmK(|\nrmK|)}]$, the stationary joint MGF of a set $\nrmK\subseteq \{1,2,\cdots,N\}$ of age processes is given by
\begin{align}\label{theorem:MGF_SA_eq1}
\overset{{\rm SA} }{M} (\nbs) = \sum_{\bar{q} \in \ncalQ}{\bar{v}^{(\nbs)}_{\bar{q},\nrmK}} = \mu^{|\nrmK|} \left(\prod_{i=1}^{|\nrmK|}{\lambda_{\nrmK(i)}}\right)\left(\dfrac{\mu}{\lambda + \mu}\right)\left(\lambda + \mu - \sum_{j=1}^{|\nrmK|}{s_{\nrmK(j)}}\right) \sum_{\nrmP \in \mathcal{P}(\nrmK)}{\dfrac{C'(\nrmP)}{C(\nrmP)}},
\end{align}
where $C'(\nrmP)$ is defined as
\begin{align}\label{theorem:MGF_SA_eq2}
C'(\nrmP) = \frac{\lambda_{\nrmP(|\nrmP|)} + \mu}{\mu} \times \frac{1}{\mu + \lambda_{\nrmP(|\nrmP|)} - s_{\nrmP(|\nrmP|)}} \times \prod_{i=1}^{|\nrmP|-1}{\dfrac{\mu + \lambda_{\nrmP(i)} - \sum_{j=i+1}^{|\nrmP|}{s_{\nrmP(j)}}}{\mu + \lambda_{\nrmP(i)} - \sum_{j=i}^{|\nrmP|}{s_{\nrmP(j)}}}}.
\end{align}
\end{theorem}
\begin{IEEEproof}
See Appendix \ref{app:theorem:MGF_SA}.
\end{IEEEproof}
\begin{cor}\label{cor:SA_marginal}
Under the LCFS-SA queueing discipline, the marginal stationary MGF of source $k$'s AoI is given by
\begin{align}\label{cor:SA_marginal_eq1}
\overset{{\rm SA} }{M} (\bar{s}_{k}) = \dfrac{\rho_k \left(1 + \rho_k\right)\left(1 + \rho - \bar{s}_k\right)}{\left(1 + \rho\right)\left(1 + \rho_k - \bar{s}_k\right)\left[\left(1 - \bar{s}_k\right)\left(\rho - \bar{s}_k\right) - \rho_{-k}\right]},
\end{align}
where $k \in 1:N$.
\end{cor}
\begin{cor}\label{cor:SA_twosources}
For $k_1, k_2 \in 1:N$, the stationary joint MGF of the two AoI processes $x_{k_1}(t)$ and $x_{k_2}(t)$ under the LCFS-SA queueing discipline is given by
\begin{align}\label{cor:SA_twosources_eq1}
\overset{{\rm SA} }{M} (\bar{s}_1,\bar{s}_2)= \nonumber& \dfrac{\rho_{k_1} \rho_{k_2}\big[1 + \rho - (\bar{s}_{k_1} + \bar{s}_{k_2})\big]}{(1 + \rho)\Big[\big[\rho - (\bar{s}_{k_1} + \bar{s}_{k_2})\big]\big[1 - (\bar{s}_{k_1} + \bar{s}_{k_2})\big] - \rho_{-\{k_1,k_2\}}\Big]} \\&\times \sum_{i \in \{k_1,k_2\}} \dfrac{(1 + \rho_i)(1 + \rho_{-i} - \bar{s}_i)}{(1 + \rho_i - \bar{s}_i)\big[1 + \rho_{-i} - (\bar{s}_{k_1} + \bar{s}_{k_2})\big]\big[(1 - \bar{s}_i)(\rho - \bar{s}_i) - \rho_{-i}\big]}.
\end{align}
\end{cor}
\begin{prop}\label{prop:MGF_SA_twosources}
For $k_1, k_2 \in 1:N$, the correlation coefficient of the two AoI processes $x_{k_1}(t)$ and $x_{k_2}(t)$ under the LCFS-SA queueing discipline is given by
\begin{align}\label{prop:MGF_SA_twosources_eq1}
\overset{\rm SA}{\rm Cor} = \dfrac{- \rho_{k_1} \rho_{k_2} g(\rho_{k_1},\rho_{k_2})}{(\rho_{k_1} + \rho_{k_2}) (1+ \rho_{k_1}) (1 + \rho_{k_2})\sqrt{f(\rho_{k_1}) f(\rho_{k_2})}},
\end{align}
where $g(\rho_{k_1},\rho_{k_2})$ and $f(\rho_i)$ are respectively given by:
\begin{align}
g(\rho_{k_1},\rho_{k_2}) \nonumber&= \rho_{k_1}^2\rho_{k_2}^2\big[(\rho_{k_1} + \rho_{k_2}) + 2(1 + \rho)^2\big] + \rho_{k_1} \rho_{k_2} (\rho_{k_1} + \rho_{k_2}) \big[2(\rho_{k_1} + \rho_{k_2}) + 3\rho^2 + 6\rho + 5 \big]\\&+ (\rho_{k_1} + \rho_{k_2})^3 + 2(\rho_{k_1} + \rho_{k_2})^2 (\rho^2 + 2\rho + 2) + (\rho_{k_1} + \rho_{k_2}) (3\rho^2 + 6\rho + 4) + 2 (1 + \rho)^2,
\end{align}
\begin{align}
\nonumber f(\rho_i) = \rho_i^3 (\rho + \rho_{-i}) + \rho_i^2 \big[\rho^3 (\rho + 2) + \rho_{-i} (2\rho^2 + 9\rho + 8)\big] + \rho_i \big[&\rho (2\rho + 1)(1 + \rho)^2 + \rho_{-i} (2\rho + 3) \\&+ \rho_{-i}^2 (3\rho + 4)\big] + (1 + \rho)^4.
\end{align}
\end{prop}
\begin{IEEEproof}
See Appendix \ref{app:prop:MGF_SA_twosources}.
\end{IEEEproof}
\begin{cor}\label{cor_SA}
When $N=2$, the correlation coefficient of the two AoI processes $x_1(t)$ and $x_2(t)$ under the LCFS-SA queueing discipline is given by
\begin{align}\label{cor_SA_eq1}
\overset{\rm SA}{\rm Cor} = \dfrac{- \rho_1 \rho_2 g(\rho_1,\rho_2)}{\rho (1+ \rho_1) (1 + \rho_2)\sqrt{f'(\rho_1,\rho_2) f'(\rho_2,\rho_1)}},
\end{align}
where $g(\rho_1,\rho_2)$ and $f'(y,z)$ are respectively given by:
\begin{align}
g(\rho_1,\rho_2) = \rho_1^2\rho_2^2(\rho + 2)(2\rho + 1) + \rho_1 \rho_2 \rho (1 + \rho)(3\rho + 5) + 2 (1 + \rho)^4,
\end{align}
\begin{align}
f'(y,z) = z^3 y + y^2 z(2\rho^2 + 7\rho + 4) + y z(\rho^2 + 6\rho + 3) + y^2 \rho^3 (\rho + 2) \nonumber&+ y \rho (2\rho^3 + 6\rho^2 + 4\rho + 1) \\&+ (1 + \rho)^4.
\end{align}
\end{cor}
\begin{figure}[t!]
\centering
\includegraphics[width=\textwidth]{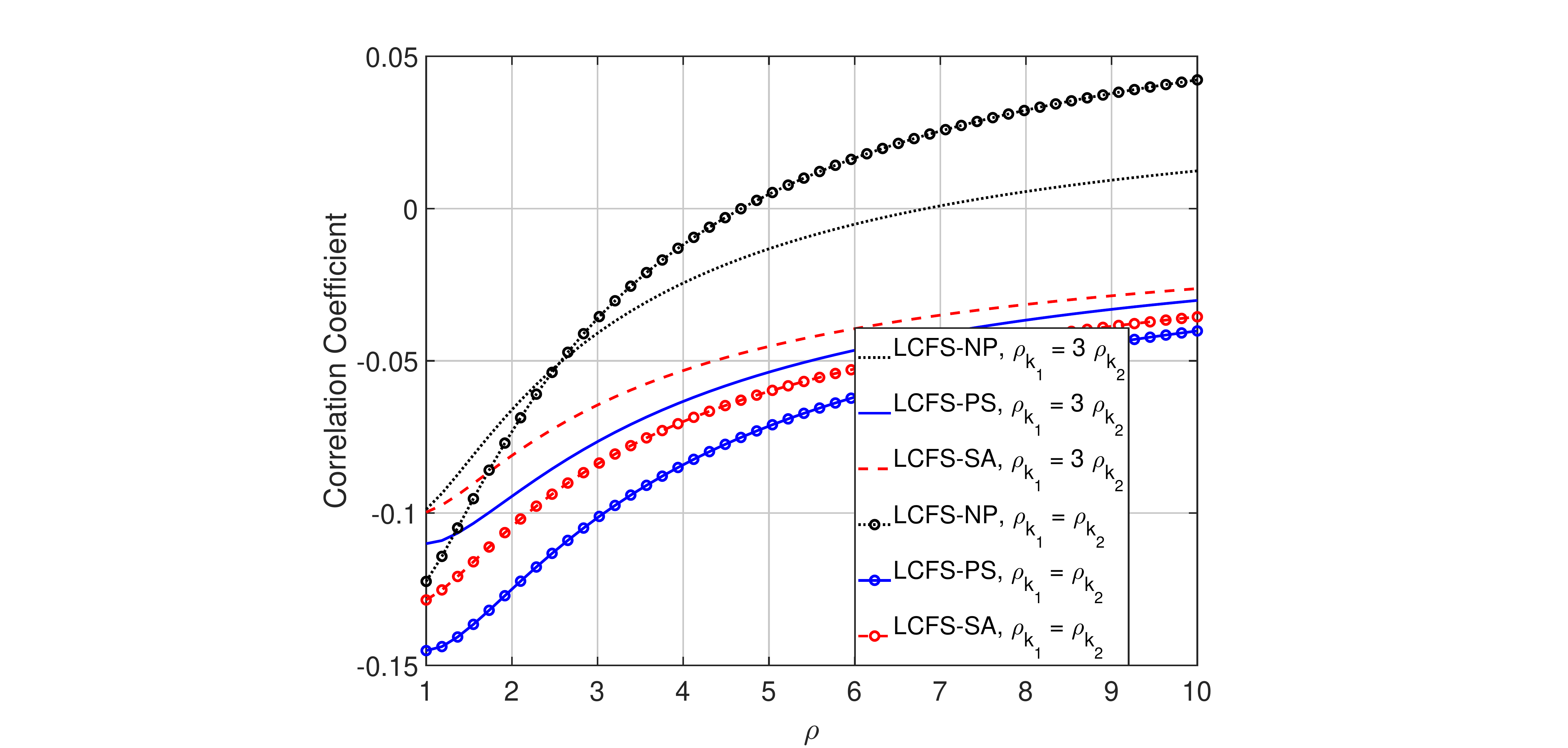}
\caption{Correlation coefficient of the two AoI processes $x_{k_1}(t)$ and $x_{k_2}(t)$ as a function of $\rho$ when $N > 2$ and $\rho_{-\{k_1,k_2\}} = 0.1 \rho$.}
\label{f:gencase_a}
\end{figure}
\begin{figure}[h!]
\centering
\includegraphics[width=\textwidth]{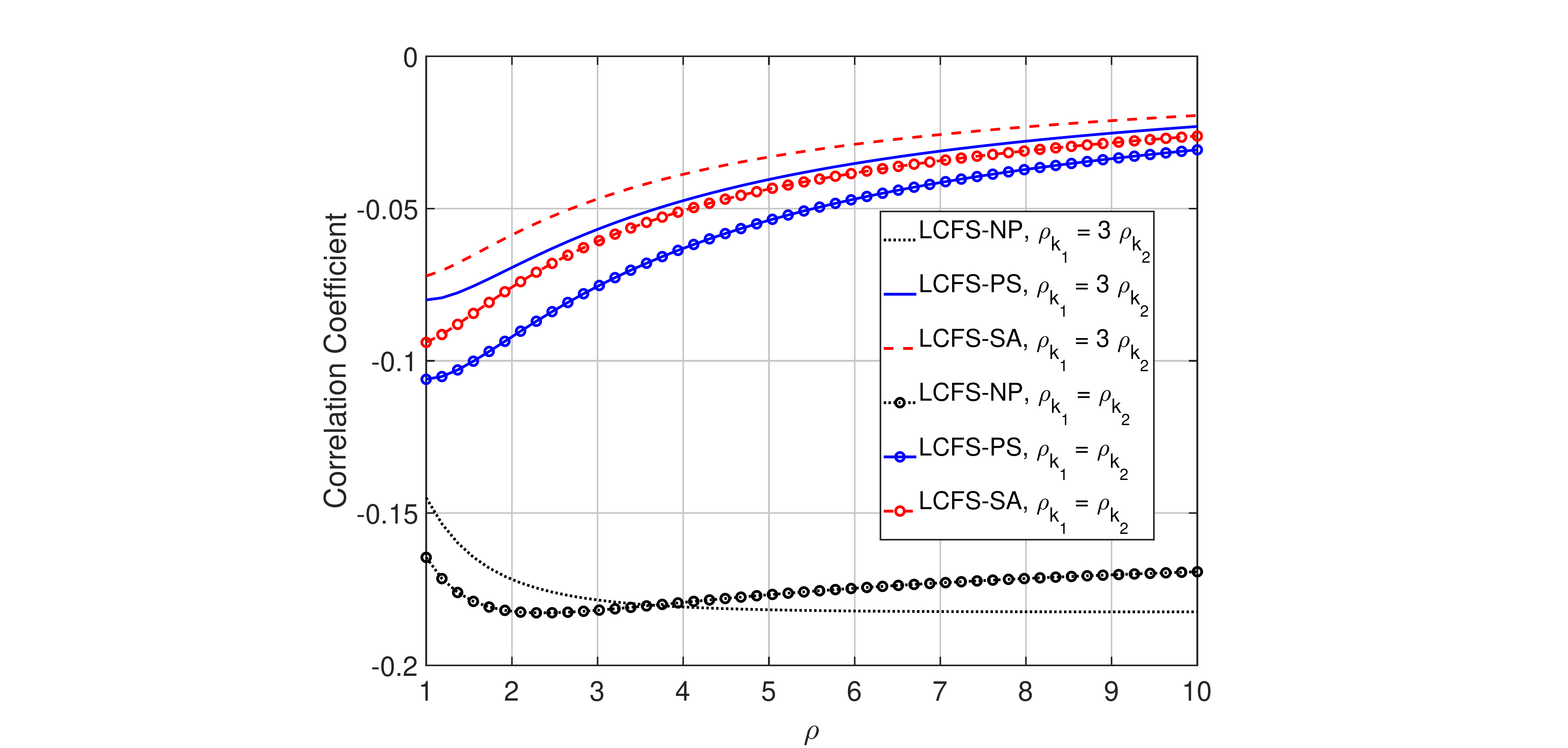}
\caption{Correlation coefficient of the two AoI processes $x_{k_1}(t)$ and $x_{k_2}(t)$ as a function of $\rho$ when $N > 2$ and $\rho_{-\{k_1,k_2\}} = 0.3 \rho$.}
\label{f:gencase_b}
\end{figure}
\begin{figure}[t!]
\centering
\includegraphics[width=\textwidth]{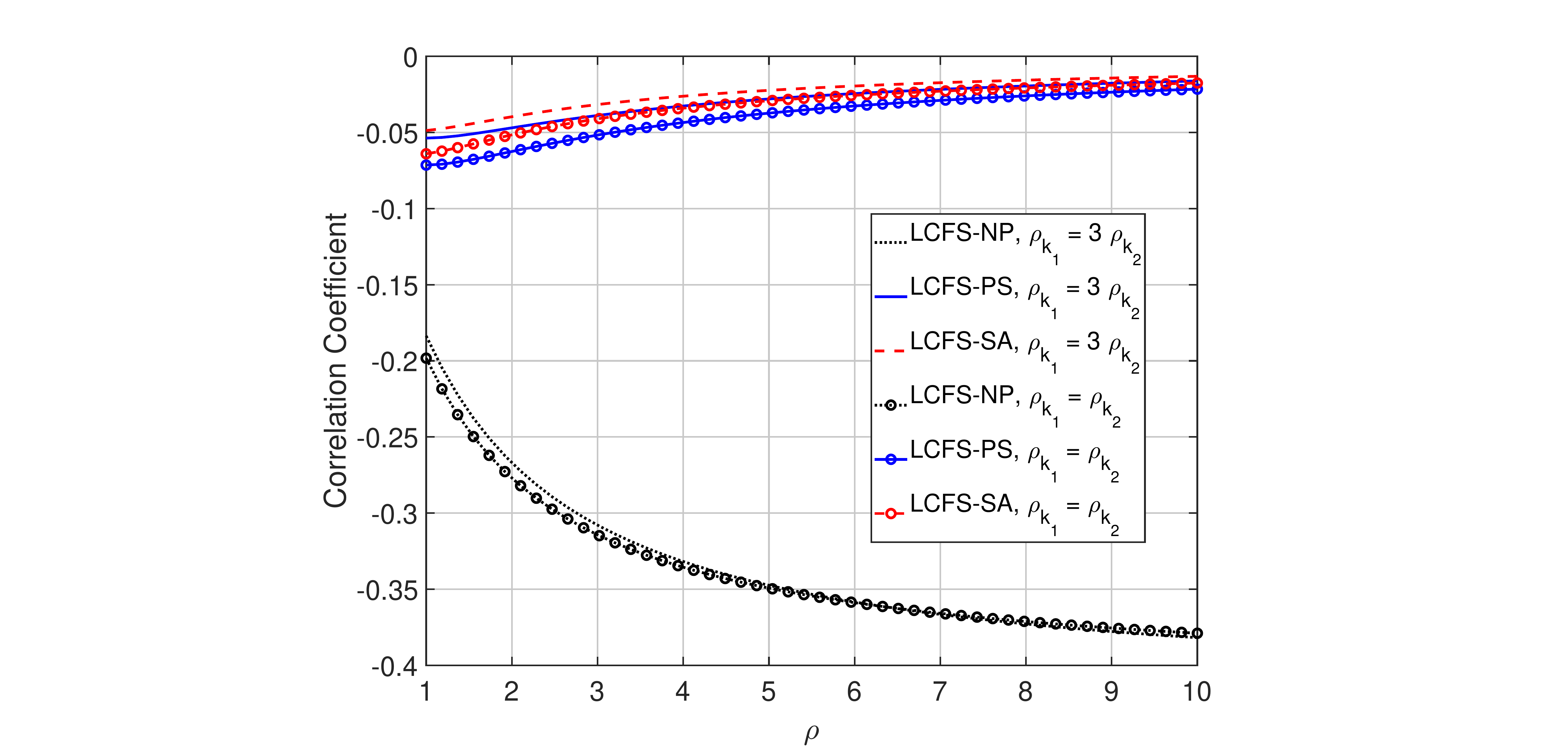}
\caption{Correlation coefficient of the two AoI processes $x_{k_1}(t)$ and $x_{k_2}(t)$ as a function of $\rho$ when $N > 2$ and $\rho_{-\{k_1,k_2\}} = 0.5 \rho$.}
\label{f:gencase_c}
\end{figure}
\subsection{Additional Discussion and Insights}\label{subsec:insights}
\begin{figure}[h!]
\centering
\includegraphics[width=\textwidth]{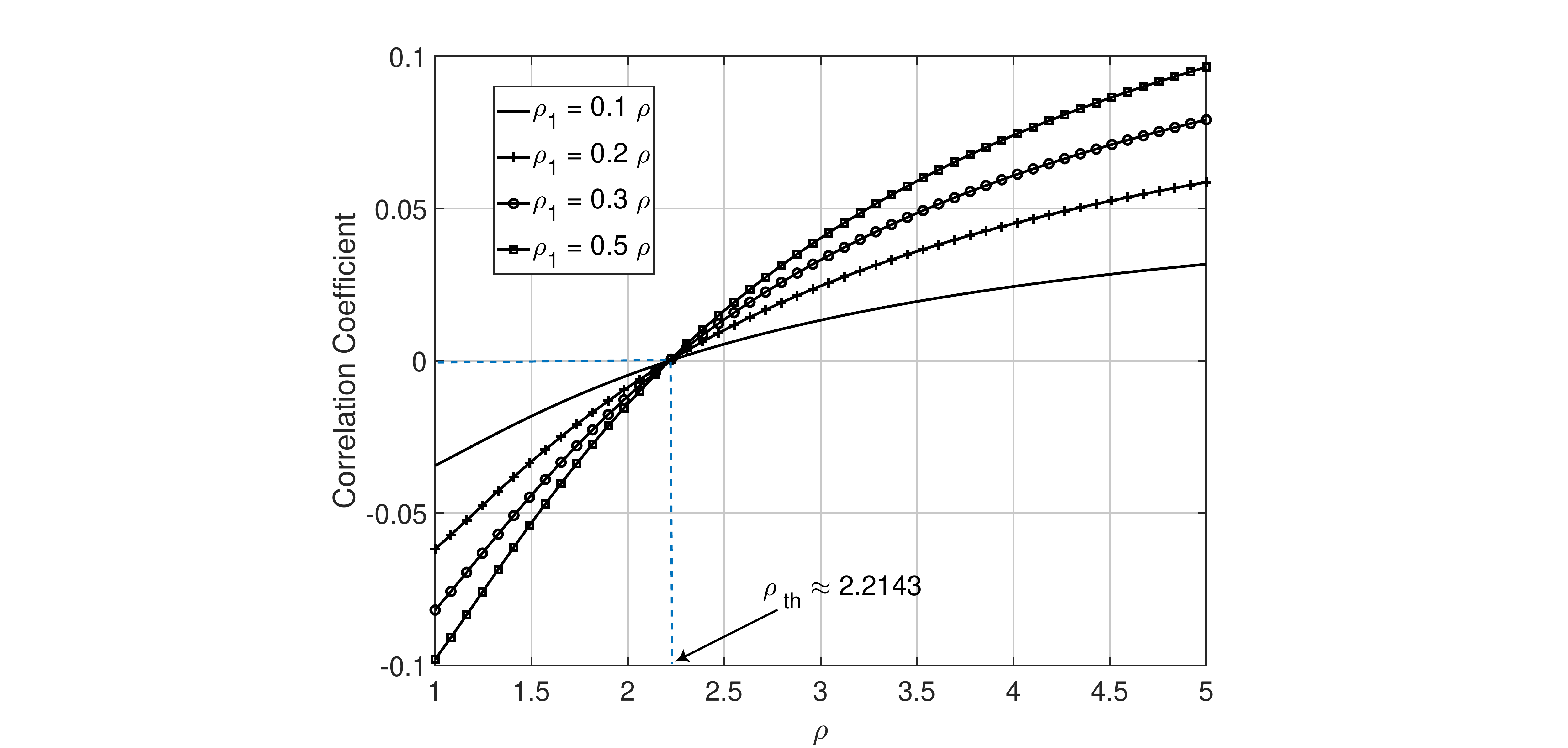}
\caption{Correlation coefficient of the two AoI processes $x_1(t)$ and $x_2(t)$ as a function of $\rho$ under the LCFS-NP queueing discipline when $N = 2$.}
\label{f:corr_NP}
\end{figure}

\begin{figure}[h!]
\centering
\includegraphics[width=\textwidth]{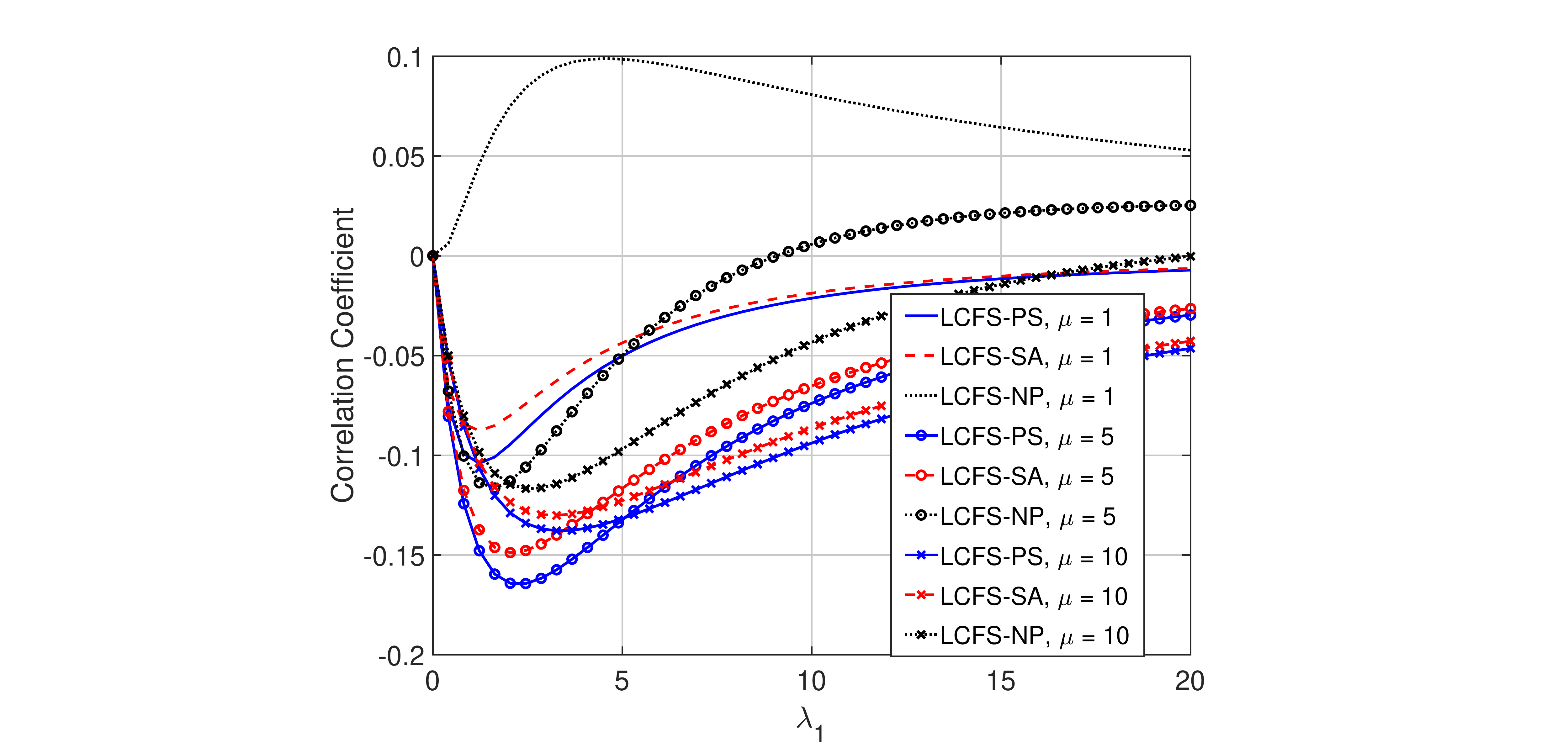}
\caption{Correlation coefficient of the two AoI processes $x_1(t)$ and $x_2(t)$ as a function of $\lambda_1$ for a fixed $\lambda_2 = 2$, $N = 2$, and different values of $\mu$.}
\label{f:corr}
\end{figure}

Now we will list some additional insights the can be obtained from the expressions derived in Section \ref{sec:MGF_analysis}. Please recall that several insights obtained from our SHS-based framework developed in this paper have already been presented in Section \ref{sec:joint_analysis}. First, we note from Propositions \ref{prop:MGF_NP_twosources}-\ref{prop:MGF_SA_twosources} that while the two age processes $x_{k_1}(t)$ and $x_{k_2}(t)$ are negatively correlated under preemptive in service queueing disciplines (LCFS-PS and LCFS-SA) for any choice of values of the system parameters, they may be positively correlated under the non-preemptive queueing discipline (LCFS-NP). This can also be observed from Figs. \ref{f:gencase_a}-\ref{f:gencase_c}. Further, when $N = 2$, one can deduce from Corollary \ref{cor_NP} that there exists a threshold value $\rho_{\rm th} \approx 2.2143$ of $\rho$ above which the two age processes $x_1(t)$ and $x_2(t)$ are positively correlated under the LCFS-NP queueing discipline, as shown in Fig. \ref{f:corr_NP}. This follows from the fact that the term $[\rho^3 - 2(2\rho + 1)]$ in (\ref{cor_NP_eq1}) is monotonically increasing for $\rho > \frac{2}{\sqrt{3}}$, and it equals to zero at $\rho \approx 2.2143$. Further, we observe from Figs. \ref{f:gencase_a}-\ref{f:gencase_c} and Fig. \ref{f:corr} that the source-aware preemption in service slightly reduces the negative correlation of the two age processes compared to the source-agnostic one.
\section{Conclusion}\label{sec:con}
In this paper, we developed an SHS-based general framework to facilitate the study of joint distributional properties of an arbitrary set of AoI/age processes in a network. In particular, a system of first-order linear differential equations was first derived for the temporal evolution of both the joint moments and the joint MGFs for an arbitrary set of age processes. Afterwards, we characterized the conditions under which the derived differential equations are asymptotically stable, which in turn enabled the characterization of the stationary joint moments and joint MGFs for the AoI processes forming the set of interest. We demonstrated the generality of our framework by recovering several existing results as its special cases. An interesting insight obtained from our analysis is that the existence of the stationary joint first moments guarantees the existence of the stationary joint higher order moments and MGFs.

As an application of our framework, we obtained closed-from expressions of the stationary joint MGF in multi-source updating systems under several queueing disciplines including non-preemptive and source-agnostic/source-aware preemptive in service queueing disciplines. Using these MGF expressions, we derived closed-form expressions of the correlation coefficient between any two arbitrary AoI processes in the system. Our derived correlation coefficient expressions demonstrated that while any two AoI processes are negatively correlated under preemptive in service queueing disciplines for any choice of values of the system parameters, they may be positively correlated under the non-preemptive queueing discipline. For instance, for a two-source updating system, there exists a threshold value of server utilization in the non-preemptive queueing discipline above which the two age processes are positively correlated. Further, we numerically demonstrated that the source-aware preemption in service slightly reduces the negative correlation of the two age processes compared to the source-agnostic one. 
 
 The generality of our analytical framework stems from the fact that it allows one to understand the joint distributional properties for an arbitrary set of AoI processes in a broad range of system settings under any arbitrary queueing discipline. This, in turn, opens the door for the use of our framework in the future to investigate the stationary joint moments and MGFs of age processes for a variety of queuing disciplines/status updating systems that have only been analyzed in terms of the marginal moments and MGFs until now. 
\appendix
%\begin{figure*}[t!]
%\subfloat[]{\includegraphics[width=\textwidth]{gencase_a.eps}%
%\label{f:gencase_a}} \vfil
%\subfloat[]{\includegraphics[width=\textwidth]{gencase_b.eps}%
%\label{f:gencase_b}} \vfil
%\subfloat[]{\includegraphics[width=\textwidth]{gencase_c.eps}%
%\label{f:gencase_c}} 
% \caption{Correlation coefficient of the two AoI processes $x_{k_1}(t)$ and $x_{k_2}(t)$ as a function of $\rho$ when $N > 2$. We use: i) $\rho_{-\{k_1,k_2\}} = 0.1 \rho$ in the top figure, ii) $\rho_{-\{k_1,k_2\}} = 0.3 \rho$ in the middle figure, and iii) $\rho_{-\{k_1,k_2\}} = 0.5 \rho$ in the bottom figure.}\label{f:gencase}
%\end{figure*} 
%%%%%
\subsection{Proof of Lemma~\ref{lemma:1}} \label{app:lemma:1}
To derive this result, we follow a similar approach to that in \cite{yates2020age} and \cite{hespanha2006modelling}, where the idea is to define the test functions $\{\psi(q,\nbx)\}$ whose expected values $\{\nbbE[\psi(q(t),\nbx(t))]\}$ are quantities of interest. Then, one can use the SHS framework to derive a system of differential equations for the temporal evolution of the expected values of the defined test functions. Since we are interested in the joint analysis of age processes in this paper, we define the following two classes of test functions
\begin{align}\label{test_mom}
\psi^{(\nbm)}_{\bar{q},\nrmK}(q,\nbx) = \prod_{j=1}^{|\nrmK|}{x_{\nrmK(j)}^{[\nbm]_j}} \delta_{\bar{q},q}, \forall \bar{q} \in \ncalQ, 
\end{align}
\begin{align}\label{test_mgf}
\psi^{(\nbs)}_{\bar{q},\nrmK}(q,\nbx) = {\rm exp}\left[\sum_{j=1}^{|\nrmK|}{[\nbs]_j x_{\nrmK(j)}}\right] \delta_{\bar{q},q}, \forall \bar{q} \in \ncalQ.
\end{align}

Clearly, taking the expectation of the two classes of test functions in (\ref{test_mom}) and (\ref{test_mgf}) gives $\{v^{(\nbm)}_{\bar{q},\nrmK}(t)\}$ and $\{v^{(\nbs)}_{\bar{q},\nrmK}(t)\}$, respectively. Now, we apply the SHS mapping $\psi \to L\psi$ (known as the extended generator) to every test function in (\ref{test_mom}) and (\ref{test_mgf}). Since the test functions defined above are time-invariant, it follows from \cite[Theorem~1]{hespanha2006modelling} that the extended generator of the considered piecewise linear SHS with linear reset maps is given by
\begin{align}\label{extendgen}
L\psi(q,\nbx) = \frac{\partial \psi(q,\nbx)}{\partial \nbx} {\bf 1}^{\rm T} + \underbrace{\sum_{l \in \ncalL}{\lambda^{(l)}(q)\big[\psi(q'_l,\nbx\nbA_l) - \psi(q,\nbx)\big]}}_{\theta(q,\nbx)},
\end{align}
where the row vector $\partial\psi(q,\nbx)/\partial\nbx$  denotes the gradient. Applying (\ref{extendgen}) to the test functions in (\ref{test_mom}) and (\ref{test_mgf}), we have
\begin{align}\label{extedgen_mom}
L\psi^{(\nbm)}_{\bar{q},\nrmK}(q,\nbx) = \frac{\partial \psi^{(\nbm)}_{\bar{q},\nrmK}(q,\nbx)}{\partial \nbx} {\bf 1}^{\rm T} +\theta^{(\nbm)}_{\bar{q},\nrmK}(q,\nbx),
\end{align}
\begin{align}\label{extedgen_mgf}
L\psi^{(\nbs)}_{\bar{q},\nrmK}(q,\nbx) = \frac{\partial \psi^{(\nbs)}_{\bar{q},\nrmK}(q,\nbx)}{\partial \nbx} {\bf 1}^{\rm T} +\theta^{(\nbs)}_{\bar{q},\nrmK}(q,\nbx),   
\end{align}
where
\begin{align}\label{partial_mom}
\frac{\partial \psi^{(\nbm)}_{\bar{q},\nrmK}(q,\nbx)}{\partial \nbx} = \sum_{j=1}^{|\nrmK|}{[\nbm]_j \psi_{\bar{q},\nrmK}^{(\nbm - \nbe_j)} \nbe_{\nrmK(j)}},
\end{align}
\begin{align}
\frac{\partial \psi^{(\nbs)}_{\bar{q},\nrmK}(q,\nbx)}{\partial \nbx} = \sum_{j=1}^{|\nrmK|}{[\nbs]_j \psi_{\bar{q},\nrmK}^{(\nbs)} \nbe_{\nrmK(j)}},
\end{align}
where in (\ref{partial_mom}), $\nbe_j \in \nbbR^{1 \times |\nrmK|}$ and $\nbe_{\nrmK(j)} \in \nbbR^{1 \times n}$. Now, to obtain $\theta^{(\nbm)}_{\bar{q},\nrmK}(q,\nbx)$ and $\theta^{(\nbs)}_{\bar{q},\nrmK}(q,\nbx)$, note that
\begin{align}\label{psi_m1_m2}
\psi^{(\nbm)}_{\bar{q},\nrmK}(q'_l,\nbx\nbA_l) = \prod_{j=1}^{|\nrmK|}{[\nbx\nbA_l]_{\nrmK(j)}^{[\nbm]_j}  \delta_{\bar{q},q'_l}} \a \prod_{j=1}^{|\nrmK|}{\big[\nbx^{[\nbm]_j}\nbA_l\big]_{\nrmK(j)}  \delta_{\bar{q},q'_l}},
\end{align}
\begin{align}
\psi^{(\nbs)}_{\bar{q},\nrmK}(q'_l,\nbx\nbA_l) = {\rm exp}\left[\sum_{j=1}^{|\nrmK|}{[\nbs]_j [\nbx \nbA_l]_{\nrmK(j)}}\right] \delta_{\bar{q},q'_l},
\end{align}
\begin{align}
\delta_{\bar{q},q'_l} \delta_{q_l,q} = \begin{cases}
\delta_{q_l,q},\; l \in \ncalL'_{\bar{q}},\\
0,\;\;\; {\rm otherwise}
\end{cases}   
,\; \delta_{\bar{q},q} \delta_{q_l,q} = \begin{cases}
\delta_{\bar{q},q},\; l \in \ncalL_{\bar{q}},\\
0,\;\;\; {\rm otherwise},
\end{cases}
\end{align}
where step (a) in (\ref{psi_m1_m2}) follows from the fact that $\nbA_l$ has no more than a single 1 in a column. Thus, we have
\begin{align}
\theta^{(\nbm)}_{\bar{q},\nrmK}(q,\nbx) = \sum_{l \in \ncalL'_{\bar{q}}}{\lambda^{(l)}\prod_{j=1}^{|\nrmK|}{\big[\nbx^{[\nbm]_j} \nbA_l\big]_{\nrmK(j)}} \delta_{q_l,q}} - \prod_{j=1}^{|\nrmK|}{x^{[\nbm]_j} _{\nrmK(j)}} \delta_{\bar{q},q} \sum_{l \in \ncalL_{\bar{q}}}{\lambda^{(l)}},
\end{align}
\begin{align}
\theta^{(\nbs)}_{\bar{q},\nrmK}(q,\nbx) = \sum_{l \in \ncalL'_{\bar{q}}}{\lambda^{(l)}{\rm exp}\left[\sum_{j=1}^{|\nrmK|}{[\nbs]_j [\nbx \nbA_l]_{\nrmK(j)}}\right] \delta_{q_l,q}} -{\rm exp}\left[\sum_{j=1}^{|\nrmK|}{[\nbs]_j x_{\nrmK(j)}}\right] \delta_{\bar{q},q} \sum_{l \in \ncalL_{\bar{q}}}{\lambda^{(l)}}.
\end{align}

Finally, the system of differential equations in 
(\ref{Lemma1_eq1}) and (\ref{Lemma1_eq2}) can be derived by applying Dynkin's formula \cite{hespanha2006modelling} to each test function and its associated extended generator. In particular, the Dynkin's formula can be expressed as
\begin{align}\label{formula}
\frac{{\rm d}\nbbE[\psi(q(t),\nbx(t))]}{{\rm d}t} = \nbbE[L\psi(q(t),\nbx(t))].   
\end{align}

Hence, from (\ref{exp_joint_moms_q}), (\ref{test_mom}) and (\ref{extedgen_mom}), we get
\begin{align}\label{diff_mom_single}
\dot{v}^{(\nbm)}_{\bar{q},\nrmK}(t) = \nbbE[L\psi^{(\nbm)}_{\bar{q},\nrmK}(q(t),\nbx(t))] &\nonumber= \nbbE\left[\frac{\partial \psi^{(\nbm)}_{\bar{q},\nrmK}(q(t),\nbx(t))}{\partial \nbx(t)} {\bf 1}^{\rm T}\right] + \nbbE[\theta^{(\nbm)}_{\bar{q},\nrmK}(q(t),\nbx(t))], \\&= \sum_{j=1}^{|\nrmK|}{[\nbm]_j v^{(\nbm - \nbe_j)}_{\bar{q},\nrmK}(t)} + \nbbE[\theta^{(\nbm)}_{\bar{q},\nrmK}(q(t),\nbx(t))],
\end{align}
where
\begin{align}\label{cons_A}
 \nbbE[\theta^{(\nbm)}_{\bar{q},\nrmK}(q(t),\nbx(t))]\nonumber&= \sum_{l \in \ncalL'_{\bar{q}}}{\lambda^{(l)}\nbbE\left[\prod_{j=1}^{|\nrmK|}{\big[\nbx^{[\nbm]_j}(t) \nbA_l\big]_{\nrmK(j)}} \delta_{q_l,q(t)}\right]} - \nbbE\left[\prod_{j=1}^{|\nrmK|}{x_{\nrmK(j)}^{[\nbm]_j}(t)} \delta_{\bar{q},q(t)}\right] \sum_{l \in \ncalL_{\bar{q}}}{\lambda^{(l)}},\\& \a \sum_{l \in \ncalL'_{\bar{q}}}{\lambda^{(l)}\big[\ncalV_{\bar{q}_l,|\nrmK|}^{(\nbm)}(t) \times_1 \nbA_l \times_2 \nbA_l \cdots \times_{|\nrmK|} \nbA_l\big]_{\nrmK}} - v^{(\nbm)}_{\bar{q},\nrmK}(t)\sum_{l \in \ncalL_{\bar{q}}}{\lambda^{(l)}},
\end{align}
where step (a) follows from the definition of $\ncalV_{\bar{q}_l,|\nrmK|}^{(\nbm)}(t)$ as $\left[\mathcal{V}_{\bar{q}_l,|\nrmK|}^{(\nbm)}(t)\right]_{\nrmK} = v_{\bar{q}_l,\nrmK}^{(\nbm)}(t)$ along with the fact that $[\nbA_l]_j = {\bf 0}^{\rm T}$ or $\nbe_i^{\rm T},$ where $i, j \in 1:n$. Let $\nrmZ_l = \{j \in \nrmK: [ \nbA_l]_j = 0\}$. When $\nrmZ_l \neq \varnothing$, one can easily see that $\nbbE\left[\prod_{j=1}^{|\nrmK|}{\big[\nbx^{[\nbm]_j}(t) \nbA_l\big]_{\nrmK(j)}} \delta_{q_l,q(t)}\right] = \big[\ncalV_{\bar{q}_l,|\nrmK|}^{(\nbm)}(t) \times_1 \nbA_l \times_2 \nbA_l \cdots \times_{|\nrmK|} \nbA_l\big]_{\nrmK} = 0$. On the other hand, when $\nrmZ_l = \varnothing$, $\nbbE\left[\prod_{j=1}^{|\nrmK|}{\big[\nbx^{[\nbm]_j}(t) \nbA_l\big]_{\nrmK(j)}} \delta_{q_l,q(t)}\right] = v_{\bar{q}_l,\nrmK'}^{(\nbm)}(t) =  \big[\ncalV_{\bar{q}_l,|\nrmK|}^{(\nbm)}(t) \times_1 \nbA_l \times_2 \nbA_l \cdots \times_{|\nrmK|} \nbA_l\big]_{\nrmK}$ such that $|\nrmK'| = |\nrmK|$ and $\big[\nbx^{[\nbm]_j}(t) \nbA_l\big]_{\nrmK(j)} = [\nbx(t)]^{[\nbm]_j}_{\nrmK'(j)}, \forall j \in 1 : |\nrmK|$. Substituting (\ref{cons_A}) into (\ref{diff_mom_single}) yields
\begin{align}\label{diff_mom_single_finalform}
\dot{v}^{(\nbm)}_{\bar{q},\nrmK}(t) = \sum_{j=1}^{|\nrmK|}{[\nbm]_j v^{(\nbm - \nbe_j)}_{\bar{q},\nrmK}(t)} + \sum_{l \in \ncalL'_{\bar{q}}}{\lambda^{(l)}\big[\ncalV_{\bar{q}_l,|\nrmK|}^{(\nbm)}(t) \times_1 \nbA_l \times_2 \nbA_l \cdots \times_{|\nrmK|} \nbA_l\big]_{\nrmK}} - v^{(\nbm)}_{\bar{q},\nrmK}(t)\sum_{l \in \ncalL_{\bar{q}}}{\lambda^{(l)}}.
\end{align}

Further, from (\ref{exp_joint_mgfs_q}), (\ref{test_mgf}) and (\ref{extedgen_mgf}), we get
\begin{align}\label{diff_mgf_single}
\dot{v}^{(\nbs)}_{\bar{q},\nrmK}(t) = \nbbE[L\psi^{(\nbs)}_{\bar{q},\nrmK}(q(t),\nbx(t))] &\nonumber= \nbbE\left[\frac{\partial \psi^{(\nbs)}_{\bar{q},\nrmK}(q(t),\nbx(t))}{\partial \nbx(t)} {\bf 1}^{\rm T}\right] + \nbbE[\theta^{(\nbs)}_{\bar{q},\nrmK}(q(t),\nbx(t))], \\&= \sum_{j=1}^{|\nrmK|}{[\nbs]_j} v^{(\nbs)}_{\bar{q},\nrmK}(t) + \nbbE[\theta^{(\nbs)}_{\bar{q},\nrmK}(q(t),\nbx(t))],
\end{align}
where $\nbbE\big[\theta^{(\nbs)}_{\bar{q},\nrmK}(q(t),\nbx(t))\big] =$ 
\begin{align}\label{cons_B}
\nonumber&\sum_{l \in \ncalL'_{\bar{q}}}{\lambda^{(l)}\nbbE\left[{\rm exp}\Big[\sum_{j=1}^{|\nrmK|}{[\nbs]_j [\nbx(t) \nbA_l]_{\nrmK(j)}}\Big] \delta_{q_l,q(t)}\right]} - \nbbE\left[{\rm exp}\Big[\sum_{j=1}^{|\nrmK|}{[\nbs]_j x_{\nrmK(j)}(t)}\Big] \delta_{\bar{q},q(t)}\right] \sum_{l \in \ncalL_{\bar{q}}}{\lambda^{(l)}},\\& \a \sum_{l \in \ncalL'_{\bar{q}}}{\lambda^{(l)}\big[\ncalV_{\bar{q}_l,|\nrmK|}^{(\nbs)}(t) \times_1 \nbA_l \times_2 \nbA_l \cdots \times_{|\nrmK|} \nbA_l\big]_{\nrmK}} + c_{\bar{q},\nrmK}(t) - v^{(\nbs)}_{\bar{q},\nrmK}(t)\sum_{l \in \ncalL_{\bar{q}}}{\lambda^{(l)}},
\end{align}
where step (a) follows from defining $\sum_{l \in \ncalL'_{\bar{q}}}{\lambda^{(l)}\nbbE\left[{\rm exp}\Big[\sum_{j=1}^{|\nrmK|}{[\nbs]_j [\nbx(t) \nbA_l]_{\nrmK(j)}}\Big] \delta_{q_l,q(t)}\right]} = c_{\bar{q},\nrmK}(t)$ when $\nrmZ_l \neq \varnothing, \forall l \in \ncalL'_{\bar{q}}$, and $c_{\bar{q},\nrmK}(t)$ is given by (\ref{c_app}) as
\begin{align*}
c_{\bar{q},\nrmK}(t) = \sum_{l \in \ncalL'_{\bar{q}}}{\lambda^{(l)} \sum_{\nrmZ \in 2^{\nrmK}\setminus\varnothing}{\nb1\big(\nrmZ_l = \nrmZ\big)\Big[\ncalV_{\bar{q}_l,|\nrmK\setminus\nrmZ_l|}^{(\nbs')}(t) \times_1 \nbA_l \times_2 \nbA_l \cdots \times_{|\nrmK\setminus\nrmZ_l|} \nbA_l\Big]_{\nrmK \setminus \nrmZ_l}}},
\end{align*}
where the vector $\nbs' = \left[[\nbs]_{\nrmI_l(1)}\;[\nbs]_{\nrmI_l(2)}\;\cdots\;[\nbs]_{\nrmI_l(|\nrmK \setminus \nrmZ_l|)}\right]$ such that the set $\nrmI_l$ contains the indices of the elements of $\nrmK \setminus \nrmZ_l$ inside $\nrmK$. Note that when $\nrmZ_l = \varnothing$, $\nbbE\left[{\rm exp}\Big[\sum_{j=1}^{|\nrmK|}{[\nbs]_j [\nbx(t) \nbA_l]_{\nrmK(j)}}\Big] \delta_{q_l,q(t)}\right] = v_{\bar{q}_l,\nrmK'}^{(\nbs)}(t) =  \big[\ncalV_{\bar{q}_l,|\nrmK|}^{(\nbs)}(t) \times_1 \nbA_l \times_2 \nbA_l \cdots \times_{|\nrmK|} \nbA_l\big]_{\nrmK}$. Finally, substituting (\ref{cons_B}) into (\ref{diff_mgf_single}) yields
\begin{align}\label{diff_mgf_single_finalform}
\dot{v}^{(\nbs)}_{\bar{q},\nrmK}(t) = \left[\sum_{j=1}^{|\nrmK|}{[\nbs]_j} - \sum_{l \in \ncalL_{\bar{q}}}{\lambda^{(l)}}\right] v^{(\nbs)}_{\bar{q},\nrmK}(t) + \sum_{l \in \ncalL'_{\bar{q}}}{\lambda^{(l)}\big[\ncalV_{\bar{q}_l,|\nrmK|}^{(\nbs)}(t) \times_1 \nbA_l \times_2 \nbA_l \cdots \times_{|\nrmK|} \nbA_l\big]_{\nrmK}} + c_{\bar{q},\nrmK}(t),
\end{align}
which completes the proof.
%The system of differential equations in (\ref{Lemma1_eq1}) and (\ref{Lemma1_eq2}) can be obtained by gathering the equations in (\ref{diff_mom_single_finalform}) and (\ref{diff_mgf_single_finalform}), and writing them in a matrix form.
 \hfill 
\IEEEQED

\subsection{Proof of Theorem~\ref{theorem 1}} \label{app:theorem 1}
We start the proof by combining the differential equations (\ref{Lemma1_eq1_tensor}) and (\ref{Lemma1_eq2_tensor}) of Lemma \ref{lemma:1} in a vector form as follows:
\begin{align}\label{proof_theorem1_eq1}
 \dot{\nbv}^{(\nbm)}_{\nrmK}(t) = \sum_{j=1}^{|\nrmK|}{[\nbm]_j \nbv^{(\nbm - \nbe_j)}_{\nrmK}(t)} + \nbv^{(\nbm)}_{\nrmK}(t) (\nbB_{\nrmK} - \nbD_{\nrmK}),
\end{align}
\begin{align}\label{proof_theorem1_eq2}
\dot{\nbv}^{(\nbs)}_{\nrmK}(t) =  \nbc_{\nrmK}(t) + \nbv^{(\nbs)}_{\nrmK}(t) \Big[\nbB_{\nrmK} - \nbD_{\nrmK} + \sum_{j=1}^{|\nrmK|}{[\nbs]_j}  \nbI\Big],    
\end{align}
where\\
$\nbv^{(\nbm)}_{\nrmK}(t) = \left[{\rm vec}\left(\ncalV_{0,|\nrmK|}^{(\nbm)}(t)\right)\;\cdots\;{\rm vec}\left(\ncalV_{q_{max},|\nrmK|}^{(\nbm)}(t)\right)\right],$\\
$\nbv^{(\nbs)}_{\nrmK}(t) = \left[{\rm vec}\left(\ncalV_{0,|\nrmK|}^{(\nbs)}(t)\right)\;\cdots\;{\rm vec}\left(\ncalV_{q_{max},|\nrmK|}^{(\nbs)}(t)\right)\right],$\\
$\nbD_{\nrmK} = {\rm diag}\left[d_0\nbI_{n^{|\nrmK|}},\cdots,d_{q_{max}}\nbI_{n^{|\nrmK|}}\right],\; d_{\bar{q}} = \sum_{l \in \ncalL_{\bar{q}}}{\lambda^{(l)}},$\\
$\nbc_{\nrmK}(t) = \left[{\rm vec}\left(\ncalC_{0,|\nrmK|}(t)\right)\;\cdots\;{\rm vec}\left(\ncalC_{q_{max},|\nrmK|}(t)\right)\right]$,\\
$\ncalR_{\bar{q},\nrmK}(t) = \sum_{l \in \ncalL'_{\bar{q}}}{\lambda^{(l)}\Big[\ncalV_{\bar{q}_l,|\nrmK|}^{(\nbm)}(t) \times_1 \nbA_l \times_2 \nbA_l \cdots \times_{|\nrmK|} \nbA_l\Big]}$,\\
$\hat{\ncalR}_{\bar{q},\nrmK}(t) = \sum_{l \in \ncalL'_{\bar{q}}}{\lambda^{(l)}\Big[\ncalV_{\bar{q}_l,|\nrmK|}^{(\nbs)}(t) \times_1 \nbA_l \times_2 \nbA_l \cdots \times_{|\nrmK|} \nbA_l\Big]}$,
\begin{align}
\label{eq_B_moment} &\left[{\rm vec}\left(\ncalR_{0,\nrmK}(t)\right)\;\cdots\;{\rm vec}\left(\ncalR_{q_{max},\nrmK}(t)\right)\right] = \nbv^{(\nbm)}(t) \nbB_{\nrmK},  \\
 \label{eq_B_mgf}&\left[{\rm vec}\left(\hat{\ncalR}_{0,\nrmK}(t)\right)\;\cdots\;{\rm vec}\left(\hat{\ncalR}_{q_{max},\nrmK}(t)\right)\right] = \nbv^{(\nbs)}(t) \nbB_{\nrmK}, 
\end{align}
such that ${\rm vec}(\mathcal{X})$ is the row vector resulting from concatenating the rows of $\mathcal{X}$ into a single long row, and $\nbX = {\rm diag}[x_1,\cdots,x_n]$ is a diagonal matrix with $[\nbX]_{i,j} = x_i \delta_{i,j}, \forall i,j \in 1:n$. %$\nbR_{\bar{q}}= \sum_{l \in \ncalL'_{\bar{q}}}{\lambda^{(l)} \nbA_l^{{\rm T}} \nbV_{q_l}^{(m_1,m_2)}(t) \nbA_l}$, and $\hat{\nbR}_{\bar{q}} = \sum_{l \in \ncalL'_{\bar{q}}}{\lambda^{(l)} \nbA_l^{{\rm T}} \nbV_{q_l}^{(s_1,s_2)}(t) \nbA_l}$. 
Note that we could construct the expressions in (\ref{eq_B_moment}) and (\ref{eq_B_mgf}) due to the fact that ${\rm vec}\left(\ncalR_{\bar{q},\nrmK}(t)\right)$ and ${\rm vec}\left(\hat{\ncalR}_{\bar{q},\nrmK}(t)\right)$ are linear in $\nbv^{(\nbm)}_{\nrmK}(t)$ and $\nbv^{(\nbs)}_{\nrmK}(t)$, respectively. In order to figure out the conditions under which (\ref{proof_theorem1_eq1}) is asymptotically stable, we first rewrite (\ref{proof_theorem1_eq1}) in the case where $\dot{\nbv}^{({\bf 1}_{|\nrmK|})}_{\nrmK}(t) = {\bf 0}$ as $t \rightarrow \infty$:
\begin{align}\label{proof_theorem1_eq3}
 \bar{\nbv}^{({\bf 1}_{|\nrmK|})}_{\nrmK} \nbD_{\nrmK} = \sum_{j=1}^{|\nrmK|}{\bar{\nbv}^{({\bf 1}_{|\nrmK|} - \nbe_j)}_{\nrmK}} + \bar{\nbv}^{({\bf 1}_{|\nrmK|})}_{\nrmK} \nbB_{\nrmK} ,
\end{align}
where $\bar{\nbv}^{(\nbm)}_{\nrmK} =  \underset{t \rightarrow \infty}{\rm lim}\nbv^{(\nbm)}_{\nrmK}(t)$. Now, it would be useful to state the result in \cite[Lemma 2]{yates2020age}, which is a minor variation on the Perron-Frobenius Theorem. In particular, for the following system of equations:
\begin{align}\label{yates_lemma2}
\nbv \nbD= \nbz + \nbv \nbB,
\end{align}
if $\nbB$ and $\nbD$ are non-negative matrices, $\nbz$ is strictly positive, and there exists a positive solution $\nbv$ for (\ref{yates_lemma2}), then all the eigenvalues of $\nbB - \nbD$ have strictly negative real parts. 
A key observation here is that both $\nbB_{\nrmK}$ and $\nbD_{\nrmK}$ in (\ref{proof_theorem1_eq3}) are non-negative matrices. Thus, based on (\ref{yates_lemma2}), if $\sum_{j=1}^{|\nrmK|}{\bar{\nbv}^{({\bf 1}_{|\nrmK|} - \nbe_j)}_{\nrmK}}$ is strictly positive and there exists a positive solution $\bar{\nbv}^{({\bf 1}_{|\nrmK|})}_{\nrmK}$ for (\ref{proof_theorem1_eq3}), then all the eigenvalues of $\nbB_{\nrmK} - \nbD_{\nrmK}$ have strictly negative real parts, and hence (\ref{proof_theorem1_eq1}) is asymptotically stable. Further, when all the eigenvalues of $\nbB_{\nrmK} - \nbD_{\nrmK}$ have strictly negative real parts, we observe from (\ref{proof_theorem1_eq2}) that there exists $s_0 > 0$ such that for all $\nbs \in \ncalS = \{\nbs: \sum_{j=1}^{|\nrmK|}{[\nbs]_{j}} < s_0 \}$, all the eigenvalues of $\Big[\nbB_{\nrmK} - \nbD_{\nrmK} + \sum_{j=1}^{|\nrmK|}{[\nbs]_j} \nbI\Big]$ will have strictly negative real parts, which guarantees the asymptotic stability of (\ref{proof_theorem1_eq2}) under the condition that $\nbc_{\nrmK}(t)$ converges as $t \rightarrow \infty$. Thus, what remains is to identify the conditions for strict positivity of $\sum_{j=1}^{|\nrmK|}{\bar{\nbv}^{({\bf 1}_{|\nrmK|} - \nbe_j)}_{\nrmK}}$ and convergence of $\nbc_{\nrmK}(t)$ as $t \rightarrow \infty$. To clearly see these conditions for the generic case of having an arbitrary set $\nrmK$ with $|\nrmK|$ age processes, we start with the cases of $|\nrmK| = 1$ and $|\nrmK| = 2$. For the cases of $|\nrmK| = 1$ and $|\nrmK| = 2$, (\ref{proof_theorem1_eq3}) respectively reduces to:
\begin{align}\label{proof_theorem1_eq4}
 \bar{\nbv}^{([1])}_{\{k_1\}} \nbD_{\{k_1\}} = \bar{\nbv}^{([0])}_{\{k_1\}} + \bar{\nbv}^{([1])}_{\{k_1\}} \nbB_{\{k_1\}} ,
\end{align}
\begin{align}\label{proof_theorem1_eq5}
 \bar{\nbv}^{([1 1])}_{\{k_1,k_2\}} \nbD_{\{k_1,k_2\}} = \bar{\nbv}^{([0\;1])}_{\{k_1,k_2\}} + \bar{\nbv}^{([1\;0])}_{\{k_1,k_2\}} +  \bar{\nbv}^{([1\;1])}_{\{k_1,k_2\}} \nbB_{\{k_1,k_2\}}. 
\end{align}

Note that $\bar{\nbv}^{([0])}_{\{k_1\}}$ in (\ref{proof_theorem1_eq4}) is formed by the state probabilities of the Markov chain $q(t)$, and $\nbc_{\{k_1\}}(t)$ is a function of the state probabilities of $q(t)$. Thus, having an ergodic Markov chain $q(t)$ with stationary distribution $\bar{\nbv}_{\bar{q}}^{(0)} > {\bf 0}, \bar{q} \in \ncalQ,$ guarantees both strict positivity of $\sum_{j=1}^{|\nrmK|}{\bar{\nbv}^{({\bf 1}_{|\nrmK|} - \nbe_j)}_{\nrmK}}$ and convergence of $\nbc_{\nrmK}(t)$ as $t \rightarrow \infty$ for the case of $|\nrmK| = 1$ (as has been demonstrated in \cite[Theorem 1]{yates2020age}). Now, we move to the case of $|\nrmK = 2|$. For this case, it is important to note that $\bar{\nbv}^{([0\;1])}_{\{k_1,k_2\}}$ and $\bar{\nbv}^{([1\;0])}_{\{k_1,k_2\}}$ in (\ref{proof_theorem1_eq5}) can be constructed using $\bar{\nbv}^{([1])}_{\{k_1\}}$, and $\nbc_{\{k_1,k_2\}}(t)$ is a function of $\nbv^{([s_{1}])}_{\{k_1\}}$ and the distribution of $q(t)$. Therefore, we can figure out that in this case, both strict positivity of $\sum_{j=1}^{|\nrmK|}{\bar{\nbv}^{({\bf 1}_{|\nrmK|} - \nbe_j)}_{\nrmK}}$ and convergence of $\nbc_{\nrmK}(t)$ hold when: 1) $q(t)$ is ergodic with distribution $\{\bar{\nbv}^{(0)}_{\bar{q}} > {\bf 0}\}_{\bar{q}\in \ncalQ}$, and 2) there exists a positive solution $\bar{\nbv}^{([1])}_{\{k_1\}}$ for (\ref{proof_theorem1_eq4}). Repeated application of (\ref{proof_theorem1_eq3}) yields the conditions under which (\ref{proof_theorem1_eq1}) and (\ref{proof_theorem1_eq2}) are asymptotically stable for a set $\nrmK$ with $|\nrmK| > 2$. By inspecting the analysis of the two cases of $|\nrmK| = 1$ and $|\nrmK| = 2$ above, we can deduce that these conditions are: 1) $q(t)$ is ergodic with distribution $\{\bar{\nbv}^{(0)}_{\bar{q}} > {\bf 0}\}_{\bar{q}\in \ncalQ}$, and 2) the existence of positive solutions $\{\bar{\nbv}^{({\bf 1}_{|\nrmZ|})}_{\nrmZ}\}_{|\nrmZ| \in \{1,2,\cdots, |\nrmK|\}}$ for (\ref{proof_theorem1_eq3}). This completes the proof.
%Note that $\bar{\nbv}^{(0,1)}$ and $\bar{\nbv}^{(1,0)}$ can be constructed using the stationary marginal first moments $\{\nbv^{(1)}_{\bar{q}}\}$, and according to (\ref{c_app}), $\nbc(t)$ is a function of the marginal MGFs $\{\nbv^{(s)}_{\bar{q}}(t)\}$ and the distribution of the Markov chain $\{\nbv^{(0)}_{\bar{q}}(t)\}$ satisfying (\ref{diff_marginal_mgf}) and (\ref{diff_marginal_0}), respectively. Therefore, based on \cite[Theorem 1]{yates2020age}, we can figure out that both strict positivity of $\bar{\nbv}^{(0,1)} + \bar{\nbv}^{(1,0)}$ and convergence of $\nbc(t)$ as $t \rightarrow \infty$ hold when: 1) the Markov chain $q(t)$ is ergodic with distribution $\{\bar{\nbv}^{(0)}_{\bar{q}} > {\bf 0}\}$, and 2) there exists a positive fixed point $\bar{\nbv}^{(1)}_{\bar{q}}, \forall \bar{q} \in \ncalQ$, for the marginal first moment in (\ref{diff_marginal_mom}). 
\hfill 
\IEEEQED

\subsection{Proof of Theorem~\ref{theorem:MGF_NP}} \label{app:theorem:MGF_NP}
 Using the set of transitions in Table \ref{table:NP} and (\ref{theorem 1_eq2}) in Theorem \ref{theorem 1}, $\bar{v}^{([s_{k_1}])}_{0,\{k_1\}}$ can be expressed as 
\begin{align}\label{theorem2_proof_eq1}
\left(\lambda - s_{k_1}\right) \bar{v}_{0,\{k_1\}}^{([s_{k_1}])} = \mu \left(\bar{v}_{k_1,\{0\}}^{([s_{k_1}])} + \sum_{\bar{q}=1,\bar{q}\notin\{k_1\}}^{N}{\bar{v}_{\bar{q},\{k_1\}}^{([s_{k_1}])}}\right),
\end{align}
where $\bar{v}_{k_1,\{0\}}^{([s_{k_1}])}$ and $\bar{v}_{\bar{q},\{k_1\}}^{([s_{k_1}])}$ are given by
\begin{align}\label{theorem2_proof_eq2}
\left(\mu - s_{k_1}\right) \bar{v}_{k_1,\{0\}}^{([s_{k_1}])} = \lambda_{k_1} \bar{v}^{(0)}_{0},  
\end{align}
\begin{align}\label{theorem2_proof_eq3}
 \left(\mu - s_{k_1}\right)\bar{v}_{\bar{q},\{k_1\}}^{([s_{k_1}])} = \lambda_{\bar{q}} \bar{v}_{0,\{k_1\}}^{([s_{k_1}])}, 
\end{align}
 where $\bar{q} \in 1:N$. Substituting (\ref{theorem2_proof_eq2}) and (\ref{theorem2_proof_eq3}) into (\ref{theorem2_proof_eq1}), we get
 \begin{align}\label{theorem2_proof_eq4}
\bar{v}_{0,\{k_1\}}^{([s_{k_1}])} = \dfrac{\mu \lambda_{k_1} \bar{v}^{(0)}_{0}}{(\lambda - s_{k_1})(\mu - s_{k_1}) - \mu \sum_{j=1, j\notin\{k_1\}}^{N}{\lambda_j}} \a \dfrac{\mu \lambda_{k_1} \bar{v}^{(0)}_{0}}{c_{\{k_1\}}},     
 \end{align}
 where $k_1 \in 1:N$ and step (a) follows from defining $c_{\nrmZ}$ for a set $\nrmZ \subseteq \{1,2,\cdots,N\}$ in (\ref{theorem:MGF_NP_eq2}) as 
 \begin{align*}
   c_{\nrmZ} = \left(\lambda - \sum_{j=1}^{|\nrmZ|}{s_{\nrmZ(j)}}\right) \left(\mu - \sum_{j=1}^{|\nrmZ|}{s_{\nrmZ(j)}}\right) - \mu \sum_{j=1,j\notin\nrmZ}^{N}{\lambda_j}.
 \end{align*}

 Now, using (\ref{theorem2_proof_eq4}), one can evaluate $\bar{v}_{0,\{k_1,k_2\}}^{([s_{k_1}\;s_{k_2}])}$. In particular, from (\ref{theorem 1_eq2}), $\bar{v}_{0,\{k_1,k_2\}}^{([s_{k_1}\;s_{k_2}])}$ can be expressed as 
 \begin{align}\label{theorem2_proof_eq5}
[\lambda - (s_{k_1} + s_{k_2})]\bar{v}_{0,\{k_1,k_2\}}^{([s_{k_1}\;s_{k_2}])} = \mu \Big[\bar{v}_{k_1,\{0,k_2\}}^{([s_{k_1}\;s_{k_2}])} + \bar{v}_{k_2,\{k_1,0\}}^{([s_{k_1}\;s_{k_2}])} + \sum_{\bar{q}=1,\bar{q}\notin\{k_1,k_2\}}^{N}{\bar{v}_{\bar{q},\{k_1,k_2\}}^{([s_{k_1}\;s_{k_2}])}}\Big],    
 \end{align}
 where
 \begin{align}\label{theorem2_proof_eq6}
 [\mu - (s_{k_1} + s_{k_2})] \bar{v}_{k_1,\{0,k_2\}}^{([s_{k_1}\;s_{k_2}])} = \lambda_{k_1} \bar{v}_{0,\{k_2\}}^{([s_{k_2}])},
 \end{align}
  \begin{align}\label{theorem2_proof_eq7}
 [\mu - (s_{k_1} + s_{k_2})] \bar{v}_{k_2,\{k_1,0\}}^{([s_{k_1}\;s_{k_2}])} = \lambda_{k_2} \bar{v}_{0,\{k_1\}}^{([s_{k_1}])},
 \end{align}
  \begin{align}\label{theorem2_proof_eq8}
 [\mu - (s_{k_1} + s_{k_2})] \bar{v}_{\bar{q},\{k_1,k_2\}}^{([s_{k_1}\;s_{k_2}])} = \lambda_{\bar{q}} \bar{v}_{0,\{k_1,k_2\}}^{([s_{k_1}\;s_{k_2}])}, 
 \end{align}
where $\bar{q} \in 1:N$. Thus, $\bar{v}_{0,\{k_1,k_2\}}^{([s_{k_1}\;s_{k_2}])}$ can be rewritten as 
 \begin{align}\label{theorem2_proof_eq9}
 \bar{v}_{0,\{k_1,k_2\}}^{([s_{k_1}\;s_{k_2}])} \nonumber&\a \dfrac{\mu\left(\lambda_{k_1} \bar{v}_{0,\{k_2\}}^{([s_{k_2}])} + \lambda_{k_2} \bar{v}_{0,\{k_1\}}^{([s_{k_1}])}\right)}{\Big[[\lambda - (s_{k_1} + s_{k_2})][\mu - (s_{k_1} + s_{k_2})] - \mu \sum_{j=1,j\notin\{k_1,k_2\}}^{N}{\lambda_j}\Big]},  \\
 \nonumber&\b \dfrac{\mu\left(\lambda_{k_1} \bar{v}_{0,\{k_2\}}^{([s_{k_2}])} + \lambda_{k_2} \bar{v}_{0,\{k_1\}}^{([s_{k_1}])}\right)}{c_{\{k_1,k_2\}}},\\
 &\c \mu^2 \lambda_{k_1} \lambda_{k_2} \bar{v}^{(0)}_{0} \left(\dfrac{1}{c_{\{k_1,k_2\}}c_{\{k_1\}}} + \dfrac{1}{c_{\{k_1,k_2\}}c_{\{k_2\}}}\right) \d \mu^2 \lambda_{k_1}\lambda_{k_2}\bar{v}^{(0)}_{0} \sum_{\nrmP \in \mathcal{P}(\{k_1,k_2\})}{\dfrac{1}{C(\nrmP)}},
 \end{align}
where step (a) follows from substituting (\ref{theorem2_proof_eq6})-(\ref{theorem2_proof_eq8}) into (\ref{theorem2_proof_eq5}), step (b) follows from the fact that $c_{\{k_1,k_2\}} = [\lambda - (s_{k_1} + s_{k_2})][\mu - (s_{k_1} + s_{k_2})] - \mu \sum_{j=1,j\notin\{k_1,k_2\}}^{N}{\lambda_j}$, step (c) follows from substituting (\ref{theorem2_proof_eq4}) into (\ref{theorem2_proof_eq9}), and step (d) follows from defining $C(\nrmP)$ as follows 
\begin{align}\label{theorem2_proof_eq10}
 C(\nrmP) = \prod_{i=1}^{|\nrmP|}{c_{\nrmP(i:|\nrmP|)}} = \prod_{i=1}^{|\nrmP|}\left[ \left(\lambda - \sum_{j=i}^{|\nrmP|}{s_{\nrmP(j)}}\right) \left(\mu - \sum_{j=i}^{|\nrmP|}{s_{\nrmP(j)}}\right) - \mu  \sum_{j=1, j \notin \nrmP(i:|\nrmP|)}^{N}{\lambda_{j}}\right].
 \end{align}

In order to clearly see how $\bar{v}_{0,\nrmK}^{(\nbs)}$ can be obtained for an arbitrary set $\nrmK\subseteq\{1,2,\cdots,N\}$, where $\nbs = [s_{\nrmK(1)}\;s_{\nrmK(2)}\;\cdots\;s_{\nrmK(|\nrmK|)}]$, it will be useful to further derive $\bar{v}_{0,\{k_1,k_2,k_3\}}^{([s_{k_1}\;s_{k_2}\;s_{k_3}])}$ using (\ref{theorem2_proof_eq9}). From (\ref{theorem 1_eq2}), $\bar{v}_{0,\{k_1,k_2,k_3\}}^{([s_{k_1}\;s_{k_2}\;s_{k_3}])}$ can be expressed as 
\begin{align}\label{theorem2_proof_eq11}
[\lambda - (s_{k_1} + s_{k_2} + s_{k_3})] \bar{v}_{0,\{k_1,k_2,k_3\}}^{([s_{k_1}\;s_{k_2}\;s_{k_3}])} = \mu \Big[\nonumber&\bar{v}_{k_1,\{0,k_2,k_3\}}^{([s_{k_1}\;s_{k_2}\;s_{k_3}])} + \bar{v}_{k_2,\{k_1,0,k_3\}}^{([s_{k_1}\;s_{k_2}\;s_{k_3}])} + \bar{v}_{k_3,\{k_1,k_2,0\}}^{([s_{k_1}\;s_{k_2}\;s_{k_3}])} \\&+ \sum_{\bar{q}=1,\bar{q}\notin\{k_1,k_2,k_3\}}^{N}{\bar{v}_{\bar{q},\{k_1,k_2,k_3\}}^{([s_{k_1}\;s_{k_2}\;s_{k_3}])}}\Big],
\end{align}
where
\begin{align}\label{theorem2_proof_eq12}
[\lambda - (s_{k_1} + s_{k_2} + s_{k_3})]\bar{v}_{k_1,\{0,k_2,k_3\}}^{([s_{k_1}\;s_{k_2}\;s_{k_3}])} = \lambda_{k_1} \bar{v}_{0,\{k_2,k_3\}}^{([s_{k_2}\;s_{k_3}])},
\end{align}
\begin{align}\label{theorem2_proof_eq13}
[\lambda - (s_{k_1} + s_{k_2} + s_{k_3})] \bar{v}_{k_2,\{k_1,0,k_3\}}^{([s_{k_1}\;s_{k_2}\;s_{k_3}])} = \lambda_{k_2} \bar{v}_{0,\{k_1,k_3\}}^{([s_{k_1}\;s_{k_3}])},
\end{align}
\begin{align}\label{theorem2_proof_eq14}
[\lambda - (s_{k_1} + s_{k_2} + s_{k_3})] \bar{v}_{k_3,\{k_1,k_2,0\}}^{([s_{k_1}\;s_{k_2}\;s_{k_3}])} = \lambda_{k_3} \bar{v}_{0,\{k_1,k_2\}}^{([s_{k_1}\;s_{k_2}])},
\end{align}
\begin{align}\label{theorem2_proof_eq15}
[\lambda - (s_{k_1} + s_{k_2} + s_{k_3})] \bar{v}_{\bar{q},\{k_1,k_2,k_3\}}^{([s_{k_1}\;s_{k_2}\;s_{k_3}])} = \lambda_{\bar{q}} \bar{v}_{0,\{k_1,k_2,k_3\}}^{([s_{k_1}\;s_{k_2}\;s_{k_3}])},
\end{align}
where $\bar{q} \in 1:N$. Thus, $\bar{v}_{0,\{k_1,k_2,k_3\}}^{([s_{k_1}\;s_{k_2}\;s_{k_3}])}$ can be rewritten as 
 \begin{align}\label{theorem2_proof_eq16}
 \bar{v}_{0,\{k_1,k_2,k_3\}}^{([s_{k_1}\;s_{k_2}\;s_{k_3}])} \nonumber&\a \dfrac{\mu\left(\lambda_{k_1} \bar{v}_{0,\{k_2,k_3\}}^{([s_{k_2}\;s_{k_3}])} + \lambda_{k_2} \bar{v}_{0,\{k_1,k_3\}}^{([s_{k_1}\;s_{k_3}])} + \lambda_{k_3} \bar{v}_{0,\{k_1,k_2\}}^{([s_{k_1}\;s_{k_2}])}\right)}{\Big[[\lambda - (s_{k_1} + s_{k_2} + s_{k_3})][\mu - (s_{k_1} + s_{k_2}) + s_{k_3}] - \mu \sum_{j=1,j\notin\{k_1,k_2,k_3\}}^{N}{\lambda_j}\Big]},  \\
 \nonumber&= \dfrac{\mu\left(\lambda_{k_1} \bar{v}_{0,\{k_2,k_3\}}^{([s_{k_2}\;s_{k_3}])} + \lambda_{k_2} \bar{v}_{0,\{k_1,k_3\}}^{([s_{k_1}\;s_{k_3}])} + \lambda_{k_3} \bar{v}_{0,\{k_1,k_2\}}^{([s_{k_1}\;s_{k_2}])}\right)}{c_{\{k_1,k_2,k_3\}}},\\
 &\nonumber\b \dfrac{\mu^3 \lambda_{k_1} \lambda_{k_2} \lambda_{k_3} \bar{v}^{(0)}_{0}}{c_{\{k_1,k_2,k_3\}}} \left(\sum_{\nrmP \in \mathcal{P}(\{k_2,k_3\})}{\dfrac{1}{C(\nrmP)}} + \sum_{\nrmP \in \mathcal{P}(\{k_1,k_3\})}{\dfrac{1}{C(\nrmP)}} + \sum_{\nrmP \in \mathcal{P}(\{k_1,k_2\})}{\dfrac{1}{C(\nrmP)}}\right), \\
 &\c \mu^3 \lambda_{k_1}\lambda_{k_2}\lambda_{k_3}\bar{v}^{(0)}_{0} \sum_{\nrmP \in \mathcal{P}(\{k_1,k_2,k_3\})}{\dfrac{1}{C(\nrmP)}},
 \end{align}
 where step (a) follows from substituting (\ref{theorem2_proof_eq12})-(\ref{theorem2_proof_eq15}) into (\ref{theorem2_proof_eq11}), step (b) follows from substituting (\ref{theorem2_proof_eq9}) into (\ref{theorem2_proof_eq16}), and step (c) follows from the fact that 
 \begin{align*}
\dfrac{1}{c_{\{k_1,k_2,k_3\}}}\left(\sum_{\nrmP \in \mathcal{P}(\{k_2,k_3\})}{\dfrac{1}{C(\nrmP)}} + \sum_{\nrmP \in \mathcal{P}(\{k_1,k_3\})}{\dfrac{1}{C(\nrmP)}} + \sum_{\nrmP \in \mathcal{P}(\{k_1,k_2\})}{\dfrac{1}{C(\nrmP)}}\right) = \sum_{\nrmP \in \mathcal{P}(\{k_1,k_2,k_3\})}{\dfrac{1}{C(\nrmP)}}.
 \end{align*}
 
 By inspecting the expressions of $\bar{v}_{0,\{k_1\}}^{([s_{k_1}])}$, $\bar{v}_{0,\{k_1,k_2\}}^{([s_{k_1}\;s_{k_2}])}$ and $\bar{v}_{0,\{k_1,k_2,k_3\}}^{([s_{k_1}\;s_{k_2}\;s_{k_3}])}$  in (\ref{theorem2_proof_eq4}), (\ref{theorem2_proof_eq9}) and (\ref{theorem2_proof_eq16}), respectively, one can see that repeated application of (\ref{theorem 1_eq2}) gives $\bar{v}_{0,\nrmK}^{(\nbs)}$ for an arbitrary set $\nrmK \subseteq \{1,2,\cdots,N\}$ as
 \begin{align}\label{theorem2_proof_eq17}
\bar{v}^{(\nbs)}_{0,\nrmK} = \mu^{|\nrmK|} \left(\prod_{i=1}^{|\nrmK|}{\lambda_{\nrmK(i)}}\right)\bar{v}^{(0)}_{0} \sum_{\nrmP \in \mathcal{P}(\nrmK)}{\dfrac{1}{C(\nrmP)}}.
\end{align}

Thus, the stationary joint MGF of a set $\nrmK\subseteq \{1,2,\cdots,N\}$ of age processes is given by
\begin{align}\label{theorem2_proof_eq18}
\overset{{\rm NP} }{M} (\nbs) = \sum_{\bar{q} \in \ncalQ}{\bar{v}^{(\nbs)}_{\bar{q},\nrmK}} \nonumber&\a \left(1 + \dfrac{\lambda}{\mu - \sum_{j=1}^{|\nrmK|}{s_{\nrmK(j)}}}\right) \bar{v}_{0,\nrmK}^{(\nbs)},
\\&\b\mu^{|\nrmK|} \left(\prod_{i=1}^{|\nrmK|}{\lambda_{\nrmK(i)}}\right)\left(\dfrac{\mu}{\lambda + \mu}\right)\left(1 + \dfrac{\lambda}{\mu - \sum_{j=1}^{|\nrmK|}{s_{\nrmK(j)}}}\right) \sum_{\nrmP \in \mathcal{P}(\nrmK)}{\dfrac{1}{C(\nrmP)}},
\end{align}
where step (a) follows from expressing $\bar{v}_{\bar{q},\nrmK}^{(\nbs)}$, $\bar{q} \in \mathcal{Q}/\{0\}$, as a function of $\bar{v}_{0,\nrmK}^{(\nbs)}$ using (\ref{theorem 1_eq2}), and step (b) follows from (\ref{theorem2_proof_eq17}) along with noting that $\bar{v}^{(0)}_{0} = \dfrac{\mu}{\lambda + \mu}$. This completes the proof.
\hfill 
\IEEEQED

\subsection{Proof of Proposition~\ref{prop:MGF_NP_twosources}}\label{app:prop:MGF_NP_twosources}
%From Theorem \ref{theorem:MGF_NP}, the stationary joint MGF of the two AoI processes $x_{k_1}(t)$ and $x_{k_2}(t)$ can be obtained by setting $\nrmK = \{k_1,k_2\}$ as
%\begin{align}\label{proof_prop:MGF_NP_twosources_eq1}
%\overset{{\rm NP} }{M} (\bar{s}_{k_1},\bar{s}_{k_2})= \nonumber& \dfrac{\rho_{k_1} \rho_{k_2}\big[1 + \rho - (\bar{s}_{k_1} + \bar{s}_{k_2})\big]}{(1 + \rho)\Big[\big[\rho - (\bar{s}_{k_1} + \bar{s}_{k_2})\big] \big[1 - (\bar{s}_{k_1} + \bar{s}_{k_2})\big] - \rho_{-\{k_1,k_2\}}\Big]\Big[1 - (\bar{s}_{k_1} + \bar{s}_{k_2})\Big]}\\ &\times \sum_{i\in\{k_1,k_2\}}{\dfrac{1}{(1 - \bar{s}_i)(\rho - \bar{s}_i) - \rho_{-i}}},
%\end{align}
%where $\bar{s}_{i} = \frac{s_{i}}{\mu}, i \in\{k_1,k_2\}$. 
The correlation coefficient of the two AoI processes $x_{k_1}(t)$ and $x_{k_2}(t)$ can be evaluated as 
\begin{align}\label{proof_prop:MGF_NP_twosources_eq2}
\overset{\rm NP}{\rm Cor} = \dfrac{\nbbE[x_{k_1}x_{k_2}] - \nbbE[x_{k_1}]\nbbE[x_{k_2}]}{\sqrt{\nbbE[x_{k_1}^2] - (\nbbE[x_{k_1}])^2}\sqrt{\nbbE[x_{k_2}^2] - (\nbbE[x_{k_2}])^2}}.   
\end{align}

Using Corollary \ref{cor:NP_twosources}, $\nbbE[x_{k_1}x_{k_2}]$ can be evaluated as
\begin{align}\label{proof_prop:MGF_NP_twosources_eq3}
\nbbE[x_{k_1}x_{k_2}] \nonumber&= \frac{\partial^{2}\big[\overset{\rm NP}{M}(\bar{s}_{k_1},\bar{s}_{k_2})\big]}{\mu^{2}\partial\bar{s}_{k_1} \partial\bar{s}_{k_2}} \Big|_{\bar{s}_{k_1}=0, \bar{s}_{k_2}=0},\\
&= \dfrac{\rho(1+\rho)(\rho_{k_1}+\rho_{k_2})^2 + (\rho_{k_1}+\rho_{k_2})\big[(1+\rho)^3 + 2\rho\rho_{k_1}\rho_{k_2}\big] - 2\rho_{k_1}\rho_{k_2}(1+\rho)}{\mu^2\rho_{k_1}\rho_{k_2}(1+\rho)(\rho_{k_1}+\rho_{k_2})}.
\end{align}

Further, from Corollary \ref{cor:NP_marginal}, $\nbbE[x_{k}]$ and $\nbbE[x_{k}^2]$ can be respectively evaluated as
\begin{align}\label{proof_prop:MGF_NP_twosources_eq4}
\nbbE[x_{k}] = \frac{{\rm d}\big[\overset{\rm NP}{M}(\bar{s}_{k})\big]}{\mu \;{\rm d}\bar{s}_{k}} \Big|_{\bar{s}_{k}=0} = \dfrac{1 + \rho}{\mu \rho_{k}} + \dfrac{\rho}{\mu \left(1 + \rho\right)},
\end{align}
\begin{align}\label{proof_prop:MGF_NP_twosources_eq5}
\nbbE[x_{k}^2] = \frac{{\rm d}^2\big[\overset{\rm NP}{M}(\bar{s}_{k})\big]}{\mu^2 \;{\rm d}\bar{s}_{k}^2} \Big|_{\bar{s}_{k}=0} = \dfrac{2\big[\rho_k^2\rho + \rho_k(\rho^2-1) + (1+\rho)^3\big]}{\mu^2\rho_k^2(1+\rho)},
\end{align}
where $k \in 1:N$. The final expression of $\overset{\rm NP}{\rm Cor}$ in (\ref{prop:MGF_NP_twosources_eq1}) can be derived by substituting (\ref{proof_prop:MGF_NP_twosources_eq3})-(\ref{proof_prop:MGF_NP_twosources_eq5}) into (\ref{proof_prop:MGF_NP_twosources_eq2}), followed by some algebraic simplifications. 
\hfill 
\IEEEQED

\subsection{Proof of Theorem~\ref{theorem:MGF_PS}} \label{app:theorem:MGF_PS}
Using the set of transitions in Table \ref{table:PS} and (\ref{theorem 1_eq2}) in Theorem \ref{theorem 1}, $\bar{v}^{([s_{k_1}])}_{0,\{k_1\}}$ can be expressed as
\begin{align}\label{theorem3_proof_eq1}
\left(\lambda - s_{k_1}\right) \bar{v}_{0,\{k_1\}}^{([s_{k_1}])} = \mu \left(\bar{v}_{k_1,\{0\}}^{([s_{k_1}])} + \sum_{\bar{q}=1,\bar{q}\notin\{k_1\}}^{N}{\bar{v}_{\bar{q},\{k_1\}}^{([s_{k_1}])}}\right),
\end{align}
where $\bar{v}_{k_1,\{0\}}^{([s_{k_1}])}$ and $\bar{v}_{\bar{q},\{k_1\}}^{([s_{k_1}])}$ are given by
\begin{align}\label{theorem3_proof_eq2}
\left(\mu + \lambda - s_{k_1}\right) \bar{v}_{k_1,\{0\}}^{([s_{k_1}])} = \lambda_{k_1} \sum_{\bar{q}=0}^{N}{\bar{v}^{(0)}_{\bar{q}}} = \lambda_{k_1},  
\end{align}
\begin{align}\label{theorem3_proof_eq3}
 \left(\mu + \lambda - s_{k_1}\right)\bar{v}_{\bar{q},\{k_1\}}^{([s_{k_1}])} = \lambda_{\bar{q}} \left(\bar{v}_{0,\{k_1\}}^{([s_{k_1}])} + \sum_{\bar{q}=1}^{N}{\bar{v}_{\bar{q},\{k_1\}}^{([s_{k_1}])}}\right),
\end{align}
 where $\bar{q} \in 1:N$. Thus, from summing the set of equations in (\ref{theorem3_proof_eq3}), we get 
 \begin{align}\label{theorem3_proof_eq4}
 \sum_{\bar{q}=1}^{N}{\bar{v}_{\bar{q},\{k_1\}}^{([s_{k_1}])}} = \dfrac{\lambda}{\mu - s_{k_1}} \bar{v}_{0,\{k_1\}}^{([s_{k_1}])}.
 \end{align}
 
 Substituting (\ref{theorem3_proof_eq4}) into (\ref{theorem3_proof_eq3}), $\bar{v}_{\bar{q},\{k_1\}}^{([s_{k_1}])}$ can be expressed in terms of $\bar{v}_{\bar{0},\{k_1\}}^{([s_{k_1}])}$ as
 \begin{align}\label{theorem3_proof_eq5}
 \bar{v}_{\bar{q},\{k_1\}}^{([s_{k_1}])} = \dfrac{\lambda_{\bar{q}}}{\mu - s_{k_1}} \bar{v}_{0,\{k_1\}}^{([s_{k_1}])},
 \end{align}
 where $\bar{q} \in 1:N$. From (\ref{theorem3_proof_eq1}), (\ref{theorem3_proof_eq2}) and  (\ref{theorem3_proof_eq5}), $\bar{v}_{0,\{k_1\}}^{([s_{k_1}])}$ can be rewritten as 
 \begin{align}\label{theorem3_proof_eq6}
\bar{v}_{0,\{k_1\}}^{([s_{k_1}])} \nonumber&\a \dfrac{\mu}{\lambda - s_{k_1}} \Big[\dfrac{\lambda_{k_1}}{\mu + \lambda - s_{k_1}} + \dfrac{\sum_{j=1,j\notin\{k_1\}}^{N}{\lambda_j}}{\mu - s_{k_1}} \bar{v}_{0,\{k_1\}}^{([s_{k_1}])}\Big], \\ 
&\nonumber= \dfrac{\mu \lambda_{k_1} \left(\mu - s_{k_1}\right)}{\mu + \lambda - s_{k_1}} \times \dfrac{1}{(\lambda - s_{k_1})(\mu - s_{k_1}) - \mu \sum_{j=1, j\notin\{k_1\}}^{N}{\lambda_j}},\\
&\b \dfrac{\mu \lambda_{k_1} \left(\mu - s_{k_1}\right)}{c_{\{k_1\}}\left(\mu + \lambda - s_{k_1}\right)},
 \end{align}
 where $k_1 \in 1:N$, step (a) follows from substituting (\ref{theorem3_proof_eq2}) and  (\ref{theorem3_proof_eq5}) into (\ref{theorem3_proof_eq1}), and step (b) follows the definition of $c_{\{k_1\}}$ in (\ref{theorem:MGF_NP_eq2}). Now, using (\ref{theorem3_proof_eq6}), one can evaluate $\bar{v}_{0,\{k_1,k_2\}}^{([s_{k_1}\;s_{k_2}])}$. In particular, from (\ref{theorem 1_eq2}), $\bar{v}_{0,\{k_1,k_2\}}^{([s_{k_1}\;s_{k_2}])}$ can be expressed as 
 \begin{align}\label{theorem3_proof_eq7}
[\lambda - (s_{k_1} + s_{k_2})]\bar{v}_{0,\{k_1,k_2\}}^{([s_{k_1}\;s_{k_2}])} = \mu \Big[\bar{v}_{k_1,\{0,k_2\}}^{([s_{k_1}\;s_{k_2}])} + \bar{v}_{k_2,\{k_1,0\}}^{([s_{k_1}\;s_{k_2}])} + \sum_{\bar{q}=1,\bar{q}\notin\{k_1,k_2\}}^{N}{\bar{v}_{\bar{q},\{k_1,k_2\}}^{([s_{k_1}\;s_{k_2}])}}\Big],    
 \end{align}
 where
 \begin{align}\label{theorem3_proof_eq8}
 [\mu + \lambda - (s_{k_1} + s_{k_2})] \bar{v}_{k_1,\{0,k_2\}}^{([s_{k_1}\;s_{k_2}])} = \lambda_{k_1} \sum_{\bar{q}=0}^{N}{\bar{v}_{\bar{q},\{k_2\}}^{([s_{k_2}])}},
 \end{align}
  \begin{align}\label{theorem3_proof_eq9}
 [\mu + \lambda - (s_{k_1} + s_{k_2})] \bar{v}_{k_2,\{k_1,0\}}^{([s_{k_1}\;s_{k_2}])} = \lambda_{k_2} \sum_{\bar{q}=0}^{N}{\bar{v}_{\bar{q},\{k_1\}}^{([s_{k_1}])}},
 \end{align}
  \begin{align}\label{theorem3_proof_eq10}
 [\mu + \lambda - (s_{k_1} + s_{k_2})] \bar{v}_{\bar{q},\{k_1,k_2\}}^{([s_{k_1}\;s_{k_2}])} = \lambda_{\bar{q}}\sum_{j=0}^{N} {\bar{v}_{j,\{k_1,k_2\}}^{([s_{k_1}\;s_{k_2}])}}, 
 \end{align}
where $\bar{q} \in \{1,2,\cdots,N\}$. Thus, by summing the set of equations in (\ref{theorem3_proof_eq10}), $\sum_{j=0}^{N} {\bar{v}_{j,\{k_1,k_2\}}^{([s_{k_1}\;s_{k_2}])}}$ can be expressed as
\begin{align}\label{theorem3_proof_eq11}
\sum_{j=1}^{N} {\bar{v}_{j,\{k_1,k_2\}}^{([s_{k_1}\;s_{k_2}])}} = \dfrac{\lambda}{\mu - \left(s_{k_1} + s_{k_2}\right)} \bar{v}_{0,\{k_1,k_2\}}^{([s_{k_1}\;s_{k_2}])}. 
\end{align}

 Therefore, by substituting (\ref{theorem3_proof_eq11}) into (\ref{theorem3_proof_eq10}), $\bar{v}_{\bar{q},\{k_1,k_2\}}^{([s_{k_1}\;s_{k_2}])}$ can be expressed in terms of $\bar{v}_{\bar{0},\{k_1,k_2\}}^{([s_{k_1}\;s_{k_2}])}$ as
 \begin{align}\label{theorem3_proof_eq12}
 \bar{v}_{\bar{q},\{k_1,k_2\}}^{([s_{k_1}\;s_{k_2}])} = \dfrac{\lambda_{\bar{q}}}{\mu - \left(s_{k_1} + s_{k_2}\right)} \bar{v}_{0,\{k_1,k_2\}}^{([s_{k_1}\;s_{k_2}])},
 \end{align}
 where $\bar{q} \in 1:N$. Further, from (\ref{theorem3_proof_eq4}), (\ref{theorem3_proof_eq8}) and (\ref{theorem3_proof_eq9}), $\bar{v}_{k_1,\{0,k_2\}}^{([s_{k_1}\;s_{k_2}])}$ and $\bar{v}_{k_2,\{k_1,0\}}^{([s_{k_1}\;s_{k_2}])}$ can be respectively expressed as
 \begin{align}\label{theorem3_proof_eq13}
 \bar{v}_{k_1,\{0,k_2\}}^{([s_{k_1}\;s_{k_2}])} =  \dfrac{\lambda_{k_1}\left(\mu + \lambda - s_{k_2}\right)}{\left(\mu - s_{k_2}\right)\big[\mu + \lambda - \left(s_{k_1} + s_{k_2}\right)\big]} \bar{v}_{0,\{k_2\}}^{([s_{k_2}])},
 \end{align}
 \begin{align}\label{theorem3_proof_eq14}
  \bar{v}_{k_2,\{k_1,0\}}^{([s_{k_1}\;s_{k_2}])} = \dfrac{\lambda_{k_2}\left(\mu + \lambda - s_{k_1}\right)}{\left(\mu - s_{k_1}\right)\big[\mu + \lambda - \left(s_{k_1} + s_{k_2}\right)\big]}\bar{v}_{0,\{k_1\}}^{([s_{k_1}])}.
 \end{align}

Thus, $\bar{v}_{0,\{k_1,k_2\}}^{([s_{k_1}\;s_{k_2}])}$ can be rewritten as 
 \begin{align}\label{theorem3_proof_eq15}
 \bar{v}_{0,\{k_1,k_2\}}^{([s_{k_1}\;s_{k_2}])} \nonumber&\a \dfrac{\mu\left(\dfrac{\mu - (s_{k_1} + s_{k_2})}{\mu + \lambda - (s_{k_1} + s_{k_2})}\right)\left(\dfrac{\lambda_{k_1}\left(\mu + \lambda - s_{k_2}\right)}{\mu - s_{k_2}} \bar{v}_{0,\{k_2\}}^{([s_{k_2}])} + \dfrac{\lambda_{k_2}\left(\mu + \lambda - s_{k_1}\right)}{\mu - s_{k_1}} \bar{v}_{0,\{k_1\}}^{([s_{k_1}])}\right)}{\Big[[\lambda - (s_{k_1} + s_{k_2})][\mu - (s_{k_1} + s_{k_2})] - \mu \sum_{j=1,j\notin\{k_1,k_2\}}^{N}{\lambda_j}\Big]},  \\
 \nonumber&\b \dfrac{\mu^2 \lambda_{k_1} \lambda_{k_2} \big[\mu - \left(s_{k_1} + s_{k_2}\right)\big]}{\mu + \lambda - \left(s_{k_1} + s_{k_2}\right)} \left(\dfrac{1}{c_{\{k_1,k_2\}}c_{\{k_1\}}} + \dfrac{1}{c_{\{k_1,k_2\}}c_{\{k_2\}}}\right),\\
 &= \dfrac{\mu^2 \lambda_{k_1} \lambda_{k_2} \big[\mu - \left(s_{k_1} + s_{k_2}\right)\big]}{\mu + \lambda - \left(s_{k_1} + s_{k_2}\right)} \sum_{\nrmP \in \mathcal{P}(\{k_1,k_2\})}{\dfrac{1}{C(\nrmP)}},
 \end{align}
where step (a) follows from substituting (\ref{theorem3_proof_eq12})-(\ref{theorem3_proof_eq14}) into (\ref{theorem3_proof_eq7}), and step (b) follows from substituting $\bar{v}_{0,\{k_2\}}^{([s_{k_2}])}$ and $\bar{v}_{0,\{k_1\}}^{([s_{k_1}])}$ from (\ref{theorem3_proof_eq6}) along with adding $c_{\{k_1,k_2\}}$ based on the definition in (\ref{theorem:MGF_NP_eq2}). Following similar steps as in (\ref{theorem3_proof_eq7})-(\ref{theorem3_proof_eq15}), one can obtain $\bar{v}_{0,\{k_1,k_2,k_3\}}^{([s_{k_1}\;s_{k_2}\;s_{k_3}])}$ using $\bar{v}_{0,\{k_1,k_2\}}^{([s_{k_1}\;s_{k_2}])}$ as
 \begin{align}\label{theorem3_proof_eq16}
 \bar{v}_{0,\{k_1,k_2,k_3\}}^{([s_{k_1}\;s_{k_2}\;s_{k_3}])} = \dfrac{\mu^3 \lambda_{k_1} \lambda_{k_2} \lambda_{k_3}\big[\mu - \left(s_{k_1} + s_{k_2} + s_{k_3}\right)\big]}{\mu + \lambda - \left(s_{k_1} + s_{k_2} + s_{k_3}\right)} \sum_{\nrmP \in \mathcal{P}(\{k_1,k_2,k_3\})}{\dfrac{1}{C(\nrmP)}}.
 \end{align}

 By inspecting the expressions of $\bar{v}_{0,\{k_1\}}^{([s_{k_1}])}$, $\bar{v}_{0,\{k_1,k_2\}}^{([s_{k_1}\;s_{k_2}])}$ and $\bar{v}_{0,\{k_1,k_2,k_3\}}^{([s_{k_1}\;s_{k_2}\;s_{k_3}])}$  in (\ref{theorem3_proof_eq6}), (\ref{theorem3_proof_eq15}) and (\ref{theorem3_proof_eq16}), respectively, one can see that repeated application of (\ref{theorem 1_eq2}) gives $\bar{v}_{0,\nrmK}^{(\nbs)}$ for an arbitrary set $\nrmK \subseteq \{1,2,\cdots,N\}$ as
 \begin{align}\label{theorem3_proof_eq17}
\bar{v}^{(\nbs)}_{0,\nrmK} = \mu^{|\nrmK|} \left(\prod_{i=1}^{|\nrmK|}{\lambda_{\nrmK(i)}}\right)\left(\dfrac{\mu - \sum_{j=1}^{|\nrmK|}{s_{\nrmK(j)}}}{\mu + \lambda - \sum_{j=1}^{|\nrmK|}{s_{\nrmK(j)}}}\right) \sum_{\nrmP \in \mathcal{P}(\nrmK)}{\dfrac{1}{C(\nrmP)}}.
\end{align}

Thus, the stationary joint MGF of a set $\nrmK\subseteq \{1,2,\cdots,N\}$ of age processes is given by
\begin{align}\label{theorem3_proof_eq18}
\overset{{\rm PS} }{M} (\nbs) = \sum_{\bar{q} \in \ncalQ}{\bar{v}^{(\nbs)}_{\bar{q},\nrmK}} \a \left(1 + \dfrac{\lambda}{\mu - \sum_{j=1}^{|\nrmK|}{s_{\nrmK(j)}}}\right) \bar{v}_{0,\nrmK}^{(\nbs)} \b\mu^{|\nrmK|} \left(\prod_{i=1}^{|\nrmK|}{\lambda_{\nrmK(i)}}\right)\sum_{\nrmP \in \mathcal{P}(\nrmK)}{\dfrac{1}{C(\nrmP)}},
\end{align}
where step (a) follows from expressing $\bar{v}_{\bar{q},\nrmK}^{(\nbs)}$, $\bar{q} \in \mathcal{Q}/\{0\}$, as a function of $\bar{v}_{0,\nrmK}^{(\nbs)}$ using (\ref{theorem 1_eq2}), and step (b) follows from substituting $\bar{v}_{0,\nrmK}^{(\nbs)}$ from (\ref{theorem3_proof_eq17}). This completes the proof.
\hfill 
\IEEEQED

\subsection{Proof of Proposition~\ref{prop:MGF_PS_twosources}}\label{app:prop:MGF_PS_twosources}
%From Theorem \ref{theorem:MGF_PS}, the stationary joint MGF of the two AoI processes $x_{k_1}(t)$ and $x_{k_2}(t)$ can be obtained by setting $\nrmK = \{k_1,k_2\}$ as
%\begin{align}\label{proof_prop:MGF_PS_twosources_eq1}
%\overset{{\rm PS} }{M} (\bar{s}_{k_1},\bar{s}_{k_2})=  \dfrac{\rho_{k_1} \rho_{k_2}}{\big[\rho - (\bar{s}_{k_1} + \bar{s}_{k_2})\big] \big[1 - (\bar{s}_{k_1} + \bar{s}_{k_2})\big] - \rho_{-\{k_1,k_2\}}} \sum_{i\in\{k_1,k_2\}}{\dfrac{1}{(1 - \bar{s}_i)(\rho - \bar{s}_i) - \rho_{-i}}}.
%\end{align}
%
From Corollary \ref{cor:PS_twosources}, $\nbbE[x_{k_1}x_{k_2}]$ can be evaluated as
\begin{align}\label{proof_prop:MGF_PS_twosources_eq2}
\nbbE[x_{k_1}x_{k_2}] = \frac{\partial^{2}\big[\overset{\rm PS}{M}(\bar{s}_{k_1},\bar{s}_{k_2})\big]}{\mu^{2}\partial\bar{s}_{k_1} \partial\bar{s}_{k_2}} \Big|_{\bar{s}_{k_1}=0, \bar{s}_{k_2}=0} = \dfrac{(1+\rho)^2(\rho_{k_1} + \rho_{k_2}) - 2\rho_{k_1}\rho_{k_2}}{\mu^2\rho_{k_1}\rho_{k_2}(\rho_{k_1}+\rho_{k_2})}.
\end{align}

Further, from Corollary \ref{cor:PS_marginal}, $\nbbE[x_{k}]$ and $\nbbE[x_{k}^2]$ can be respectively evaluated as
\begin{align}\label{proof_prop:MGF_PS_twosources_eq3}
\nbbE[x_{k}] = \frac{{\rm d}\big[\overset{\rm PS}{M}(\bar{s}_{k})\big]}{\mu \;{\rm d}\bar{s}_{k}} \Big|_{\bar{s}_{k}=0} = \dfrac{1+\rho}{\mu \rho_k},
\end{align}
\begin{align}\label{proof_prop:MGF_PS_twosources_eq4}
\nbbE[x_{k}^2] = \frac{{\rm d}^2\big[\overset{\rm PS}{M}(\bar{s}_{k})\big]}{\mu^2 \;{\rm d}\bar{s}_{k}^2} \Big|_{\bar{s}_{k}=0} = \dfrac{2\big[(1+\rho)^2 - \rho_k\big]}{\mu^2 \rho_k^2},
\end{align}
where $k \in 1:N$. The final expression of $\overset{\rm PS}{\rm Cor}$ in (\ref{prop:MGF_PS_twosources_eq1}) can be derived by substituting (\ref{proof_prop:MGF_PS_twosources_eq2})-(\ref{proof_prop:MGF_PS_twosources_eq4}) into (\ref{proof_prop:MGF_NP_twosources_eq2}), followed by some algebraic simplifications. 
\hfill 
\IEEEQED

\subsection{Proof of Theorem~\ref{theorem:MGF_SA}} \label{app:theorem:MGF_SA}
 Using the set of transitions in Table \ref{table:SA} and (\ref{theorem 1_eq2}) in Theorem \ref{theorem 1}, $\bar{v}^{([s_{k_1}])}_{0,\{k_1\}}$ can be expressed as 
\begin{align}\label{theorem4_proof_eq1}
\left(\lambda - s_{k_1}\right) \bar{v}_{0,\{k_1\}}^{([s_{k_1}])} = \mu \left(\bar{v}_{k_1,\{0\}}^{([s_{k_1}])} + \sum_{\bar{q}=1,\bar{q}\notin\{k_1\}}^{N}{\bar{v}_{\bar{q},\{k_1\}}^{([s_{k_1}])}}\right),
\end{align}
where $\bar{v}_{k_1,\{0\}}^{([s_{k_1}])}$ and $\bar{v}_{\bar{q},\{k_1\}}^{([s_{k_1}])}$ are given by
\begin{align}\label{theorem4_proof_eq2}
\left(\mu + \lambda_{k_1} - s_{k_1}\right) \bar{v}_{k_1,\{0\}}^{([s_{k_1}])} = \lambda_{k_1} \left(\bar{v}^{(0)}_{0} + \bar{v}^{(0)}_{k_1}\right ) \a \dfrac{\lambda_{k_1} + \mu}{\mu} \bar{v}^{(0)}_{0},  
\end{align}
\begin{align}\label{theorem4_proof_eq3}
 \left(\mu + \lambda_{\bar{q}} - s_{k_1}\right)\bar{v}_{\bar{q},\{k_1\}}^{([s_{k_1}])} = \lambda_{\bar{q}} \left(\bar{v}_{0,\{k_1\}}^{([s_{k_1}])} + \bar{v}_{\bar{q},\{k_1\}}^{([s_{k_1}])}\right), 
\end{align}
 where $\bar{q} \in 1:N$ and step (a) in (\ref{theorem4_proof_eq2}) follows from the fact that $\bar{v}^{(0)}_{k_1} = \frac{\lambda_{k_1}}{\mu}\bar{v}^{(0)}_{0}$. Substituting (\ref{theorem4_proof_eq2}) and (\ref{theorem4_proof_eq3}) into (\ref{theorem4_proof_eq1}), we get
 \begin{align}\label{theorem4_proof_eq4}
\bar{v}_{0,\{k_1\}}^{([s_{k_1}])} \nonumber&= \dfrac{\mu \lambda_{k_1} \bar{v}^{(0)}_{0}\left(\mu - s_{k_1}\right)}{(\lambda - s_{k_1})(\mu - s_{k_1}) - \mu \sum_{j=1, j\notin\{k_1\}}^{N}{\lambda_j}} \times \dfrac{\frac{\lambda_{k_1} + \mu}{\mu}}{\mu + \lambda_{k_1} - s_{k_1}},\\
&=\dfrac{\mu \lambda_{k_1} \bar{v}^{(0)}_{0}\left(\mu - s_{k_1}\right)}{c_{\{k_1\}}} \times \dfrac{\frac{\lambda_{k_1} + \mu}{\mu}}{\mu + \lambda_{k_1} - s_{k_1}},   
 \end{align}
 where $k_1 \in 1:N$. Now, using (\ref{theorem4_proof_eq4}), one can evaluate $\bar{v}_{0,\{k_1,k_2\}}^{([s_{k_1}\;s_{k_2}])}$. In particular, from (\ref{theorem 1_eq2}), $\bar{v}_{0,\{k_1,k_2\}}^{([s_{k_1}\;s_{k_2}])}$ can be expressed as 
 \begin{align}\label{theorem4_proof_eq5}
[\lambda - (s_{k_1} + s_{k_2})]\bar{v}_{0,\{k_1,k_2\}}^{([s_{k_1}\;s_{k_2}])} = \mu \Big[\bar{v}_{k_1,\{0,k_2\}}^{([s_{k_1}\;s_{k_2}])} + \bar{v}_{k_2,\{k_1,0\}}^{([s_{k_1}\;s_{k_2}])} + \sum_{\bar{q}=1,\bar{q}\notin\{k_1,k_2\}}^{N}{\bar{v}_{\bar{q},\{k_1,k_2\}}^{([s_{k_1}\;s_{k_2}])}}\Big],    
 \end{align}
 where
 \begin{align}\label{theorem4_proof_eq6}
 [\mu + \lambda_{k_1} - (s_{k_1} + s_{k_2})] \bar{v}_{k_1,\{0,k_2\}}^{([s_{k_1}\;s_{k_2}])} = \lambda_{k_1} \left(\bar{v}_{0,\{k_2\}}^{([s_{k_2}])} + \bar{v}_{k_1,\{k_2\}}^{([s_{k_2}])}\right) \a \lambda_{k_1} \left(1 + \dfrac{\lambda_{k_1}}{\mu - s_{k_2}}\right)\bar{v}_{0,\{k_2\}}^{([s_{k_2}])},
 \end{align}
  \begin{align}\label{theorem4_proof_eq7}
 [\mu + \lambda_{k_2} - (s_{k_1} + s_{k_2})] \bar{v}_{k_2,\{k_1,0\}}^{([s_{k_1}\;s_{k_2}])} = \lambda_{k_2} \left(\bar{v}_{0,\{k_1\}}^{([s_{k_1}])} + \bar{v}_{k_2,\{k_1\}}^{([s_{k_1}])}\right) \a \lambda_{k_2} \left(1 + \dfrac{\lambda_{k_2}}{\mu - s_{k_1}}\right)\bar{v}_{0,\{k_1\}}^{([s_{k_1}])},
 \end{align}
  \begin{align}\label{theorem4_proof_eq8}
 [\mu + \lambda_{\bar{q}} -(s_{k_1} + s_{k_2})] \bar{v}_{\bar{q},\{k_1,k_2\}}^{([s_{k_1}\;s_{k_2}])} = \lambda_{\bar{q}} \left(\bar{v}_{0,\{k_1,k_2\}}^{([s_{k_1}\;s_{k_2}])} + \bar{v}_{\bar{q},\{k_1,k_2\}}^{([s_{k_1}\;s_{k_2}])}\right), 
 \end{align}
where $\bar{q} \in \{1,2,\cdots,N\}$ and step (a) in (\ref{theorem4_proof_eq6}) and (\ref{theorem4_proof_eq7}) follows from substituting $\bar{v}_{k_1,\{k_2\}}^{([s_{k_2}])}$ and $\bar{v}_{k_2,\{k_1\}}^{([s_{k_1}])}$ from (\ref{theorem4_proof_eq3}). Thus, $\bar{v}_{0,\{k_1,k_2\}}^{([s_{k_1}\;s_{k_2}])}$ can be rewritten as 
 \begin{align}\label{theorem4_proof_eq9}
 \nonumber&\bar{v}_{0,\{k_1,k_2\}}^{([s_{k_1}\;s_{k_2}])} \a \dfrac{\mu\big[\mu - \left(s_{k_1} + s_{k_2}\right)\big]\left(\dfrac{\lambda_{k_1}\left(1 + \frac{\lambda_{k_1}}{\mu - s_{k_2}}\right)}{\big[\mu + \lambda_{k_1} - \left(s_{k_1} + s_{k_2}\right)\big]} \bar{v}_{0,\{k_2\}}^{([s_{k_2}])} + \dfrac{\lambda_{k_2}\left(1 + \frac{\lambda_{k_2}}{\mu - s_{k_1}}\right)}{\big[\mu + \lambda_{k_2} - \left(s_{k_1} + s_{k_2}\right)\big]} \bar{v}_{0,\{k_1\}}^{([s_{k_1}])}\right)}{\Big[[\lambda - (s_{k_1} + s_{k_2})][\mu - (s_{k_1} + s_{k_2})] - \mu \sum_{j=1,j\notin\{k_1,k_2\}}^{N}{\lambda_j}\Big]},  \\
 \nonumber&\b \dfrac{\mu^2 \lambda_1 \lambda_2 \bar{v}^{(0)}_{0} \big[\mu - \left(s_{k_1} + s_{k_2}\right)\big]}{c_{\{k_1,k_2\}}}\times \Bigg[\frac{1}{c_{\{k_1\}}} \times \dfrac{\frac{\lambda_{k_1} + \mu}{\mu} \left(\mu - \lambda_{k_2} - s_{k_1}\right)} {\left(\mu + \lambda_{k_1} - s_{k_1}\right)\big[\mu + \lambda_{k_2} - \left(s_{k_1} + s_{k_2}\right)\big]} \\
 \nonumber&+ \frac{1}{c_{\{k_2\}}} \times \dfrac{\frac{\lambda_{k_2} + \mu}{\mu} \left(\mu - \lambda_{k_1} - s_{k_2}\right)} {\left(\mu + \lambda_{k_2} - s_{k_2}\right)\big[\mu + \lambda_{k_1} - \left(s_{k_1} + s_{k_2}\right)\big]} \Bigg],\\
 & \c \mu^2 \lambda_{k_1}\lambda_{k_2}\bar{v}^{(0)}_{0} \big[\mu - \left(s_{k_1} + s_{k_2}\right)\big] \sum_{\nrmP \in \mathcal{P}(\{k_1,k_2\})}{\dfrac{C'(\nrmP)}{C(\nrmP)}},
 \end{align}
where step (a) follows from substituting (\ref{theorem4_proof_eq6})-(\ref{theorem4_proof_eq8}) into (\ref{theorem4_proof_eq5}), step (b) follows from substituting $\bar{v}_{0,\{k_2\}}^{([s_{k_2}])}$ and $\bar{v}_{0,\{k_1\}}^{([s_{k_1}])}$ from (\ref{theorem4_proof_eq4}) along with the definition of $c_{\{k_1,k_2\}}$ in (\ref{theorem:MGF_NP_eq2}), and step (c) follows from defining $C'(\nrmP)$ in (\ref{theorem:MGF_SA_eq2}) as 
\begin{align*}
C'(\nrmP) = \frac{\lambda_{\nrmP(|\nrmP|)} + \mu}{\mu} \times \frac{1}{\mu + \lambda_{\nrmP(|\nrmP|)} - s_{\nrmP(|\nrmP|)}} \times \prod_{i=1}^{|\nrmP|-1}{\dfrac{\mu + \lambda_{\nrmP(i)} - \sum_{j=i+1}^{|\nrmP|}{s_{\nrmP(j)}}}{\mu + \lambda_{\nrmP(i)} - \sum_{j=i}^{|\nrmP|}{s_{\nrmP(j)}}}}.
\end{align*}

 Following similar steps as in (\ref{theorem4_proof_eq5})-(\ref{theorem4_proof_eq9}), one can obtain $\bar{v}_{0,\{k_1,k_2,k_3\}}^{([s_{k_1}\;s_{k_2}\;s_{k_3}])}$ using $\bar{v}_{0,\{k_1,k_2\}}^{([s_{k_1}\;s_{k_2}])}$ as
 \begin{align}\label{theorem4_proof_eq10}
 \bar{v}_{0,\{k_1,k_2,k_3\}}^{([s_{k_1}\;s_{k_2}\;s_{k_3}])} = \mu^3 \lambda_{k_1}\lambda_{k_2} \lambda_{k_3}\bar{v}^{(0)}_{0} \big[\mu - \left(s_{k_1} + s_{k_2} + s_{k_3}\right)\big]\sum_{\nrmP \in \mathcal{P}(\{k_1,k_2,k_3\})}{\dfrac{C'(\nrmP)}{C(\nrmP)}}.
 \end{align}

 By inspecting the expressions of $\bar{v}_{0,\{k_1\}}^{([s_{k_1}])}$, $\bar{v}_{0,\{k_1,k_2\}}^{([s_{k_1}\;s_{k_2}])}$ and $\bar{v}_{0,\{k_1,k_2,k_3\}}^{([s_{k_1}\;s_{k_2}\;s_{k_3}])}$  in (\ref{theorem4_proof_eq4}), (\ref{theorem4_proof_eq9}) and (\ref{theorem4_proof_eq10}), respectively, one can see that repeated application of (\ref{theorem 1_eq2}) gives $\bar{v}_{0,\nrmK}^{(\nbs)}$ for an arbitrary set $\nrmK \subseteq \{1,2,\cdots,N\}$ as
 \begin{align}\label{theorem4_proof_eq11}
\bar{v}^{(\nbs)}_{0,\nrmK} = \mu^{|\nrmK|} \left(\prod_{i=1}^{|\nrmK|}{\lambda_{\nrmK(i)}}\right)\bar{v}^{(0)}_{0}\left(\mu - \sum_{j=1}^{|\nrmK|}{s_{\nrmK(j)}}\right) \sum_{\nrmP \in \mathcal{P}(\nrmK)}{\dfrac{C'(\nrmP)}{C(\nrmP)}}.
\end{align}

Thus, the stationary joint MGF of a set $\nrmK\subseteq \{1,2,\cdots,N\}$ of age processes is given by
\begin{align}\label{theorem4_proof_eq12}
\overset{{\rm SA} }{M} (\nbs) = \sum_{\bar{q} \in \ncalQ}{\bar{v}^{(\nbs)}_{\bar{q},\nrmK}} \nonumber&\a \left(1 + \dfrac{\lambda}{\mu - \sum_{j=1}^{|\nrmK|}{s_{\nrmK(j)}}}\right) \bar{v}_{0,\nrmK}^{(\nbs)},\\ &\b\mu^{|\nrmK|} \left(\prod_{i=1}^{|\nrmK|}{\lambda_{\nrmK(i)}}\right)\left(\dfrac{\mu}{\lambda + \mu}\right)\left(\lambda + \mu - \sum_{j=1}^{|\nrmK|}{s_{\nrmK(j)}}\right) \sum_{\nrmP \in \mathcal{P}(\nrmK)}{\dfrac{C'(\nrmP)}{C(\nrmP)}},
\end{align}
where step (a) follows from expressing $\bar{v}_{\bar{q},\nrmK}^{(\nbs)}$, $\bar{q} \in \mathcal{Q}/\{0\}$, as a function of $\bar{v}_{0,\nrmK}^{(\nbs)}$ using (\ref{theorem 1_eq2}), and step (b) follows from substituting $\bar{v}_{0,\nrmK}^{(\nbs)}$ from (\ref{theorem4_proof_eq11}) along with the fact that $\bar{v}^{(0)}_{0} = \frac{\mu}{\lambda + \mu}$. This completes the proof.
\hfill 
\IEEEQED

\subsection{Proof of Proposition~\ref{prop:MGF_SA_twosources}}\label{app:prop:MGF_SA_twosources}
%From Theorem \ref{theorem:MGF_SA}, the stationary joint MGF of the two AoI processes $x_{k_1}(t)$ and $x_{k_2}(t)$ can be obtained by setting $\nrmK = \{k_1,k_2\}$ as
%\begin{align}\label{proof_prop:MGF_SA_twosources_eq1}
%\overset{{\rm SA} }{M} (\bar{s}_1,\bar{s}_2)= \nonumber& \dfrac{\rho_{k_1} \rho_{k_2}\big[1 + \rho - (\bar{s}_{k_1} + \bar{s}_{k_2})\big]}{(1 + \rho)\Big[\big[\rho - (\bar{s}_{k_1} + \bar{s}_{k_2})\big]\big[1 - (\bar{s}_{k_1} + \bar{s}_{k_2})\big] - \rho_{-\{k_1,k_2\}}\Big]} \\&\times \sum_{i \in \{k_1,k_2\}} \dfrac{(1 + \rho_i)(1 + \rho_{-i} - \bar{s}_i)}{(1 + \rho_i - \bar{s}_i)\big[1 + \rho_{-i} - (\bar{s}_{k_1} + \bar{s}_{k_2})\big]\big[(1 - \bar{s}_i)(\rho - \bar{s}_i) - \rho_{-i}\big]}.
%\end{align}
%
From Corollary \ref{cor:SA_twosources}, $\nbbE[x_{k_1}x_{k_2}]$ can be evaluated as
\begin{align}\label{proof_prop:MGF_SA_twosources_eq2}
\nbbE[x_{k_1}x_{k_2}] = \frac{\partial^{2}\big[\overset{\rm SA}{M}(\bar{s}_{k_1},\bar{s}_{k_2})\big]}{\mu^{2}\partial\bar{s}_{k_1} \partial\bar{s}_{k_2}} \Big|_{\bar{s}_{k_1}=0, \bar{s}_{k_2}=0} = \dfrac{\sum_{j=0}^3{\alpha_j (\rho_{k_1}\rho_{k_2})^j}}{\mu^2\rho_{k_1}\rho_{k_2}(1 + \rho_{k_1})^2 (1 + \rho_{k_2})^2 (1 + \rho) (\rho_{k_1}+\rho_{k_2})},
\end{align}
\begin{align*}
&\alpha_3 = -2(1+\rho), \\&\alpha_2 =  (\rho_{k_1} + \rho_{k_2}) \Big[-(2+\rho)(\rho_{k_1} + \rho_{k_2}) + (\rho^3 + 5\rho^2 + 5\rho - 1)\Big], \\&\alpha_1 =   -(2\rho+3)(\rho_{k_1}+\rho_{k_2})^3 + (\rho_{k_1} + \rho_{k_2})^2(2\rho^3 + 9\rho^2 + 9\rho -1) + 2(\rho_{k_1} + \rho_{k_2})\rho(\rho+2)^2 - 2(1+\rho), \\&\alpha_0 = (1+\rho) (\rho_{k_1} + \rho_{k_2})(1 + \rho_{k_1} + \rho_{k_2})\Big[-(\rho_{k_1}+\rho_{k_2})^2 + (\rho_{k_1}+\rho_{k_2})\big[(1+\rho)^2 + \rho\big] + (1+\rho)^2\Big].
\end{align*}

Further, from Corollary \ref{cor:SA_marginal}, $\nbbE[x_{k}]$ and $\nbbE[x_{k}^2]$ can be respectively evaluated as
\begin{align}\label{proof_prop:MGF_SA_twosources_eq3}
\nbbE[x_{k}] = \frac{{\rm d}\big[\overset{\rm SA}{M}(\bar{s}_{k})\big]}{\mu \;{\rm d}\bar{s}_{k}} \Big|_{\bar{s}_{k}=0} = \dfrac{(1+\rho)^2(1+\rho_k) + \rho_k \rho_{-k}}{\mu \rho_k (1 + \rho_k)(1+\rho)},
\end{align}
\begin{align}\label{proof_prop:MGF_SA_twosources_eq4}
\nonumber&\nbbE[x_{k}^2] = \frac{{\rm d}^2\big[\overset{\rm SA}{M}(\bar{s}_{k})\big]}{\mu^2 \;{\rm d}\bar{s}_{k}^2} \Big|_{\bar{s}_{k}=0}, \\&= \dfrac{2\Big[-\rho_k^3(3+2\rho) + \rho_k^2(\rho^3+4\rho^2+2\rho-2) + \rho_k(1+\rho)(2\rho^2+5\rho+1) + (1+\rho)^3\Big]}{\mu^2\rho_k^2(1+\rho_k)^2(1+\rho)},
\end{align}
where $k \in 1:N$. The final expression of $\overset{\rm SA}{\rm Cor}$ in (\ref{prop:MGF_SA_twosources_eq1}) can be derived by substituting (\ref{proof_prop:MGF_SA_twosources_eq2})-(\ref{proof_prop:MGF_SA_twosources_eq4}) into (\ref{proof_prop:MGF_NP_twosources_eq2}), followed by some algebraic simplifications. 
\hfill 
\IEEEQED

\bibliographystyle{WiOpt_2022.bbl}
\bibliography{ref}
\end{document}